\documentclass[aps,physrev,twocolumn,unsortedaddress]{revtex4-2}
\usepackage{graphicx} 
\usepackage{booktabs} 
\usepackage{amsmath} 
\usepackage{siunitx}
\usepackage{soul}
\usepackage{multirow}
\usepackage{float} 
\usepackage{xurl}
\usepackage{hyperref}
\usepackage[normalem]{ulem}
\hypersetup{
    colorlinks=true,
    linkcolor=black,
    citecolor=black,
    filecolor=black,
    urlcolor=black
}

\usepackage{pdfpages}

\makeatletter
\AtBeginDocument{\let\LS@rot\@undefined}
\makeatother

\begin{document}

\title{Atomic-level study of dose-dependent lattice thermal conductivity degradation in irradiated tungsten}

\author{Mary K. Chessey}
\author{José Julio Gutiérrez Moreno}
\email{julio.gutierrez@bsc.es}
\affiliation{Barcelona Supercomputing Center (BSC), Barcelona, Spain}
\author{Sajjan Singh Mehta}
\affiliation{oloBion SL, Parc Científic de Barcelona, Barcelona, Spain}
\author{Paolo Settembri}
\author{Arda Erbasan}
\author{Dominik Freinberger}
\author{Marina García Romero}
\affiliation{Barcelona Supercomputing Center (BSC), Barcelona, Spain}
\author{Mervi J. Mantsinen}
\affiliation{Barcelona Supercomputing Center (BSC), Barcelona, Spain}
\affiliation{ICREA, Barcelona, Spain}

\begin{abstract}
Plasma-facing components (PFCs) in future fusion reactors must withstand extreme environments, making tungsten (W) a prime candidate material due to its high thermal conductivity, high melting temperature, low sputtering yield, low tritium absorption, and other desirable characteristics. However, neutron irradiation introduces complex nanostructural defects that severely degrade these critical properties. In this work, we use Non-Equilibrium Molecular Dynamics (NEMD) simulations coupled with the Creation Relaxation Algorithm (CRA) to investigate the lattice thermal conductivity ($\kappa$) of W across a wide range of irradiation doses. Systems treated with different doses up to 1.8 canonical displacements per atom (cdpa) concentrated in a target region of 250k atoms, and larger systems treated with doses up to 0.4 cdpa in a target region of 2M atoms, were simulated to generate and study diverse defect structures, including vacancies and self-interstitial atom (SIA) clusters. Although the phononic contribution to $\kappa$ is typically negligible in most metals, it represents a substantial fraction in W, being rather sensitive to defect-induced scattering. By correlating spatially-resolved temperature gradients with local defect concentrations, we show that even small changes in defect type and distribution significantly affect thermal transport. We employ Machine Learning (ML) models to develop a quantitative relationship between the defect features and the resulting thermal conductivity degradation. The ML analysis identifies the density of vacancies as the dominant structural factor controlling thermal transport. We found a significant, nonlinear suppression of $\kappa$ with increasing defect density. Predictive multi-defect mapping built upon our ML model resolved a step-like degradation transition at a critical vacancy threshold of $0.05\ \mathrm{at.\%}$, beyond which phonon scattering completely saturated and capped the maximum relative lattice thermal conductivity degradation at approximately $30\%$. This integrated simulation and ML approach provides a powerful, predictive framework for quantifying the complex relationship between structure and thermal conductivity in irradiated fusion materials.
\end{abstract}

\maketitle

\section{Introduction}
\label{intro}

Fusion energy research and development have the potential to provide clean, safe, and abundant power, contribute to climate change mitigation strategies, and increase global energy reliability \cite{Mohamed2024Global,Nicholas2021Re-examining,Lerede2023Analysis}.
The need for plasma-facing component (PFC) materials that have high thermal conductivity, high melting temperatures, low sputtering yield, low tritium absorption, and other desirable characteristics for the harsh environment near the fusion plasma is an active and pressing area of investigation \cite{knaster_materials_2016}.
Tungsten (W) has been chosen as a plasma-facing armor material for projects such as ITER (world’s largest experimental tokamak, designed to prove the feasibility of fusion energy), and is a key material for fusion research \cite{you_european_2016, hirai_use_2016, rieth_review_2011}.
Therefore, a deeper understanding of the electronic structure and fundamental properties of W is continuously pursued to optimize its performance in various demanding technological applications \cite{kalha2021lifetimeW}.

Neutron irradiation of PFC, such as the W strike plates on divertors in tokamaks, is known to cause defects associated with alterations to material properties like thermal conductivity, affecting the performance of the component \cite{abernethy_predicting_2017, hasegawa2014neutron, reali2024finite, mason_estimate_2021}.
PFC will degrade during the operation of future fusion power plants, but understanding exactly how and when these defects will render PFCs unable to perform their basic functions remains a matter of urgent investigation \cite{eurofusion_european_2018}. To address this question and advance towards the next generation of reactors, experimental and computational approaches complement each other; computational methods provide insights where experimental data are limited or impossible to collect. Also, simulations guide mission parameters for experiment, such as the IFMIF-DONES and IFMIF (International Fusion Materials Irradiation Facility) programs, while reducing time, resources and costs relative to experiment alone \cite{eurofusion_european_2018,zsakai2019ifmif,bernardi2022ifmif}. This acceleration will be further enabled by next-generation high-performance computing (HPC), which will allow for more accurate and scalable atomistic modeling of key fusion materials \cite{gutierrez-moreno-exascale-HPC-materials, garcia-mantsinen-exascale-fusion-2025, lasa2024development}.

Classical molecular dynamics (MD) is a well-established atomic-scale computational approach to study the properties and behavior of fusion materials in the vicinity of extended defect structures. For example, complex material behaviors in extreme environments, such as evolution of W grain boundaries caused by helium (He) bubble growth, have been discovered and documented using classical MD \cite{hatton2024he}.
MD has been widely used to study thermal conductivity ($\kappa$) of fusion materials because efficient heat transfer by PFC to cooling mechanisms of the reactor prevents overheating of the plasma-facing surface, extending the lifetime of the component. 

One category of ways to estimate $\kappa$ computationally is by non-equilibrium molecular dynamics (NEMD). NEMD has been used to examine models of defects in W in the presence of a temperature gradient, including defective structures such as grain boundaries \cite{fu_calculation_2012, ding_simulation_2021, fu_effect_2013}, spherical voids and He bubbles \cite{hu_thermal_2017, ding_simulation_2021, zhang_study_2020}, and point defects such as self-interstitial atoms (SIA) and He atoms \cite{martinez_thermal_2021}. While there are also previous studies that estimated $\kappa$ of such structures using the Green-Kubo EMD method \cite{mason_estimate_2021,zhang2023}, NEMD allows for the direct observation of local thermal resistance across heterogeneous defect structures, like vacancies and SIAs. Moreover, NEMD also offers a computational advantage, as the larger system sizes required for NEMD are more efficiently parallelized than the extremely long trajectories needed to converge heat flux autocorrelation functions that are required for the Green-Kubo method.

Using MD, the estimated $\kappa$ represents the phononic, or lattice, contribution to the overall thermal conductivity of the material, which is the sum of lattice and electronic terms. In metals, the lattice contribution is typically much smaller than the electronic contribution, but this ratio depends on factors such as temperature, pressure, and presence of defects.\cite{thermoreview} Further, in W, the lattice contribution to $\kappa$ is likely the largest of all metals \cite{chen2019understanding}, estimated to range from 10\% \cite{hu_thermal_2017,fu_calculation_2012,zhang_study_2020} to above 30\% \cite{chen2019understanding,bhatt_transition_2023,zhang_pressure_2022} of the total value of $\kappa$ at ambient temperature and pressure. Because the characteristic wavelengths of heat-carrying phonons are highly sensitive to lattice periodicity, lattice thermal conductivity undergoes more significant relative degradation than the electronic component when the crystal symmetry is disrupted by defect-induced scattering.
Crucially, as established by Mason et al. \cite{mason_estimate_2021}, the degradation pathways for the lattice and electronic components in irradiated metals are intrinsically coupled. Both transport mechanisms are governed by highly correlated scattering cross-sections that scale concurrently with the accumulation of nano- and microstructural defects. Therefore, while models often approximate metallic thermal transport as entirely electronic, a rigorous atomistic assessment of the lattice component is vital. Investigating lattice degradation not only sheds light on an often-overlooked fraction of the conductivity, but also serves as a useful proxy for the broader degradation trends of the entire system. Estimation of the degradation of lattice $\kappa$ due to scattering of heat-conducting phonons in the presence of defects in the crystal structure is therefore an active area of investigation for fusion materials.

Simulation of defects that are characteristic of neutron irradiation of W has been achieved using a variety of approaches, including the Creation Relaxation Algorithm (CRA) \cite{derlet_microscopic_2020}. CRA was developed to generate extended defect structures that approximate realistic high-dose damage structures well, while dramatically reducing the computational cost relative to full cascade simulations alone \cite{mason_parameter-free_2021}. CRA has been shown to reproduce overall defect evolution and clustering behavior of SIA similar to the more expensive simulations, but vacancy clustering is not well approximated by CRA without subsequent cascade annealing \cite{granberg_effect_2023,wu2023simulation}. Nevertheless, the algorithm effectively captures the structural disorder and varied local atomic environments characteristic of a prototypical irradiated metal, providing a rich topological landscape to unveil the underlying structure-property relationships and identify factors governing the reduction of $\kappa$ in these complex structures. CRA-generated final relaxed microstructures have been structurally validated to converge to the same high-dose steady states produced by computationally intensive, overlapping MD collision cascades \cite{mason_parameter-free_2021}, which have been shown to directly replicate the nanoscale strain fields and dislocation environments observed in experimental TEM micrographs of irradiated tungsten \cite{Mason2024_ActaMat_TEM}.

In this work, we present a computational framework combining the simplicity offered by CRA to generate high-dose defect structures in irradiated W, with large-scale NEMD simulations, coupled with machine learning (ML) analysis to establish a novel, predictive relationship between defect characteristics and the degradation of lattice thermal conductivity. The structure of the paper continues with descriptions of the methods for defect creation, estimation of $\kappa$, identification of resultant defect concentrations, and data modeling. Results are then presented about the behavior and character of defect structures that were achieved, and ML model outcomes with insights regarding the effect of each feature on predictions of lattice thermal conductivity.

\section{Methods}
\label{methods}

\begin{figure}
    \centering
    \includegraphics[width=1.0\linewidth]{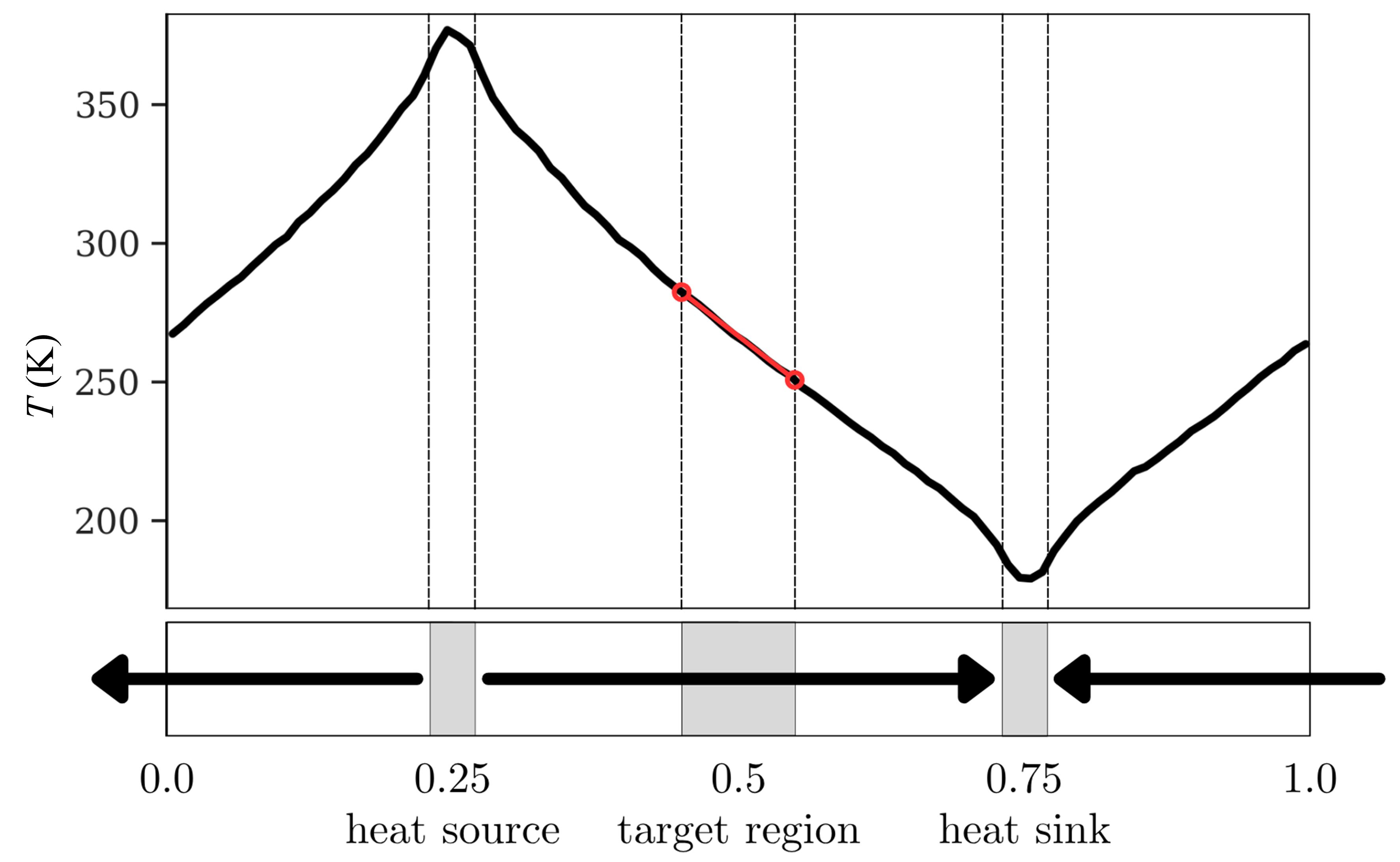}
    \caption{NEMD ``direct" method for estimation of $\kappa$. Arrows indicate the direction of heat flow ($\pm x$). Position is given as fraction of box length. An example temperature profile is plotted for a small system without defects.}
    \label{fig:sim-cell-with-temp}
\end{figure}

The NEMD ``direct" method was implemented to estimate $\kappa$ of defective body-centered cubic (BCC) single crystals of W, using the classical MD code LAMMPS, version 29 Oct 2020 \cite{thompson_lammps_2022}. Fig. \ref{fig:sim-cell-with-temp} shows a diagram of the simulation setup, which includes heat source and heat sink regions surrounding a central target region, evenly spaced along the simulation cell in the $x$-dimension. This arrangement generates the needed temperature gradient through the target region as a result of the applied heat flux, which allows for the estimation of $\kappa$ of the material within the target region. Simulations were carried out on the JFRS-1 supercomputer.

Thirty-two ``small" systems with target regions of 250 thousand atoms, and 7 ``large" systems with target regions of 2M atoms, were prepared. The cube-shaped target region of each system was positioned at the center of the long box, so that the target region occupied a fraction of the length in $x$ (parallel to the [1 0 0] crystal axis), and the full width and depth in $y$ and $z$. 3D periodic boundary conditions were applied. For small systems, total box dimensions were 500 × 50 × 50 conventional cubic BCC unit cells (1582.6 {\AA} × 158.26 {\AA} × 158.26 {\AA}), for a total of 2.5M atoms, including only $10\%$ in the target region. Large systems were prepared with 2000 x 100 x 100 unit cells (6337.6 {\AA} × 316.88 {\AA} × 316.88 {\AA}), for a total of 40M atoms, $5\%$ of which were in the target region. The use of extremely large system sizes was required not just to accommodate realistic extended defects and their associated strains, but also to minimize errors arising from finite-size effects in the temperature profile. In fact, in smaller cells, the proximity of the heat source and sink can lead to nonlinearities and temperature jumps at the boundaries.

The atomic interactions of W were described by the semi-empirical, Embedded Atom Model (EAM) interatomic potential developed by Mason et al. \cite{mason_empirical_2017}. This potential was optimized to reproduce the complex energetics of vacancy clusters, the dominant scattering centers for heat carriers in our system, while correcting for the surface energy underestimations common in earlier EAM models. By using a $7^{\text{th}}$ order polynomial cutoff, the potential ensures the analytical smoothness of force derivatives, which is essential for stable lattice dynamics and the statistical convergence of transport coefficients in long-duration MD. Since our methodology focused on the thermal transport of pre-existing defect structures rather than the short-range many-body dynamics of collision cascades, this potential provides a more computationally tractable and physically targeted alternative to high-dimensional machine-learning interatomic potentials (MLIPs). In fact, the superior HPC parallel efficiency of EAM potentials ensured the feasibility of our study, as the extensive sampling required to converge thermal gradients would be prohibitively expensive using MLIPs.

\subsection{Defect creation}
\label{method-defect-creation}

In the static model (at 0 K), defects were introduced using the Creation Relaxation Algorithm (CRA), entailing creation of defects by insertion of Frenkel pairs (FP) followed by stress-driven relaxation under isotropic zero pressure  \cite{derlet_microscopic_2020,mason_observation_2020,mason_parameter-free_2021,granberg_effect_2023,granberg_molecular_2021,ma_athermal_2023}.
CRA was carried out within a target region of 250k lattice atoms in small systems, and of 2M atoms in large systems. Target regions were centred at $x=0.5L\textsubscript{x}$, where $L\textsubscript{x}$ was the length of the box in the $x$-direction.

By CRA, one FP was inserted in the target region by selecting an atom at random, then changing the position of the selected atom to a new location chosen at random, also within the target region. This process generated one self-interstitial atom (SIA) in the selected atom's new location, and left behind one point vacancy (Vac) in the original lattice site of the selected atom. In the small systems, 1000 FPs were inserted in the target region before each relaxation, iteratively until the desired cumulative dose was achieved. In large systems, 2000 FPs were inserted in the target region before each relaxation.

To describe dose independently from the size of the box, the canonical displacement per atom (\textit{cdpa}) was calculated by dividing the total number of FPs inserted by the number of lattice sites in the target region \cite{derlet_microscopic_2020}. For comparison, the \textit{cdpa} introduced in the small systems with each iteration of CRA was 0.004, while in the large systems it was 0.001. Thirty-one defective small systems were prepared with cumulative \textit{cdpa} ranging from 0.004 to 1.8. Due to the heavier computational cost of relaxing large systems and the more conservative implementation of CRA, six defective large systems were prepared with cumulative \textit{cdpa} ranging from 0.01 to 0.4. The resultant structures spontaneously evolved via interstitial dislocation loop nucleation and coalescence into extended dislocation structures, due to the relaxation of the internal stress field that develops at higher doses \cite{mason_observation_2020}. These relaxed, static defective models served as the starting point for heating and time evolution needed to reveal thermal properties of irradiated W.

\subsection{Non-Equilibrium Molecular Dynamics}
\label{method-MD}

All systems were thermalized to 300 K at a pressure of 1 bar using the Nosé-Hoover thermostat and barostat (NPT) for 400 ps (small systems) or 3 ns (large systems), with a temperature damping parameter of 0.01 ps, a pressure damping parameter of 0.1 ps, and a time step of 1 fs. Thermalization was followed by a system trajectory consistent with the canonical ensemble (NVT) for 200 ps. Equilibration was confirmed qualitatively by visual inspection of key thermodynamic variables behaving as minor variations around a constant average over time, such as pressure per atom, potential energy, temperature, and volume. An example set of plots illustrating convergence can be found in the Supplementary Material (SM) \cite{supp}.

The equilibrated structures were subjected to a heat flux ($J$) in the $x$-direction using the asymmetric version of the enhanced heat exchange algorithm (\verb|fix ehex|) \cite{wirnsberger_ehex_2015}. This algorithm acted at every time step by adding and removing energy in heat source and heat sink regions, respectively. These regions were each 65 {\AA} wide, centered at $0.25L\textsubscript{x}$ (heat source) and $0.75L\textsubscript{x}$ (heat sink).
A heat flux of $J=0.002$ eV/ps/{\AA}\textsuperscript{2} was applied to small systems and $J=0.001$ eV/ps/{\AA}\textsuperscript{2} to large systems.
These values of $J$ were found to achieve temperature profiles with sufficiently steep gradients in the target region for estimation of $\kappa$, without the temperature of the heat sink region dropping near 0 K.
To achieve steady state, the heat flux was applied to a trajectory consistent with the microcanonical ensemble (NVE) for 600 ps in small systems and 3.5 ns in large systems.
Plots illustrating convergence can be found in the SM.
Once steady state was achieved, NVE trajectories were continued for 200 ps in small systems and 100 ps in large systems, recorded at 20 ps time intervals.
The choice of trajectory duration allowed for time sampling of a sufficiently long trajectory for meaningful results without excessive computational cost.
We note that, unlike EMD approaches, NEMD imposes a heat flux that could in principle bias defect state or migration during the NVE trajectory. To assess this, time evolution of defect characteristics was tracked, and is summarized in section \ref{results}; mobility of SIA and dislocations was largely confined to the relaxation and thermalization stages in the preparation of the systems, while vacancies remained immobile, as expected at 300 K.

After MD simulations were completed, temperature profiles were calculated from the average kinetic energy of the atoms within evenly spaced bins along the $x$-dimension of the box. Bin widths were selected to contain sufficient atoms per bin to reduce noise in the temperature profiles, and to be narrow enough to remain sensitive to local variations in defect concentration across doses. Small systems were divided into 100 bins (15.8 {\AA} wide with 25k atoms each, and 10 bins in the target region). Large systems were divided into 1000 bins (6.3 {\AA} wide with 40k atoms each, and 50 bins in the target region). The greater $y$ and $z$ dimensions of the large systems allowed for narrower bin widths while increasing the number of atoms per bin. Temperature profiles for each system were averaged for time steps sampled during steady state, representing 10 time steps (200 ps) for small systems and 5 time steps (100 ps) for large systems.

Local lattice thermal conductivity ($\kappa$) of each bin was calculated using the classical Fourier heat transfer law,

\begin{equation} \label{eq-kappa}
\kappa = \frac{J}{\partial T / \partial x} \ ,
\end{equation}

\noindent where $J$ is the flux of thermal energy and $\partial T / \partial x$ is the temperature gradient \cite{kittel2018introduction}.
The temperature gradient was calculated using second order accurate central differences implemented by the Python NumPy library (\verb|numpy.gradient|).
A cubic B-spline was first fitted to noisy temperature profiles to ensure a smoother gradient, resulting in $\kappa$ values that were less responsive to sharp variations in temperature profiles.
Smoothing was implemented using the \verb|make_splrep| function from the Interpolation sub-package in the Python Scipy library.
A smoothing condition (\textit{s}) of 5 was used for small systems and \textit{s}=250 for large systems.

The possible impact of variations in time of $\partial T / \partial x$ on results was minimized using two strategies. First, we ensured that steady state $T$ profiles (and therefore $\partial T / \partial x$ and $\kappa$ profiles) were converged before sampling NVE trajectories.
Second, $T$ profiles at different times during steady state were averaged together (over 10 and 5 time steps for small and large systems, respectively). Both of these techniques reduce the impact of small scale time variations, increasing confidence in findings using the NEMD approach in systems containing various defects.

The impact of defects on $\kappa$ was expressed as the dimensionless thermal conductivity ratio $\kappa/\kappa\textsubscript{m}$ \cite{hu_thermal_2017}, where $\kappa\textsubscript{m}$ was the average lattice thermal conductivity of the target region in the defect-free reference system (\textit{cdpa}=0) for each size. The values of \(\kappa\textsubscript{m}\) are included in Table \ref{tab:average_kappa_and_temp}.

\subsection{Defect analysis}
\label{method-defect-analysis}

\begin{figure}
    \centering
    \includegraphics[width=0.8\linewidth]{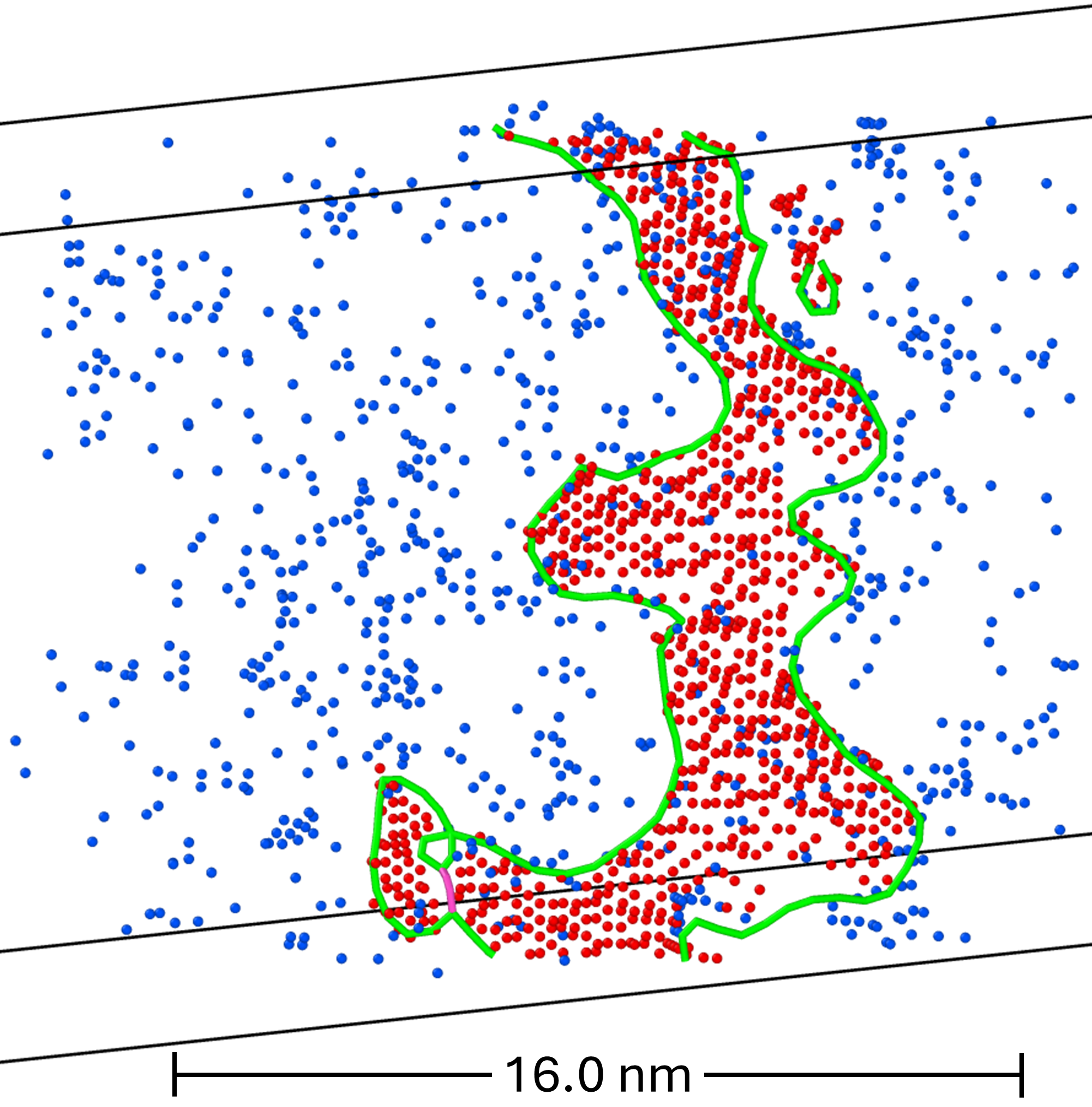}
    \caption{The target region of a small system with \textit{cdpa}=0.12. Positions of SIA (red) and vacancies (blue) are shown. Dislocation loops (Burgers vector $1/2<111>$ in green, $<100>$ in pink) enclose the SIA cluster.}
    \label{fig:OVITO}
\end{figure}

Defective structures were analyzed using the Dislocation Extraction Algorithm (DXA) and Wigner-Seitz (WS) in OVITO, version 3.11.3 \cite{ovito, stukowski_automated_2012} (an example is shown in Fig. \ref{fig:OVITO}). Defect data, including key characteristics of point defects (position and type being SIA or Vac) and dislocations (Burgers vector, number of vertices, positions of vertices, open segment or closed loop) are available in a FAIR repository (Public DOI will be generated after acceptance, this temporary link is provided for revision purposes \url{https://dataverse.bsc.es/previewurl.xhtml?token=c52134b2-5063-4d58-98d8-f6500cd1ec21}).
DXA identified dislocations in the defective structures relative to a reference BCC crystal. It characterized the size and orientation of dislocation cores enclosed by Burgers circuit paths, determining the associated Burgers vectors by the local deviation from the reference structure.
A summary of resultant dislocation characteristics can be found in section \ref{results-dislocations}.

\begin{figure}
    \centering
    \includegraphics[width=1.0\linewidth]{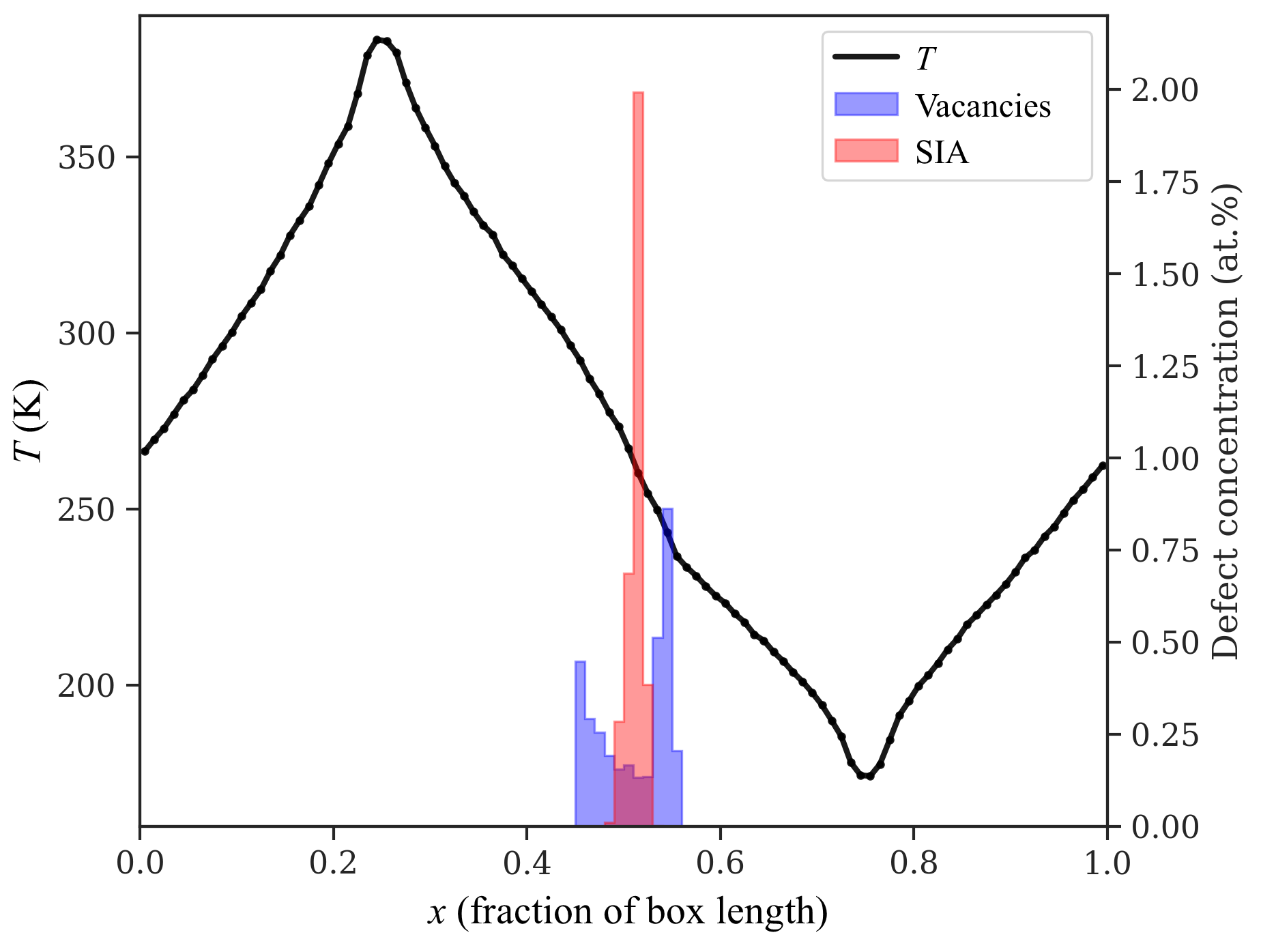}
    \caption{A temperature profile of a small system with \textit{cdpa}=0.12 in the target region, overlaid on a histogram of point defect locations identified by WS in OVITO (SIA in red, vacancies in blue).}
    \label{fig:defect-histogram-temp}
\end{figure}

Next, WS identified the positions of SIA and vacancies by comparison of atomic positions in the defective structure to Voronoi cells around lattice sites of a defect-free reference structure. See Fig. \ref{fig:defect-histogram-temp} for an example defect distribution observed using WS.
In large systems, high defect concentration led to the creation of new planes of atoms normal to the $x$-direction (parallel to heat flux). Carrying out WS against a reference structure with an $x$-displacement of 1.6 {\AA} revealed the locations of true point defects while accounting for the new planes of atoms.
 yield directly comparable defect concentration profiles for analysis against the time-averaged temperature profiles, point defect positions were time-averaged across the steady state trajectory. Averages were calculated using identical time ranges as for temperature profiles (10 time steps during 200 ps for small systems and 5 time steps during 100 ps for large systems).

To summarize, the complete dataset consisted of features describing (i) the overall systems and (ii) the localized bins along the $x$-dimension from in and immediately surrounding the target regions of all systems. 
Thirty-two small systems of varying doses each yielded 12 bins that were 15.83 {\AA} wide, containing 25k atoms (384 observations from small systems, total), and seven large systems of different doses each yielded 56 bins that were 6.34 {\AA} wide, containing 40k atoms (392 observations from large systems, total).
Data from small and large systems were analyzed separately.
For each, observations were compiled detailing applied dose as \textit{cdpa}, $x$-position of the center of the bin, number of self-interstitial atoms (N\textsubscript{SIA}), number of vacancies (N\textsubscript{Vac}), average temperature (\textit{T}), and calculated lattice thermal conductivity ($\kappa$).
Numbers of defects were also represented as defect concentrations (at.\%). Using these features, the properties and behavior of irradiated W were investigated by fitting various models to the data.

\subsection{Data analysis and modeling}
\label{method-data-analysis}

Initial analysis began with Ordinary Least Squares (OLS) multiple linear regression, predicting average $\kappa$ ($\bar{\kappa}$) of W within a single bin, given the independent variables $1/T$, $N\textsubscript{Vac}$, and $N\textsubscript{SIA}$. Parameters of this model are available in the SM. This simplistic approach offered a basic quantitative interpretation of the relationship of the independent variables with $\kappa$. To complement this analysis and quantify nonlinear structure–property relationships, Machine Learning (ML) methods were applied next.
Correlation analyses were used to assess dependencies among variables, helping to identify multicollinearity and informing feature selection strategies. Predictive analyses used four input variables: applied dose as \textit{cdpa}, $T$, $N\textsubscript{Vac}$ and $N\textsubscript{SIA}$, setting $\kappa$ as the target. To expand the original four features into a more comprehensive set, quadratic terms (\textit{e.g.} $T^2$) and two-way interaction terms (\textit{e.g.} $N\textsubscript{Vac} \cdot N\textsubscript{SIA}$) were also considered. To minimize feature interdependence and mitigate multicollinearity arising from higher-order terms, all features were first centered on zero by subtracting their respective mean values, before the expansion was performed. Standard scaling was then applied to the resulting feature set before training ML models, such that each feature had a mean of zero and a standard deviation of one. This prevented features with larger value ranges from disproportionately influencing the model, and ensured that all predictors contributed equally to the distance calculation.

Multiple distinct ML approaches, namely Random Forest Regression \cite{breiman2001random}, Gradient Boosting Regressor (GBM), extreme gradient boosting (XGBoost) \cite{chen2016xgboost}, and LightGBM \cite{ke2017lightgbm}, were implemented to capture the complex, nonlinear relationships between variables, along with the confidence intervals for predictions. To ensure robust evaluation and prevent overfitting, all models underwent 5-fold Cross-Validation (CV), where the training $\text{(80\% of data)}$ and validation $\text{(20\%)}$ sets were dynamically adjusted across iterations, and partitioned using a fixed random seed to guarantee reproducibility. For small systems, training data consisted of 307 observations and ML model predictions were validated against the remaining 77. For large systems, models were trained on 314 observations and predictions were validated against the remaining 78.

Hyperparameter optimization was conducted using the Optuna framework with 200 iterations per model, using a Tree-Structured Parzen Estimator (TPE) approach \cite{akiba2019optuna}. Root mean square (RMSE), mean absolute errors (MAE), and coefficients of determination ($R\textsuperscript{2}$) were used as evaluation metrics. SHAP (SHapley Additive exPlanations) analyses provided further transparency and scientific insights into ML predictions, assigning values that quantified the importance and effect of each feature on the predictions \cite{shap-paper}. 

The structural degradation of \(\kappa\) across a continuous multi-defect domain was mapped using the optimized XGBoost model. A two-dimensional coordinate mesh was constructed by systematically varying $N\textsubscript{Vac}$ and $N\textsubscript{SIA}$. To carry out this mapping, all other features were held constant: $cdpa$ was set to 0.00 to avoid the influence of the complex correlations between inserted and surviving defects, and temperature was set to $T = 300\text{ K}$. Raw prediction matrices were smoothed with a Gaussian filter to mitigate step-wise features from space partitioning. For the final representation, absolute defect counts were
expressed as atomic concentrations (at.\%) relative to the total number of lattice sites per bin for large systems.

\section{Results and discussion}
\label{results}

Results are presented in four subsections and disaggregated by small and large systems. First, a discussion of lattice thermal conductivity estimation in the perfect system (section \ref{results-thermal-cond-pristine}), and general observations about the overall behavior of defect concentrations and structures at high doses (section \ref{results-defects-trends-structures}), are reported. Next, the fit parameters of the models of the data are presented (section \ref{data-model}). Finally, ML models are compared and interpreted (section \ref{ML_analyses}).

\subsection{Thermal conductivity of pristine systems}
\label{results-thermal-cond-pristine}

Key information describing the simulated systems is summarized in Table \ref{tab:average_kappa_and_temp}. In our simulations, we verified that the temperature gradients in time remained linear and steady throughout the production phase, with all systems demonstrating excellent energy conservation and convergence of primary thermodynamic variables. As expected for NEMD simulations, the measured thermal conductivity exhibited a significant dependence on system size (box dimensions). For the pristine small system, we obtained $\kappa = 15.73 \text{ W/mK}$, while the large system yielded $\kappa = 21.11 \text{ W/mK}$; the two system sizes provided different but consistent results, and sampling different system sizes was essential for estimating and accounting for finite-size effects. Moreover, it should be noted that while both the length and the in-plane dimensions were increased between the two systems, the cross-sections were maintained at scales sufficiently large to ensure that the resulting estimation of bulk properties remained accurate and was not dominated by lateral confinement.

To estimate $\kappa$ of a macroscopic (infinite) system, we applied a linear extrapolation to the limit ($1/L \to 0$), where the $y$-intercept defines a bulk thermal conductivity of $\kappa_\infty = 22.90 \text{ W/mK}$. This extrapolation confirmed that the pristine large system captured $>90\%$ of the infinite-limit conductivity, representing a highly converged state. Furthermore, the slope of the linear fit is given by $\vert{}m\vert{} = 495.5 \text{ \AA} \times 22.90 \text{ W/mK} = 11.35 \text{ W/mK} \cdot \text{\AA}$. Using the relation $\lambda_{\text{eff}} = \vert{}m\vert{} / \kappa_\infty$, this slope yields an effective phonon mean free path of $\lambda_{\text{eff}} \approx 495.5 \text{ \AA}$. Therefore, given that our largest simulation box was roughly 12 times longer than the calculated $\lambda_{eff}$, the system was safely within the diffusive transport regime. This ensured a robust and converged baseline for the pristine material, providing a reliable reference for evaluating thermal degradation caused by structural defects.

Typically, semi-empirical potentials have been used for evaluating lattice transport, particularly when investigating irradiation-induced defects that require large-scale ensembles. Previous NEMD studies reported values of $\kappa$ ranging from $15.0$ to $17.3 \text{ W/m}\cdot\text{K}$ \cite{fu_calculation_2012, hu_thermal_2017}, depending on the potential, system size, and crystallographic orientation relative to heat flux. Subsequent refinements in potential parameters, larger simulation sizes \cite{zhang_study_2020}, and the most recent testing of equilibrium methods \cite{zhang2023}, led to higher estimates, around $21 \text{ W/m}\cdot\text{K}$. Our result ($22.90 \text{ W/m}\cdot\text{K}$) aligns closely with these established benchmarks.

While modern alternatives exist, the use of EAM potentials remains uniquely feasible for simulating the evolution of irradiated defective structures. Recent advances in ab initio calculations and MLIPs have provided evidence of a significantly higher intrinsic lattice contribution for W. DFT estimates based on the Boltzmann Transport Equation (BTE) have yielded values of lattice $\kappa = 46 \text{ W/m}\cdot\text{K}$ \cite{chen2019understanding}, with some simulations exceeding $60 \text{ W/m}\cdot\text{K}$ \cite{zhang_pressure_2022}. Similarly, MLIP-based Green-Kubo predictions have reported values of $60 \pm 9 \text{ W/m}\cdot\text{K}$ \cite{bhatt_transition_2023}. These higher values are often attributed to anharmonic phonon scattering captured by high-fidelity electronic structure calculations. However, these methods are primarily optimized for pristine or small-scale periodic systems and do not yet offer the versatility required for the complex, non-equilibrium defective environments and long trajectories required for the present investigation. 

We emphasize that comparing these numerical results to experimental data remains a significant challenge, as the lattice contribution is typically overwhelmed by the electronic component. While electrons are the primary heat carriers in metals, the lattice contribution in W is anomalously high compared to other transition metals, leading to complex transport behaviors under extreme conditions. Experimental estimates for W historically relied on the magnetothermal resistance method \cite{Wagner72}, in which large magnetic fields are applied at cryogenic temperatures to saturate the electronic magnetoresistance, providing values that are now seen as conservative lower bounds. At room temperature, researchers must rely on analytical models of experimental total conductivity. These models have suggested a range of values, with estimates around $40 \text{ W/m}\cdot\text{K}$ \cite{stojanovic2010thermal,Wang2020}. This discussion highlights the ongoing discrepancy between traditional interatomic potentials and newer theoretical frameworks, further emphasizing the difficulty of defining an absolute experimental baseline for the isolated lattice contribution to thermal conductivity in W.

\begin{table}[H]
    \centering
    \caption{Summary of systems studied. $N$ is the number of systems. $N\textsubscript{atoms}$ is the number of lattice atoms in the target region. $T$ gives the range of temperatures in the target region of the defect-free (\textit{cdpa}=0) system. \(\kappa\textsubscript{m}\) is the average value of \(\kappa\) in the target region of the defect-free system. \(\kappa\textsubscript{m}\) in the macroscopic limit is an extrapolation.}
    \setlength{\tabcolsep}{4pt}
    \begin{tabular}{llclc}
        \toprule
        System size & $N$ & $N\textsubscript{atoms}$ & $T$ (K) & \(\kappa\textsubscript{m}\) (W/m·K) \\
        \midrule
        Small & 32 & 250k & 248 - 283 & 15.73 \\
        Large & 7 & 2M & 272 - 299 & 21.11 \\
        Macroscopic & - & $\infty$ & $\sim$ 300 & 22.90 \\
        \bottomrule
    \end{tabular}
    \label{tab:average_kappa_and_temp}
\end{table}

\subsection{Resultant defect trends and structures}
\label{results-defects-trends-structures}

Results of defect analyses are presented in three groups of related observations. First, the nonlinear behavior of total defect concentration as a function of irradiation dose ($cdpa$) is described. Second, the characteristics of spatial distributions of point defects (SIA and vacancies) and SIA clusters are presented. Third, an overall description of dislocation properties and trends in large systems is provided.

\subsubsection{Defect saturation at high doses}
We found that defect concentration saturated at high irradiation doses. Specifically, in small systems, defect concentration increased to a maximum around 0.65 at.\% with increasing dose up to about 0.25 cdpa. Defect concentration then decreased with increasing dose until leveling off to 0.2 at.\% above 0.5 cdpa, within the range of doses tested (up to $cdpa=1.8$).
In large systems, similar behavior was observed, with increasing defect concentration up to a maximum of 0.4 at.\% at 0.025 cdpa, and then leveling off to 0.25 at.\% above 0.1 cdpa. See plots of defect concentration vs \textit{cdpa} in the SM \cite{supp}.

This observation of a nonlinear relationship between defect concentration and applied dose is consistent with prior work implementing CRA \cite{derlet_microscopic_2020, granberg_molecular_2021, granberg_effect_2023}.
The phenomenon can be understood in the following way. Mobility of SIA during the stress-driven relaxations in CRA, and again after thermalization to 300 K, enabled self-healing, whereby FP were annihilated when mobile SIA became bonded to vacant lattice sites. At higher and higher doses, the naturally occurring self-healing behavior of W prevented ever-increasing defect concentrations.

It should be noted that CRA-only simulations tend to overestimate the number of defects associated with a given dose, in comparison with CRA followed by cascade annealing, and with experiment, but that the overall behavior of the effect of dose on thermal properties is the same for all, namely, a significantly reduced but saturated effect above 0.1 dpa \cite{mason_estimate_2021}. Our methods accurately reproduced the expected defect saturation, but we further consider the impact of this method on our results. 
This known tendency of the CRA-only approach is tied to dose ($cdpa$) as the point of comparison with other simulation methods and with experiment. However, we were able to draw meaningful information from these computationally more affordable simulations in two ways: (i) by considering SIA and vacancy counts as independent features alongside dose in our analyses, enabling us to study the relationship between $\kappa$ and actual presence of surviving defects directly, and (ii) by implementing NEMD (as opposed to EMD, such as Green-Kubo), which allowed for the detection of inhomogeneities in $\kappa$ in the simulation box, caused by local variations in defect concentrations.

\subsubsection{Spatial distribution of defects}
\begin{figure}
    \centering
    \includegraphics[width=1.0\linewidth]{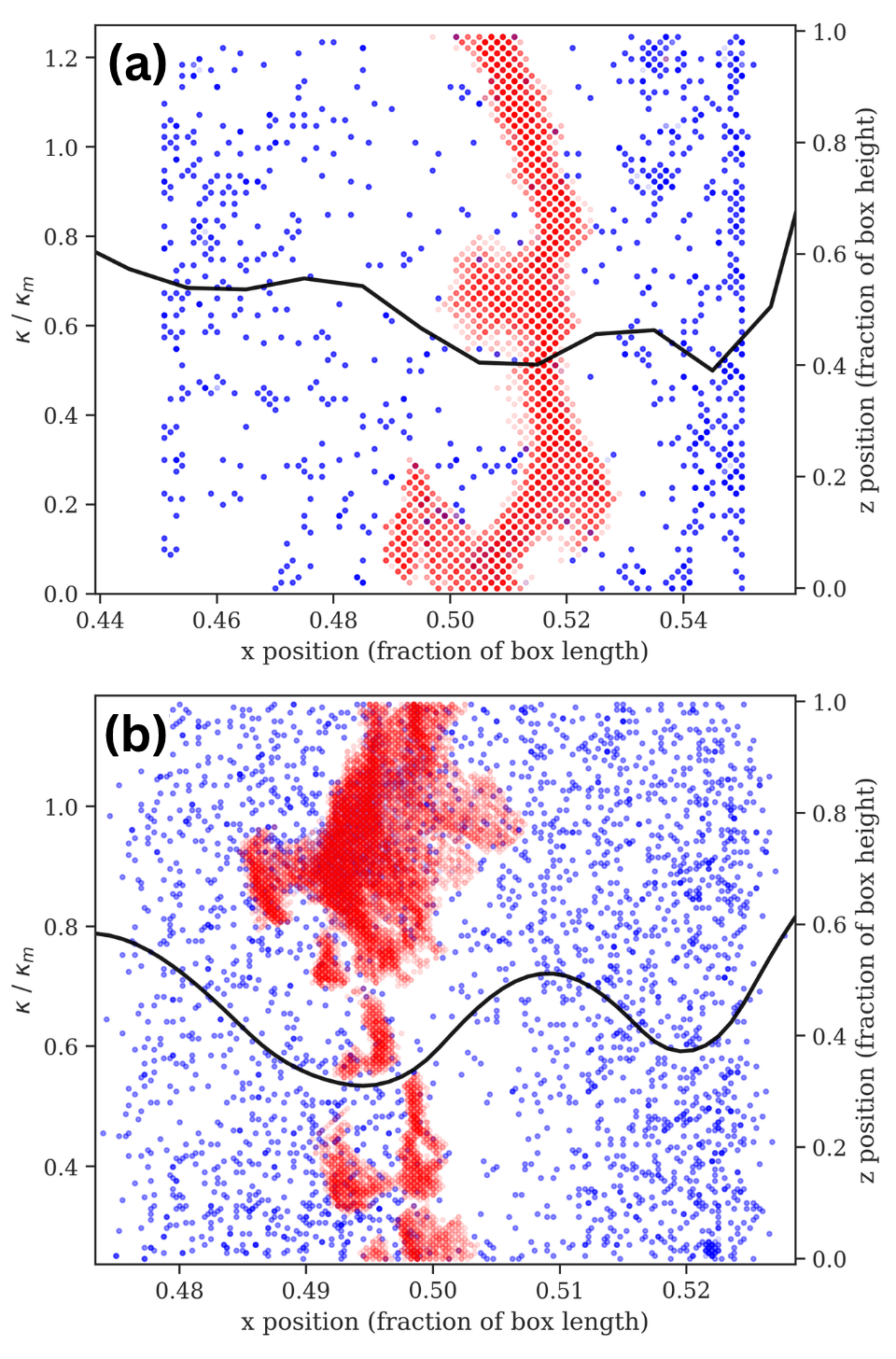}
    \caption{Target regions of (a) a small system with \textit{cdpa}=0.12 and (b) a large system with \textit{cdpa}=0.2. Locations of SIA (red) and vacancies (blue) are marked. The dimensionless \(\kappa/\kappa\textsubscript{m}\) profile is given by the black line. Marker opacity indicates temporal stability: fully opaque markers represent stationary defects, while semi-transparent markers denote mobile defects captured during a fraction of the trajectory.  Opacity was summed for ten and five 20-ps time steps during steady state for small and large systems, respectively. $\kappa_m$ represents the average thermal conductivity of a defect-free reference system.}
    \label{fig:kappa-scatterplot-overlay}
\end{figure}

Next, results pertaining to defect concentrations within each system are presented. Fig. \ref{fig:kappa-scatterplot-overlay} shows a scatterplot of the positions of point defects in example target regions for one small and one large system, with position-dependent dimensionless thermal conductivity ($\kappa/\kappa\textsubscript{m}$) overlaid.
See the SM for plots of defect locations and $\kappa/\kappa\textsubscript{m}$ for all irradiated systems \cite{supp}.
Figures \ref{fig:OVITO}, \ref{fig:defect-histogram-temp}, and \ref{fig:kappa-scatterplot-overlay} all illustrate a characteristic defect distribution consisting of a large cluster of SIA in a field of primarily isolated point vacancies.
Fewer vacancies were observed in the immediate vicinity of the large SIA cluster.

This defect distribution is consistent with prior MD work implementing CRA \cite{mason_estimate_2021,mason_observation_2020}.
The formation of large SIA clusters is driven by high interstitial mobility in W at 300 K \cite{martinez_thermal_2021} and a negative interaction energy that favors clustering over isolation. As shown in Fig. \ref{fig:kappa-scatterplot-overlay}, local minima in the $\kappa/\kappa_m$ profile correspond to these regions of high defect concentration, highlighting the localized impact of SIA and vacancies on thermal transport.

The mobility of these large SIA clusters during steady state is visible in Fig. \ref{fig:kappa-scatterplot-overlay}, where markers with higher opacity indicate point defects that remained in one position across multiple time steps. This visualization illustrates limited diffusion of largely stable SIA clusters. Further details regarding sizes and positions of dislocation segments and loops are presented in the next section, \ref{results-dislocations}.

The observed distribution of isolated vacancies is consistent with low-energy, high-dose irradiation regimes \cite{wu2023simulation}. In our FP-insertion approach, the initial vacancy distribution remained largely static during thermalization to 300 K, as vacancies are practically immobile at room temperature on MD timescales. While high-energy primary knock-on atoms (PKA $>$ 300 eV) are known to trigger local temperature spikes that redistribute vacancies into larger clusters \cite{granberg_effect_2023, wu2023simulation}, our results reflect the stability of the point-defect-dominant regime. This behavior aligns with prior work identifying the onset of significant vacancy mobility in W only at much higher temperatures, specifically in the 1700–2400 K range \cite{kong2025temperature}. Furthermore, the lower vacancy concentration observed near SIA clusters is attributed to self-healing; as mobile SIAs approach stationary vacancies, they recombine to restore the BCC lattice site, thereby eliminating the FP, as recently demonstrated in an MLIP-based atomistic study of W \cite{suarez2025early}.

\subsubsection{Dislocation characteristics in large systems}
\label{results-dislocations}

In the seven large systems, six of which contained defects, DXA of all time steps independently yielded 574 dislocations. When combining independent dislocations within each system based on similar numbers of vertices along the line, positions of vertices, and parallel Burgers vectors, this number was reduced to 86 unique dislocations that were tracked across multiple time steps. Of these 86 unique dislocations, 78 had 1/2$<111>$ Burgers vectors (energetically favored), and 8 had $<100>$ dislocations. On average, the number of vertices was larger for 1/2$<111>$ than for $<100>$ type dislocations, at 37.01 and 13.22 vertices per dislocation, respectively. This difference also aligns with the expectation that more favorable defects, having a lower formation energy, tend to be larger.

Of the 86 unique dislocations, 49 were open segments (with starting and ending vertices at different positions) and 37 were closed loops (where the starting and ending vertices were located at the same position during at least one time step). These closed loops were visibly very common when viewing defective structures following DXA in OVITO. The higher frequency of open dislocation segments may relate to the large, unevenly shaped SIA clusters containing multiple dislocation cores, which tended to be enclosed by multiple open Burgers circuit paths.

When comparing resultant characteristics such as number of unique dislocations, ratio of 1/2$<111>$ to $<100>$ Burgers vectors, ratio of open segments to closed loops, and persistence of a uniquely identifiable defect structure across multiple time steps, against the independent variable of irradiation dose, no clear trend emerged. This could be a result of the saturation of defect concentrations at high doses, where 4 of 6 irradiated large systems were at doses in the nonlinear regime, and 2 were at doses above the saturation threshold, making them difficult to distinguish from each other in terms of final physical attributes.

The possibility of defect evolution and diffusion of mobile 1/2$<111>$ dislocations was explored for large systems. The 78 unique dislocations of this type were described by the average 3D position of their vertices at each time step. The average net displacement of this 3D position for all 1/2$<111>$ dislocations was calculated and compared to that of the remaining 8 dislocations. As expected, the average net displacement was higher for 1/2$<111>$ type dislocations (5.89 {\AA}) than for $<100>$ type dislocations (1.16 {\AA}), consistent with their higher diffusion capability at 300 K. Considering that the bin width in the $x$-dimension (6.34 {\AA}) was greater than either of these values, the likely impact of diffusion on NEMD calculations was estimated as low. Further, the observation that the average $x$-width of 1/2$<111>$ type dislocations (33.79 {\AA}) was more than five times the average net displacement, we concluded that significant overlap in the position of extended defect structures existed for the amount that they moved during the steady state trajectory. These findings, along with time-averaging temperature profiles and defect concentration profiles, aid the interpretation of NEMD results relating defects to thermal conductivity degradation.

\subsection{Parameter estimation}
\label{data-model}

\begin{figure*}
    \centering
    \includegraphics[width=1.0\linewidth]{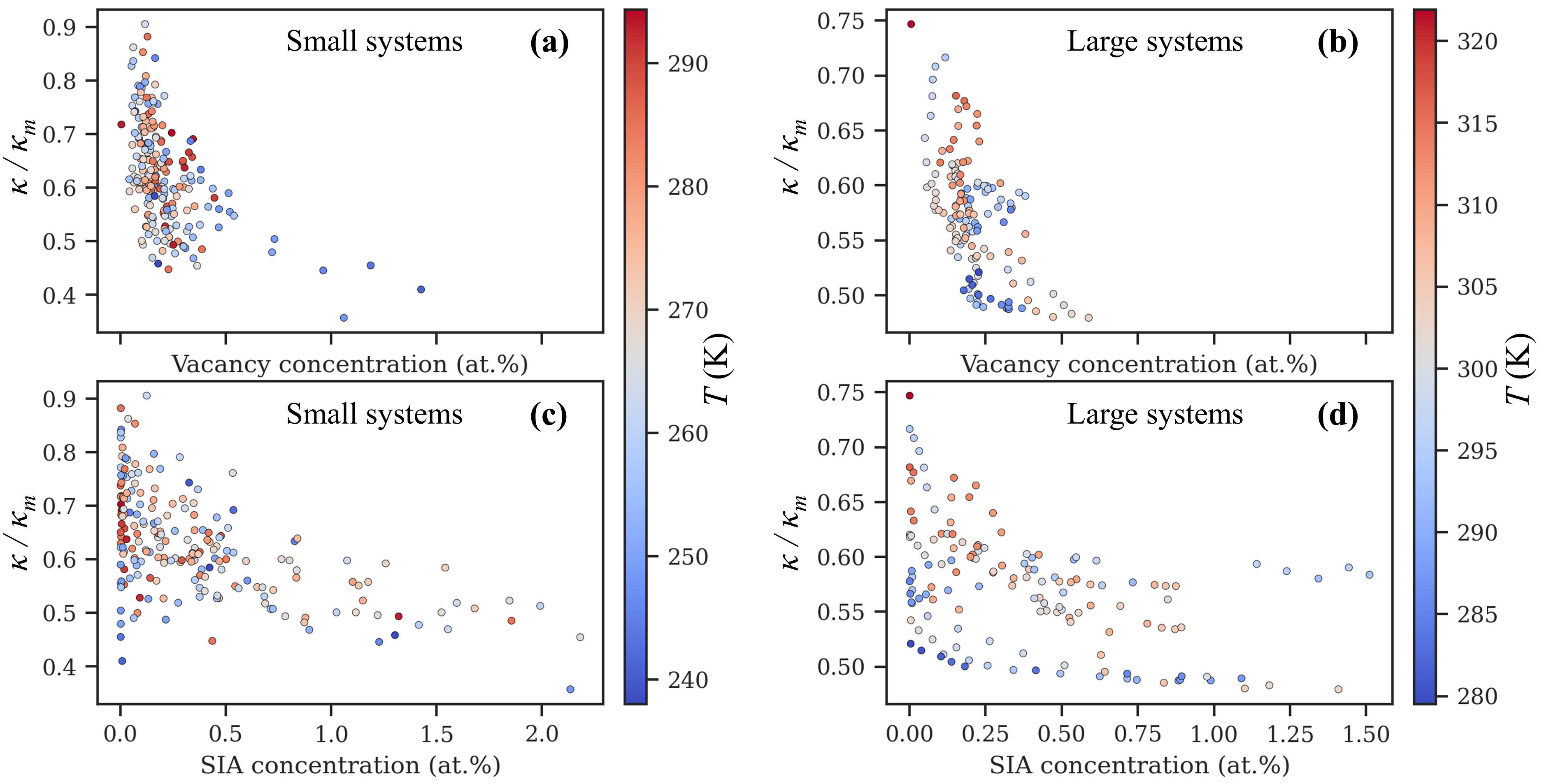}
    \caption{Dimensionless $\kappa/\kappa\textsubscript{m}$ vs vacancy concentration in (a) small and (b) large systems, and $\kappa/\kappa\textsubscript{m}$ vs SIA concentration in (c) small and (d) large systems. Thirty-two small systems each had 12 bins containing 25k atoms each, within and bordering the target region (N=384). Seven large systems each had 56 bins containing 40k atoms each, within and bordering the target region (N=392). Points representing bins with zero defects were hidden.}
    \label{fig:scatterplots-defects-temp}
\end{figure*}

With resultant defective structures showing good alignment with prior work, the complete datasets containing features of the bins in the target regions of small and large systems were then modeled. Fig. \ref{fig:scatterplots-defects-temp} shows plots of $\kappa/\kappa\textsubscript{m}$ vs defect concentrations.

The relationships between observables were explored quantitatively using a variety of modeling approaches. The initial OLS multiple linear regression predicted the dependent variable $\kappa$ as a function of independent variables $1/T$, $N\textsubscript{Vac}$, and $N\textsubscript{SIA}$. The model for small systems $(R\textsuperscript{2}=0.528$) revealed that all three independent variables had statistically significant correlations with $\kappa$ ($p$-value $< 0.01$), with decreased $\kappa$ associated with higher temperatures and higher defect concentrations. The model for large systems $(R\textsuperscript{2}=0.642$) yielded statistically significant results only for $1/T$ and $N\textsubscript{Vac}$, with the impact of $N\textsubscript{SIA}$ not captured by the multiple linear fit. Results of parameter estimation from the OLS linear models are summarized in the SM. Due to the low explainability offered by linear modeling, further analyses were carried out.

\begin{figure}
    \centering
    \includegraphics[width=0.8\linewidth]{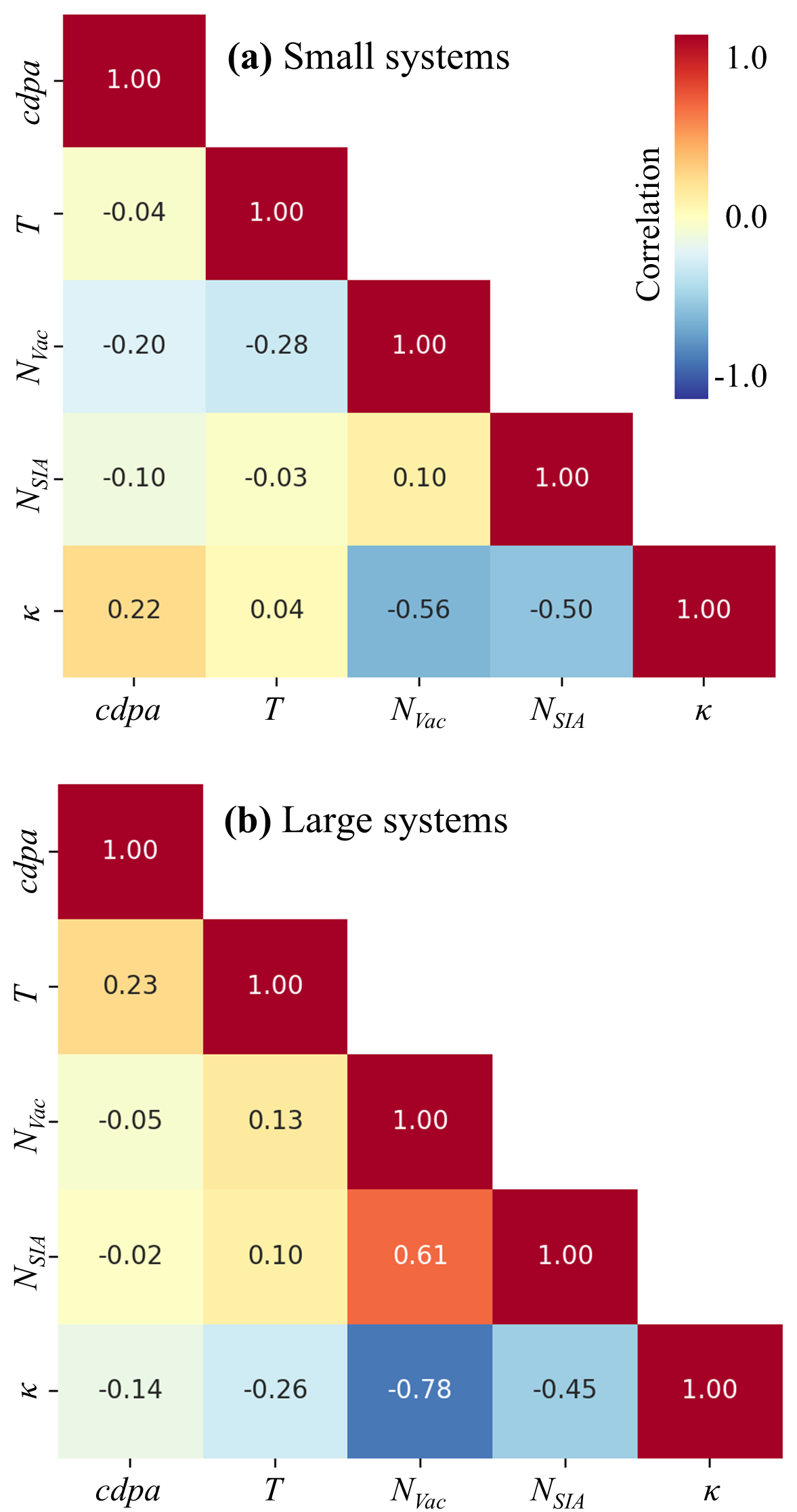}
    \caption{Heatmap showing linear relationships, measured by Pearson correlation coefficients for all pairwise combinations, between \textit{cdpa}, $T$, $N\textsubscript{Vac}$, $N\textsubscript{SIA}$, and $\kappa$ for (a) small and (b) large systems. To avoid redundancy, only the lower triangle of the matrix is displayed.}
    \label{fig:corr-matrix}
\end{figure}

To further explore multicollinearity and nonlinearity, and to lay the foundation for Machine Learning (ML) regression techniques, correlation analysis was used to assess the linear dependencies among variables, helping to inform feature selection strategies.
Correlation matrices were computed among input features, such as the initial dose (\textit{cdpa}), the number of surviving defects after annealing (N\textsubscript{Vac}, N\textsubscript{SIA}), and the temperature ($T$), as well as the target variable, $\kappa$.
The heatmaps in Fig. \ref{fig:corr-matrix} show the correlations between variables for (a) small and (b) large systems.
The thermal conductivity is, in both cases, strongly and negatively correlated with N\textsubscript{Vac} (-0.56 and -0.78 for small and large systems, respectively) and N\textsubscript{SIA} (-0.50 and -0.45 for small and large systems, respectively).
The results show a weak correlation between N\textsubscript{Vac} and N\textsubscript{SIA} (0.1) in small and a strong correlation (0.61) in large systems.
This indicates that the smaller model, with shorter average diffusion distances, promoted more rapid migration than the larger model.
Therefore, defect types tended to migrate and recombine independently upon annealing, while surviving defects were randomly distributed throughout the cell.
In the large model, the strong correlation with \(\kappa\) indicates that defects were not randomly distributed, but that they tended to be found in proximity to each other.
In other words, SIA tended to form clusters or loops, while vacancies tended to be left more at the edges of the target region, due to self-healing in the vicinity of the large SIA cluster.

Whereas one could expect some direct correlation between, for example, the initial and surviving defects, or \(\kappa\) and $T$, the bivariate correlation matrix revealed relatively weak linear relationships ($r < 0.3$). This indicates that these variables were not well described by a simple linear relationship, but rather arose from the cumulative effect of the entire set of predictors, in this case demonstrating sensitivity to model specification and sampling variation. To ensure a robust comparison between system sizes, we performed a supplementary analysis restricting the small-system dataset to $\text{cdpa}$ values $\le 0.4$, matching the range covered by the large-system data. These tests produced consistent results, suggesting that our conclusions are robust against variations in the underlying data sampling.

\subsection{Structure-property mapping via ML}
\label{ML_analyses}

Given that correlations observed in the Pearson correlation analysis suggested potentially complex physical mechanisms that linear methods alone could not adequately represent, multiple advanced ML techniques were employed to supplement our statistical analyses. The predictive modeling effort followed a rigorous pipeline involving feature engineering, cross-validated model selection, hyperparameter optimization, and detailed explainability analysis, described in the Methods (section \ref{method-data-analysis}). These models aimed not only to predict the target variable based on the four primary independent variables: $N\textsubscript{Vac}$, \textit{cdpa}, $T$, and $N\textsubscript{SIA}$, but also to consider second-degree features to capture complex material behavior. 

The first key observation was the superior accuracy of ML models trained on large systems compared to those trained on small systems (Table \ref{tab:ML-model-comparison}). While models trained on small systems achieved $R^2 \approx 0.8$, the large-system models consistently reached $R^2 > 0.94$. This improvement is attributed to large systems providing a statistically more robust foundation for regression analysis. When comparing specific algorithms, LightGBM occasionally outperformed XGBoost; however, these variations were minor, with all ML models trained on large systems consistently maintaining $R^2 > 0.94$.

Given the limited statistical significance of ML models of small systems using both ranges of applied dose (full and biased, or 1.8 and 0.4 cdpa, respectively), we restricted all subsequent investigations to the range generated for large systems (observations for small and large systems with doses up to 0.4 cdpa). Comparative analysis of four ML regressors highlighted XGBoost as the most robust and generalizable model, with LightGBM as a close second. XGBoost demonstrated consistency by performing best across small and large simulation box sizes (highest $R^2=0.979$, and lowest $MAE=0.527$). 

\begin{table}
    \centering
    \caption{ML model comparison based on prediction error metrics, ranked from best performance (top) to worst performance (bottom), where better performance is given by higher $R^2$ and lower $MAE$.}
    \label{tab:ML-model-comparison}
    \begin{tabular}{l c c c}
        \toprule
        \multicolumn{4}{c}{\textbf{Small systems}} \\
        \cmidrule(lr){1-4}
        \textbf{Model} & $\mathbf{R^2}$ & \textbf{RMSE} & \textbf{MAE} \\
        \midrule
        XGBoost & 0.778 & 0.953 & 0.779 \\
        LightGBM & 0.774 & 0.960 & 0.782 \\
        Gradient Boosting & 0.772 & 0.966 & 0.770 \\
        Random Forest & 0.763 & 0.985 & 0.818 \\
        \midrule[\heavyrulewidth] 
        \multicolumn{4}{c}{\textbf{Large systems}} \\ 
        \cmidrule(lr){1-4}
        \textbf{Model} & $\mathbf{R^2}$ & \textbf{RMSE} & \textbf{MAE} \\
        \midrule
        XGBoost & 0.979 & 0.527 & 0.337 \\
        LightGBM & 0.975 & 0.574 & 0.325 \\
        Gradient Boosting & 0.969 & 0.647 & 0.401 \\
        Random Forest & 0.946 & 0.849 & 0.486 \\
        \bottomrule
    \end{tabular}
\end{table}

\begin{figure}
    \centering
    \includegraphics[width=0.8\linewidth]{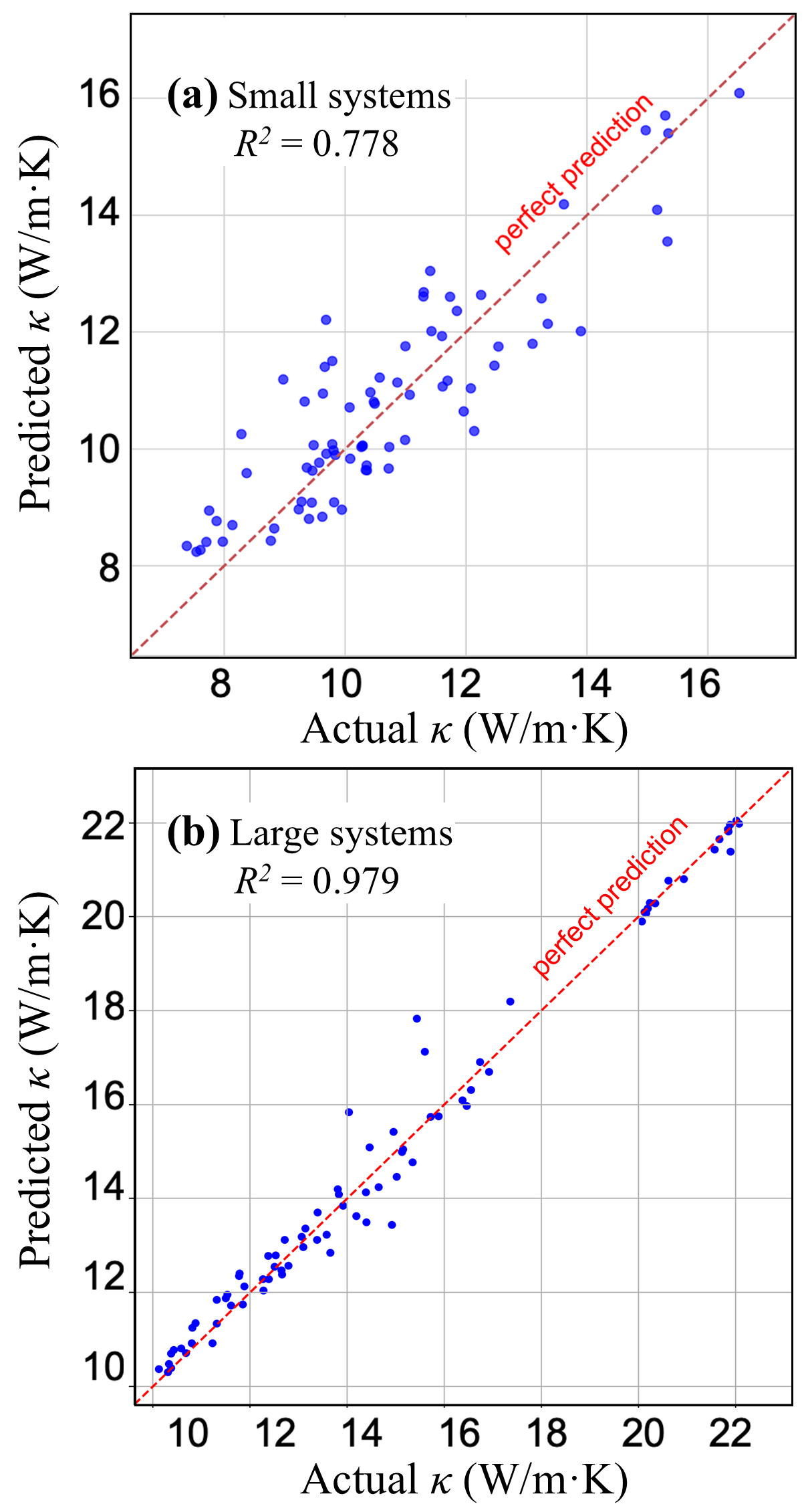}
    \caption{Prediction of \(\kappa\) for the XGBoost model: (a) small and (b) large systems.}
    \label{fig:XGBoost}
\end{figure}

Nevertheless, LightGBM was also found to achieve good performance across system sizes with a lower computational cost, although in our case, this was not considered a critical issue. When re-running ML regression for testing, LightGBM would occasionally edge out XGBoost for the highest $R^2$, but variations were small, with all ML models trained on large systems consistently achieving $R^2 > 0.94$. The robust generalization capability made XGBoost our final choice for the prediction model. The comparison of actual versus predicted values for XGBoost is displayed in Figure \ref{fig:XGBoost}. The comparison for all models can be found in the SM.

\begin{figure*}
    \centering
    \includegraphics[width=0.8\linewidth]{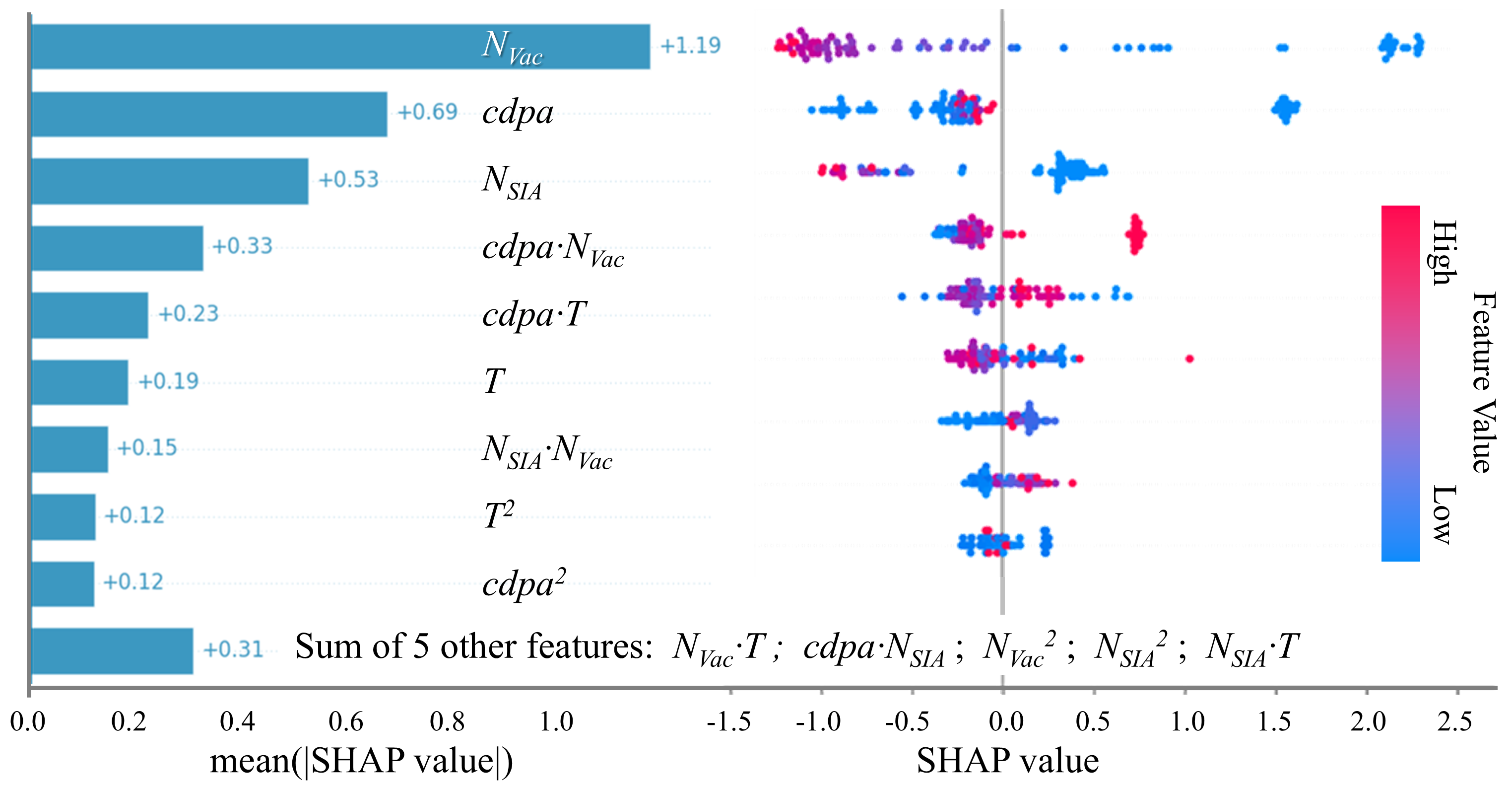}
    \caption{SHAP features extracted from the XGBoost model of large systems. The left panel presents the mean absolute SHAP values, which quantify the average contribution of each feature to the model’s output. The five least-contributing features are listed in order from higher to lower mean absolute SHAP value. The right panel shows the detailed SHAP value distributions, indicating the direction and magnitude by which variations in each feature influenced predicted $\kappa$.}
    \label{fig:SHAP}
\end{figure*}

To gain a deeper understanding of the model’s internal decision-making process and to connect its predictions with physical observables, SHAP (SHapley Additive exPlanations) analyses were applied to interpret XGBoost regression outputs for thermal conductivity. Figure \ref{fig:SHAP} summarizes the results, showing the overall contribution and relative importance of each feature and their interactions.

SHAP analysis revealed that vacancy concentration (N\textsubscript{Vac}) was the dominant factor controlling \(\kappa\) predictions. This indicates that variations in vacancy concentration explained most of the model’s predictive behavior, suggesting that vacancy defects acted as strong phonon-scattering centers, significantly reducing lattice thermal conductivity. While the vacancy concentration ($N_{\text{Vac}}$) was the primary driver, ($N_{\text{SIA}}$) was also identified for its influence on predicted $\kappa$ values. Similarly to vacancies, SIA contributed to phonon scattering, confirming that both primary defect types contributed to lattice thermal conductivity degradation.

SHAP analysis identified \textit{cdpa} as the second most influential factor governing predicted $\kappa$. While the interpretation of the influence of \textit{cdpa} is non-trivial, it may serve as a quantifiable measure of the irradiation dose, or reactor operational time. As illustrated by the defect concentration vs \textit{cdpa} plots in the SM, \textit{cdpa} correlated with more defective systems (accumulated $N_{\text{Vac}}$ and $N_{\text{SIA}}$), up to a specific irradiation limit of about \textit{cdpa}=0.2 for small systems, while beyond this point, the relationship changed due to dynamic defect annihilation or saturation. However, considering that despite possible overestimation, previous studies utilizing this methodology have shown the predicted irradiation damage to be proportional to the actual irradiation cascades. As such, \textit{cdpa} can be considered a robust proxy for total accumulated irradiation \cite{mason_estimate_2021}. In this context, our model revealed the negative influence of \textit{cdpa} on $\kappa$, confirming that increased irradiation dose and operational time degrade lattice thermal conductivity. The $T$-related features presented a lower relative weight, confirming that, while temperature influences \(\kappa\), its effect is secondary to defect-related terms in our model at limited $T$ ranges. 

Beyond the primary vacancy contribution, several second-order interaction terms (\textit{e.g.} \textit{cdpa} $\cdot$ N\textsubscript{Vac}, \textit{cdpa} $\cdot$ $T$, and N\textsubscript{Vac} $\cdot$ N\textsubscript{SIA}) appeared among the top-ranked features, confirming that the model successfully learned non-additive and conditional dependencies. These higher-order combinations captured several nonlinear and multivariable effects. For instance, the correlation between predicted $\kappa$ and given \textit{cdpa} $\cdot$ N\textsubscript{Vac} indicated that the interaction between increasing irradiation damage ($cdpa$) and subsequent accumulation of vacancies (N\textsubscript{Vac}) and structure amorphisation, played a discernible role in the overall reduction in \(\kappa\).

The SHAP summary (Fig. \ref{fig:SHAP}, right panel) showed that variables such as N\textsubscript{Vac}, N\textsubscript{SIA} and T, high feature values (red points) were predominantly associated with negative SHAP, meaning that higher defect concentrations and higher temperatures tended, as expected, to reduce \(\kappa\). This is in line with the phonon transport picture, where higher defect densities and thermal effects tend to reduce conductivity. On the other hand, low values of these features (shown in blue) are associated with positive SHAP contributions, reflecting that pristine or weakly damaged configurations at lower temperatures yielded higher \(\kappa\) values, as expected for crystalline systems with long phonon mean free paths.

Overall, the SHAP analysis validated the physical trends of the ML model, which showed how \(\kappa\) decreased with increasing defect concentration and $T$, while helping to establish clear structure-property relationships and quantify these dependencies, which are presented as nonlinear and feature-interactive. These results highlight the complexity beyond thermal and defect-driven scattering processes observed in non-equilibrium simulations, demonstrating that temperature and concentration of defects are tightly correlated.

\begin{figure}
    \centering
    \includegraphics[width=0.8\linewidth]{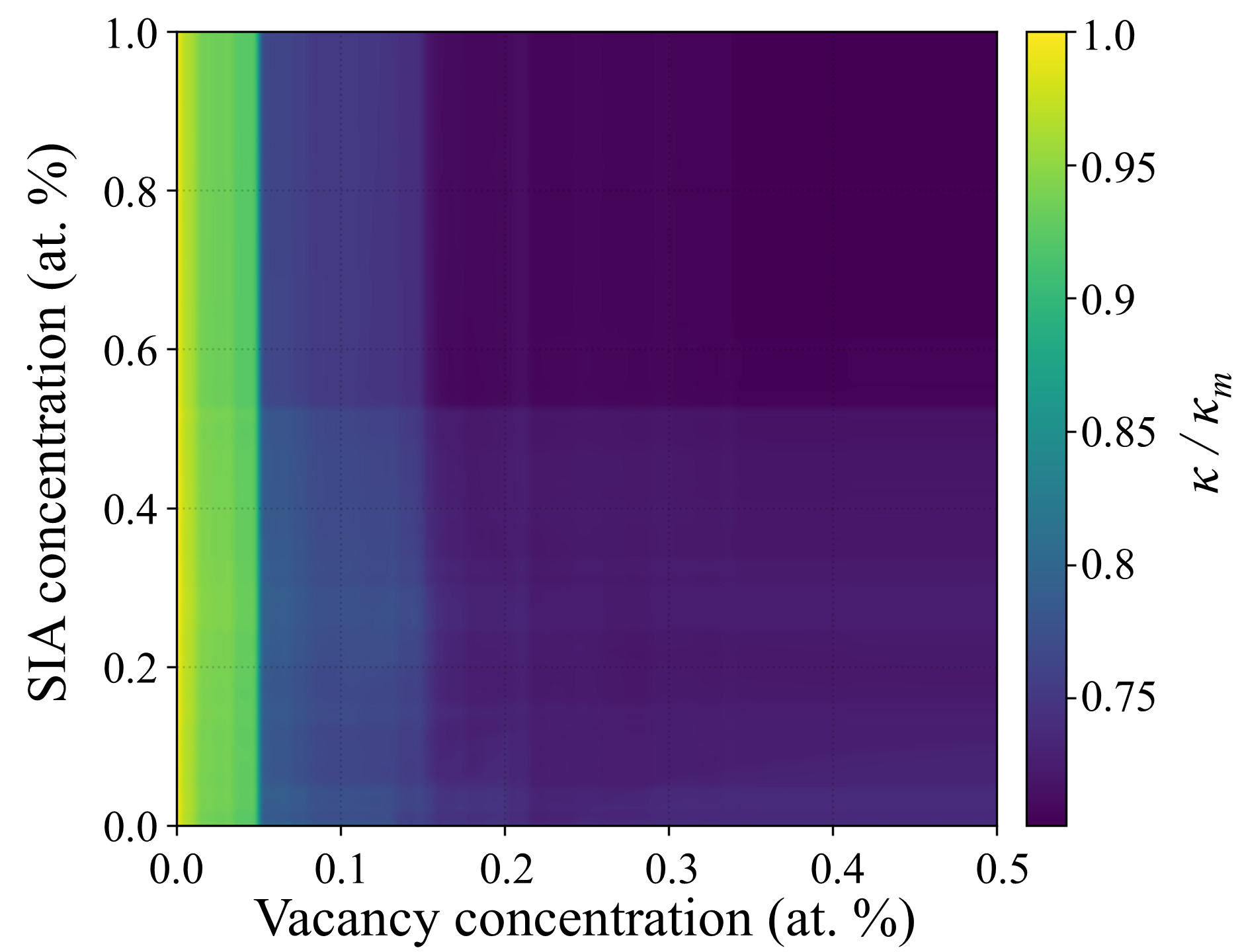}
    \caption{ML-driven mapping of dimensionless $\kappa/\kappa\textsubscript{m}$ as a function of point defect concentrations at $T = 300\text{ K}$. The 1.0 on the color scale (in yellow) represents $\kappa/\kappa\textsubscript{m}$ for the defect-free lattice state in the large system.}
    \label{fig:Tcondmap}
\end{figure}

To translate these global SHAP dependencies into a more physically interpretable map, Figure \ref{fig:Tcondmap} was generated to represent the coupled defect landscape. Rather than exhibiting a simple linear decay, a clear step-like transition in degradation rate is visible at a critical vacancy concentration of $0.05\ \mathrm{at.\%}$ (corresponding to 20 vacancies within a $40k$-atom domain), followed by a secondary drop around $0.15\ \mathrm{at.\%}$. Below these critical thresholds, $\kappa$ undergoes rapid and progressive drops where individual defects seem to act as isolated perturbation centers. Beyond these limits, however, the system transitioned into a fully saturated regime, where the conductivity was no longer affected, hitting a stable floor of roughly $70\%$ retention, which means that, independently of the concentration of surviving defects, maximum degradation was capped at $30\%$. In fact, although our training dataset covered much larger defect areas, no further reduction in $\kappa$ was observed past these boundaries. This plateau indicates a clear physical regime change, most probably due to an already severe disruption in the phonon transport pathways, where, no matter what we do with further defect accumulation, the scattering cross-section remains fully saturated.

This nonlinear evolution became even more pronounced when examining the coupled coexisting defect map, contrasting with conventional analytical approximations in which scattering rates from $N\textsubscript{Vac}$ and $N\textsubscript{SIA}$ are sometimes assumed to be purely additive. Looking into the multi-defect interference landscape, we observed that for low $N\textsubscript{SIA}$ concentration regimes, the underlying lattice remained relatively unperturbed, preserving higher conduction efficiency. On the other hand, a distinct horizontal transition was visible near an SIA concentration of around $0.5\ \mathrm{at.\%}$. Beyond this threshold, the marginal reduction in $\kappa$ dropped significantly, and the system was driven directly into its lowest saturation zone. Beyond offering descriptive insights into these atomistic phenomena, these quantitative multi-variable topologies provide a bridge for multiscale material modeling. By capturing the simultaneous, nonlinear dependencies of $\kappa$ across continuous defect and dose fractions, our results yielded explicit, dose-dependent scaling maps moving beyond purely qualitative descriptions.

\section{Conclusions}
\label{conclusion}

In this study, an integrated atomistic simulation approach was employed to analyze the thermal conductivity of irradiated W. The Creation Relaxation Algorithm (CRA) for defect generation was combined with Non-Equilibrium Molecular Dynamics (NEMD) simulations and Machine Learning (ML) analysis to quantitatively investigate the degradation of lattice thermal conductivity ($\kappa$). Simulated samples ranged a set of doses relevant to fusion reactor operation. We successfully generated and analyzed large-scale defect nanostructures, which is essential for capturing the formation and clustering of complex defects critical to understanding thermal transport degradation.

Our findings confirmed a significant, nonlinear suppression of $\kappa$, demonstrating that the high-dose irradiation environment drastically reduced the capability of W for heat dissipation, which is a critical issue for plasma-facing component (PFC) performance. The conductivity was found to rapidly drop by 50\% with increasing vacancy concentrations up to 0.25\% before the rate of decrease moderated, highlighting the non-trivial relationships between dose, defect evolution, and thermal transport properties. Machine Learning (ML) models coupled with SHAP analyses indicated that $\kappa$ was primarily controlled by defect concentration (especially that of vacancies), which was also affected by the irradiation dose and temperature. Our interpretations exposed the nonlinear and synergistic nature of these dependencies, which would be difficult to obtain from the simulation of smaller or less complex defect structures, or from conventional regression alone. This demonstrates the utility of explainable ML in extracting consistent and physically meaningful trends from complex atomistic simulations. Specifically, continuous defect mapping revealed a sharp, step-like transition at a critical vacancy concentration of $0.05\ \mathrm{at.\%}$, driving the system into a definitive physical saturation floor at approximately $70\%$ conductivity retention. Past this boundary, further defect accumulation yielded a net-zero marginal impact on transport due to the complete saturation of the phonon-scattering cross-section.

Our results offer two major contributions. First, they provide $\kappa$ data for W nanostructures relevant to future fusion reactor conditions, complete with a methodology to produce further data while explaining properties. Such a foundation could potentially be taken as input in multiscale modeling frameworks, relying on atomistic calculations to build micro- and meso-scale models, which have the power to reproduce realistic operational conditions and scales in fusion materials research \cite{lasa2024development,garcia-mantsinen-exascale-fusion-2025,kalha2021lifetimeW}. Second, the ML framework establishes a predictive capability to identify structural features that have the greatest impact on lattice thermal conductivity degradation.

In conclusion, this work provides a detailed atomic-scale mapping of dose-dependent lattice thermal conductivity degradation in irradiated tungsten. Although bulk metallic thermal transport is primarily electronic in origin, the unusually high lattice contribution in tungsten allowed our phonon transport calculations to serve as a highly correlated proxy for the material's overall thermal degradation trends. By capturing the complexity of defect-phonon scattering, our results provide an explicit physical parameterisation of the exact correction that macroscopic engineering frameworks have historically neglected. While the present study omits direct electronic transport modelling, the high-fidelity structural and transport relationships established here provide a critical foundation. Our findings could motivate future studies directly coupling our lattice-driven ML framework with electronic transport models to examine synergistic defect effects on thermal transport, and to develop further models that incorporate ML-identified structural features to improve material lifetime predictions, such as digital twins of fusion reactor components. Also, extending the framework to multicomponent systems by using new EAM potentials that describe W, H, and He interactions \cite{tirumala2025empirical} could enable more accurate, predictive simulations of lattice thermal conductivity degradation under plasma-exposure conditions.

\section*{CRediT authorship contribution statement}
\label{credit-authorship}
\textbf{M. K. Chessey}: Formal analysis, Investigation, Methodology, Visualization, Writing - original draft, Writing - review \& editing. \textbf{J. J. Gutiérrez M.}: Conceptualization, Investigation, Methodology, Supervision, Writing - original draft, Writing - review \& editing. \textbf{S. S. Mehta}: Formal analysis, Investigation, Visualization. \textbf{P. Settembri}: Investigation, Methodology, Writing - review \& editing. \textbf{A. Erbasan}: Investigation, Writing - review \& editing. \textbf{D. Freinberger}: Investigation, Writing - review \& editing. \textbf{M. García R.}: Investigation, Writing - review \& editing. \textbf{M. J. Mantsinen}: Funding acquisition, Project administration, Resources, Writing - review \& editing.

\section*{Declaration of competing interest}
\label{competing-interest}
The authors declare that they have no competing financial interests that could have influenced the study.

\section*{Data availability}
\label{data-availability}
MD simulations and defect data are openly accessible from the BSC Dataverse FAIR repository \url{https://doi.org/10.82201/VEQPJA}. The data pre- and post-processing tools, scripts for ML analyses, and resulting models are available at \url{https://github.com/juliogutierrezmoreno/Tcond_ML_NEMD} to track changes and future contributions, and in Zenodo \url{10.5281/zenodo.21891587} for permanent storage. Any additional required information will be made available upon reasonable request.

\begin{acknowledgments}
This work was developed within the framework and with the financial support of the FusionCAT project (001-P-001722) co-financed by the European Union Regional Development Fund within the framework of the ERDF Operational Program of Catalonia 2014-2020, with a grant of 50\% of the total cost eligible. P.S., A.E., and D.F. contributed to this project with the support of the Summer of HPC program in 2020 and 2021, funded by the Partnership for Advanced Computing in Europe (PRACE). M.K.C. and J.J.G.M. acknowledge their AI4S fellowships within the ``Generación D" initiative by Red.es, Ministerio para la Transformación Digital y de la Función Pública, for talent attraction (C005/24-ED CV1), funded by NextGenerationEU through PRTR. We also acknowledge the ML-driven bottom-up design of atomically-layered heterostructures for green production (MLALH) project, awarded by the EIG CONCERT-Japan call with reference PCI2023-143426 funded by MCIN/AEI/10.13039/501100011033 and the EU. Part of this work was carried out using supercomputer resources provided under the EU-JA Broader Approach collaboration in the Computational Simulation Centre of the International Fusion Energy Research Centre (IFERC-CSC). The authors thank Daniel Mason for providing a LAMMPS script implementing CRA.
\end{acknowledgments}

\bibliography{mi-biblioteca}



\section*{Supplementary material}

# Supplementary Material for: Atomic-level study of dose-dependent thermal conductivity degradation in irradiated tungsten

Mary K. Chessey and José Julio Gutiérrez Moreno*
*Barcelona Supercomputing Center (BSC), Barcelona, Spain*

Sajjan Singh Mehta
*oloBion SL, Parc Científic de Barcelona, Barcelona, Spain*

Paolo Settembri, Arda Erbasan, Dominik Freinberger, and Marina Garcia-Romero
*Barcelona Supercomputing Center (BSC), Barcelona, Spain*

Mervi J. Mantsinen
*Barcelona Supercomputing Center (BSC), Barcelona, Spain and*
*ICREA, Barcelona, Spain*

## CONTENTS



* julio.gutierrez@bsc.es

# 1. EQUILIBRIUM AND STEADY STATE

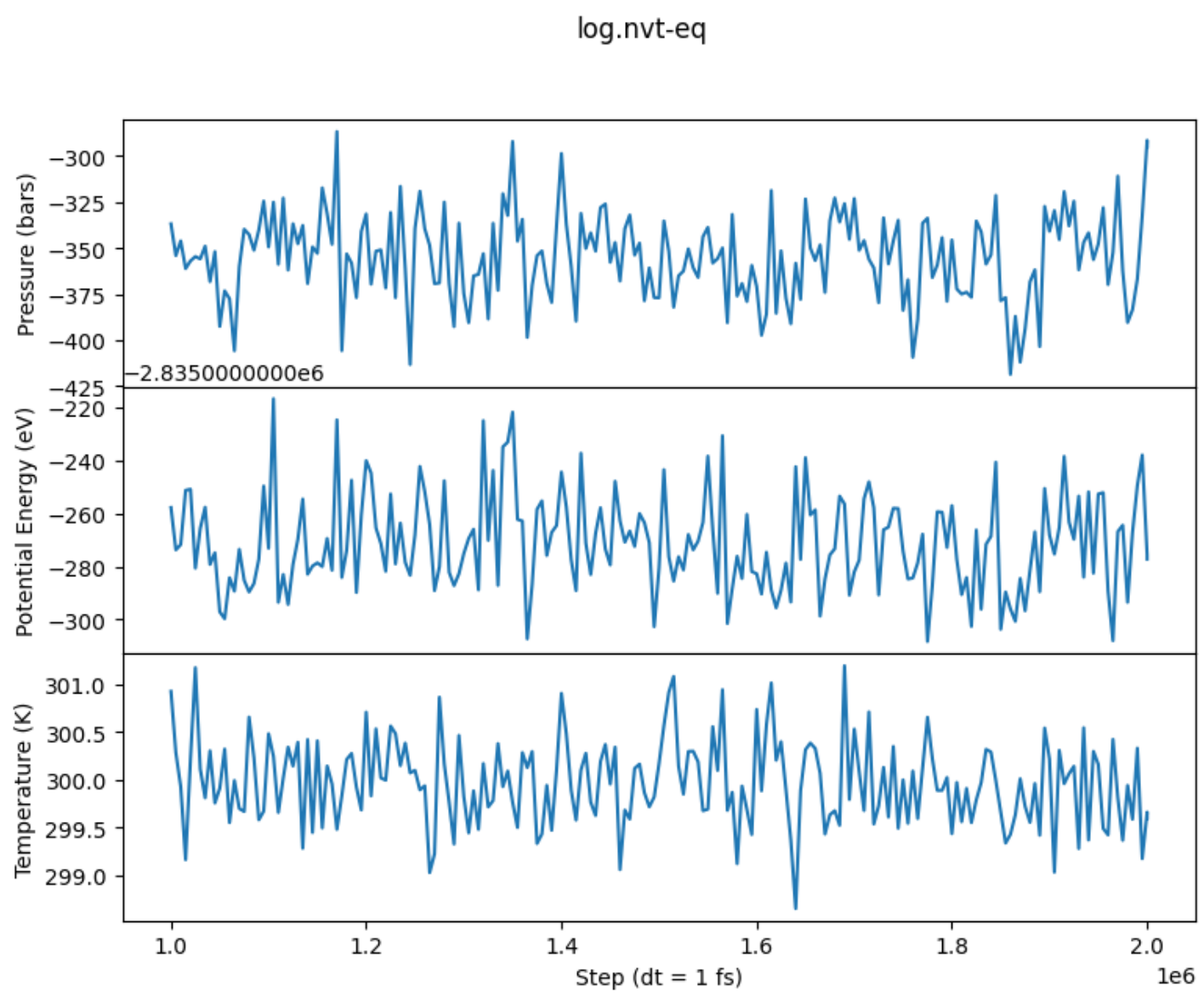


**FIG. S1**. Example illustration of equilibrium state achieved before applying heat flux. Pressure, potential energy, and temperature are plotted over time to show that these key properties varied as fluctuations about a constant average by the end of the NVT trajectory.

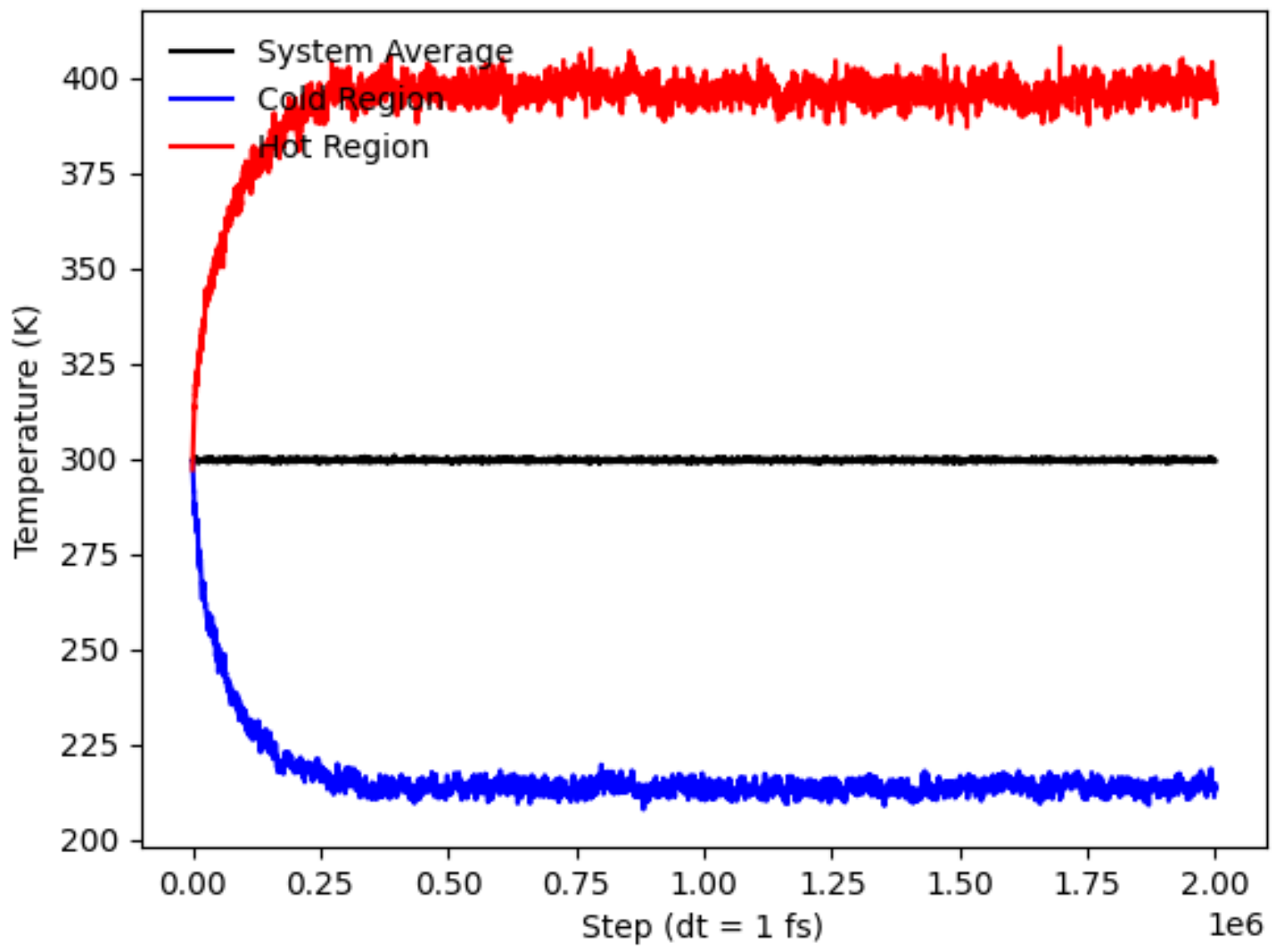


**FIG. S2**. Example illustration of steady state achieved while applying heat flux. Temperature of the heat source ("Hot") region, heat sink ("Cold") region and system average overall are plotted over time to show that these values varied as fluctuations about a constant average by the end of the NVE heating trajectory.

## 2. DEFECT CONCENTRATION SATURATION

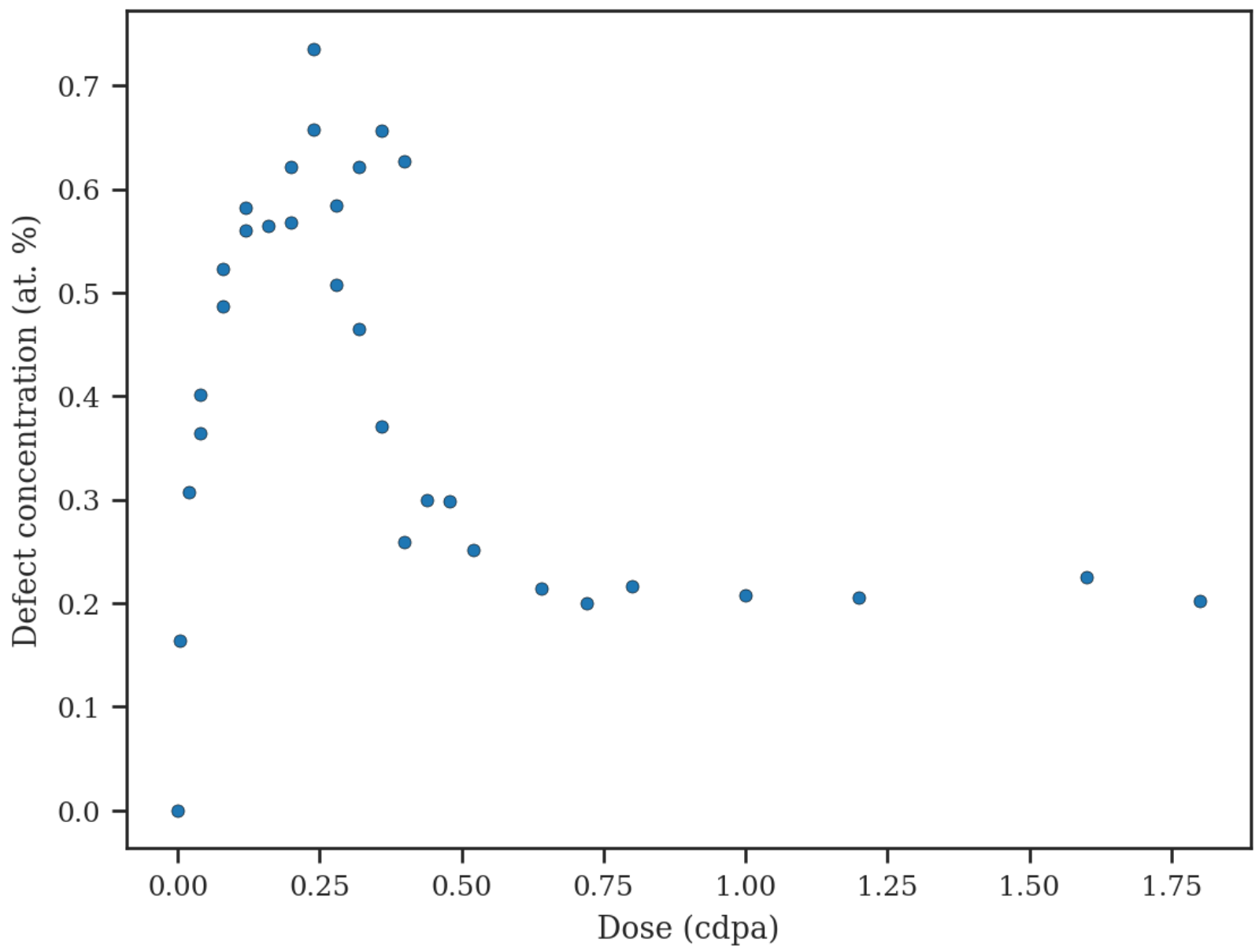


**FIG. S3**. For small systems (N=32), the average atomic percentage of point defects (sum of SIA and vacancies) observed in the target region (250k atoms) after annealing to 300 K plotted as a function of applied dose (*cdpa*)

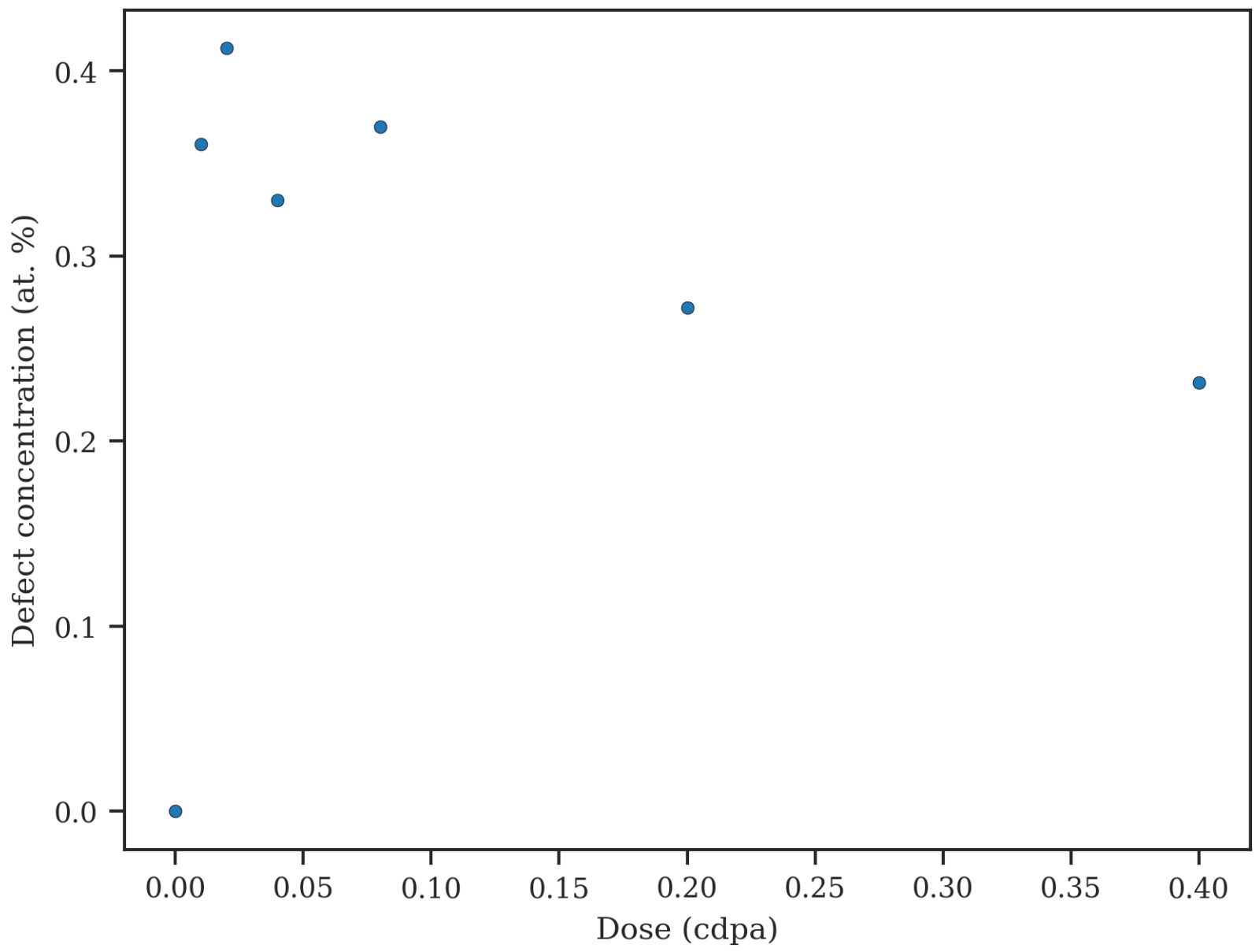


**FIG. S4**. For large systems (N=7), the average atomic percentage of point defects (sum of SIA and vacancies) observed in the target region (2M atoms) after annealing to 300 K plotted as a function of applied dose (*cdpa*)

## 3. DEFECT POSITIONS AND THERMAL CONDUCTIVITY FOR LARGE SYSTEMS

Target regions of all large irradiated systems. Locations of SIA (red) and vacancies (blue) are marked. The dimensionless $\kappa/\kappa_{\mathrm{m}}$ profile is given by the black line. Semi-transparent points indicate that the defect changed positions during the trajectory.

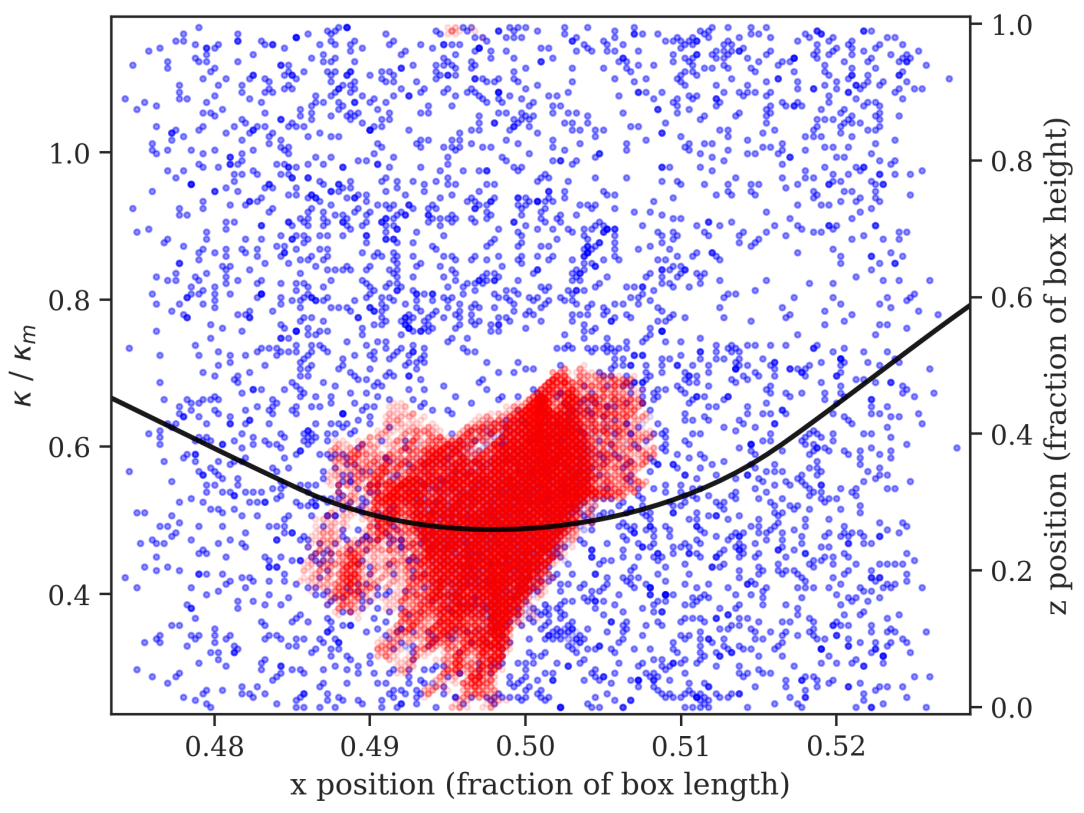


**FIG. S5**. Large system with *cdpa*=0.01

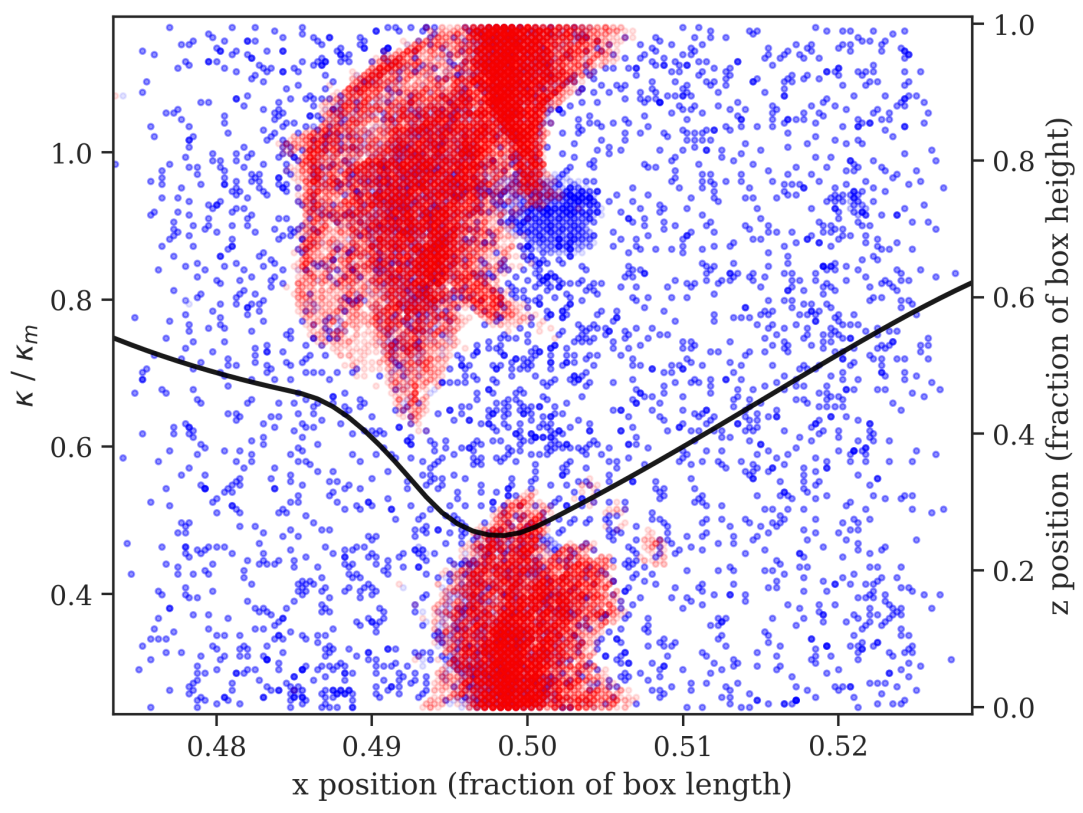


**FIG. S6**. Large system with *cdpa*=0.02

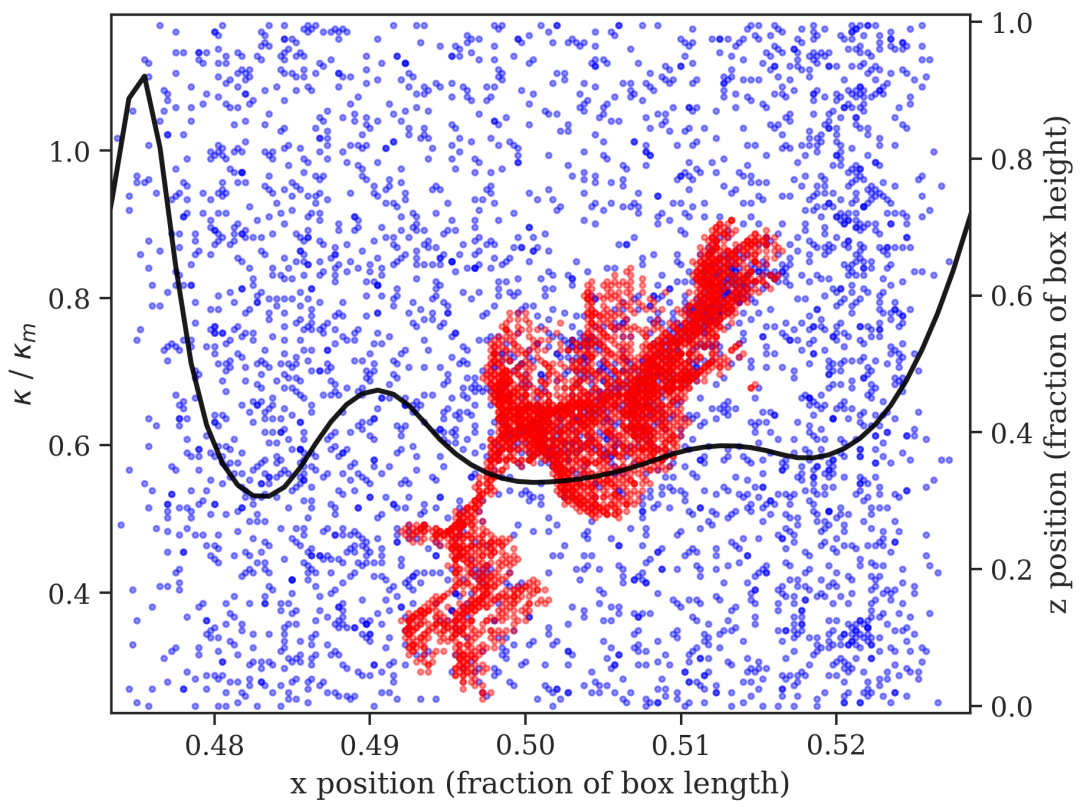


**FIG. S7**. Large system with *cdpa*=0.04

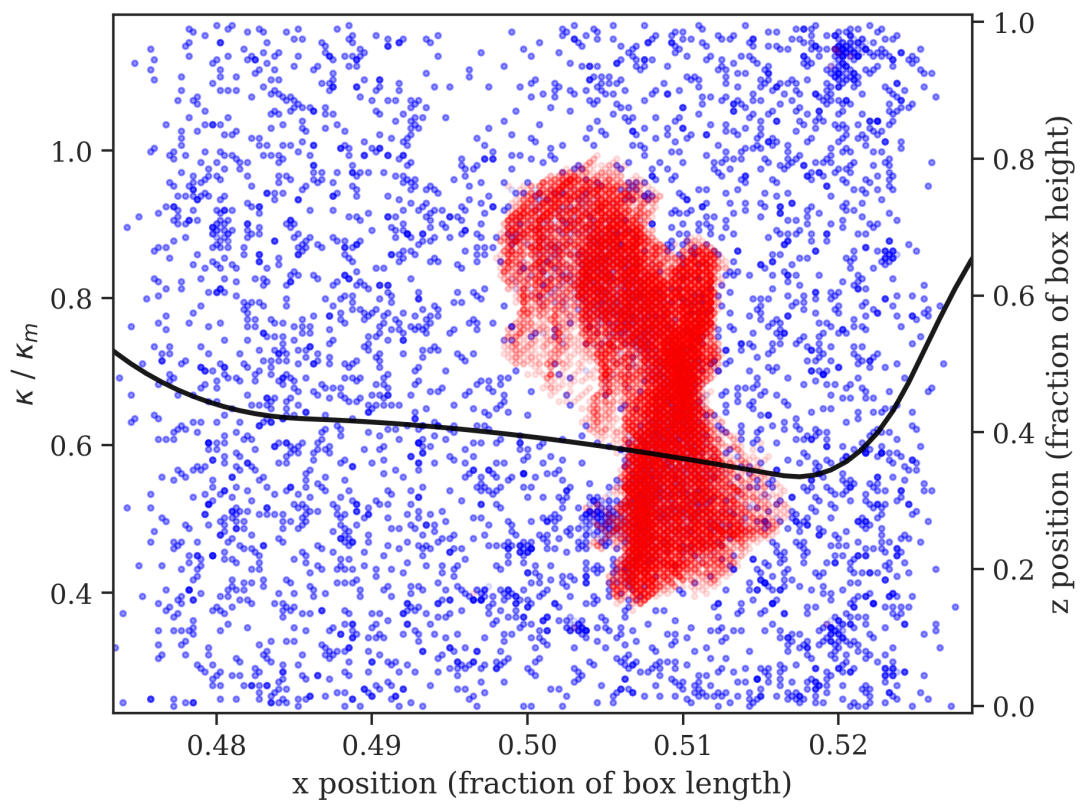


**FIG. S8**. Large system with *cdpa*=0.08

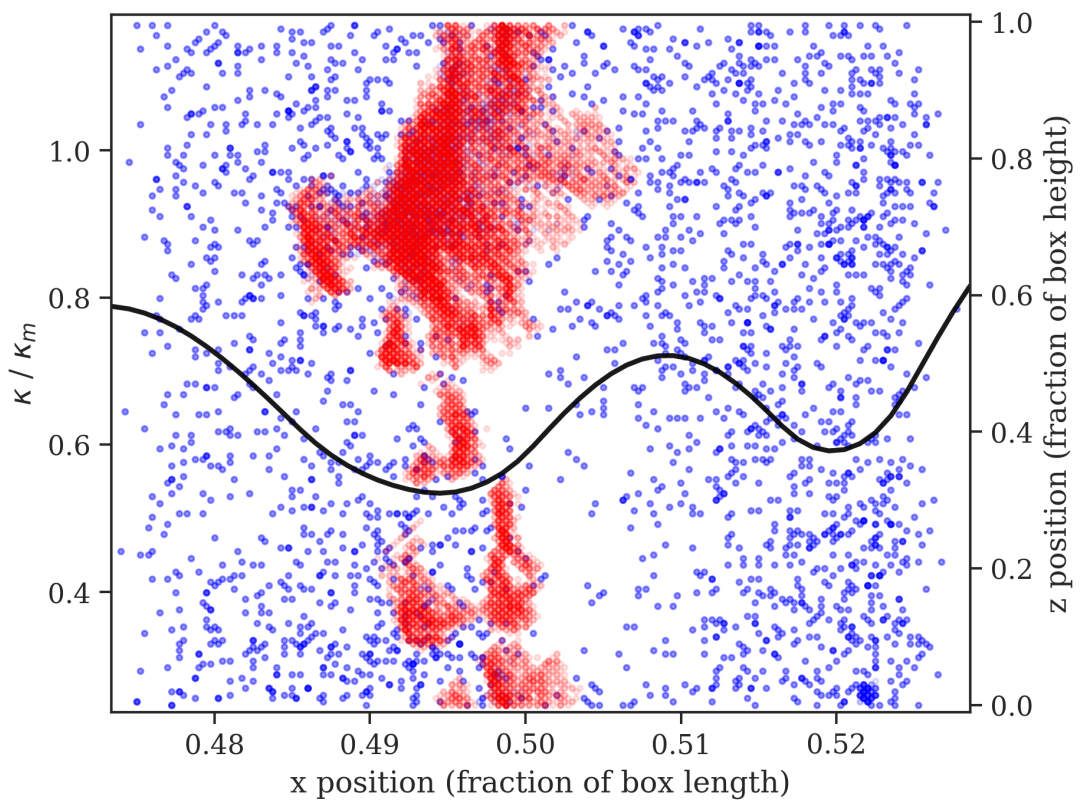


**FIG. S9**. Large system with *cdpa*=0.2

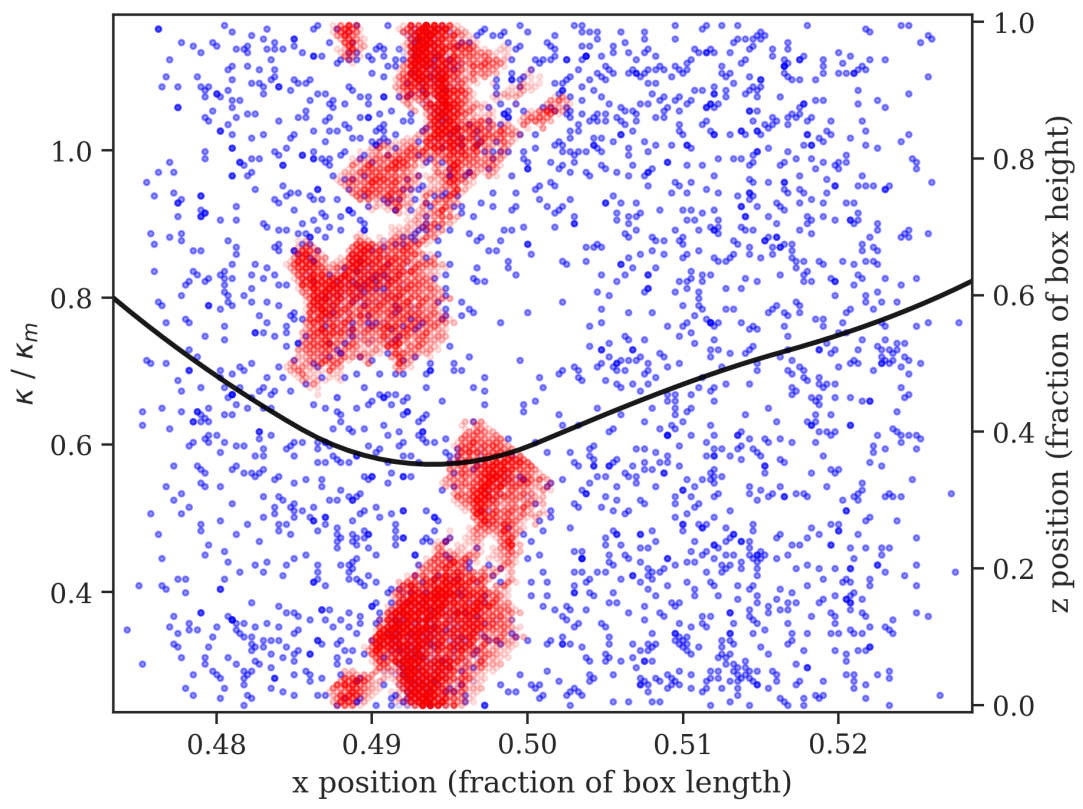


**FIG. S10**. Large system with *cdpa*=0.4

## 4. DEFECT POSITIONS AND THERMAL CONDUCTIVITY FOR SMALL SYSTEMS

Target regions of small irradiated systems. Locations of SIA (red) and vacancies (blue) are marked. The dimensionless $\kappa/\kappa_{\mathrm{m}}$ profile is given by the black line. Semi-transparent points indicate that the defect changed positions during the trajectory.

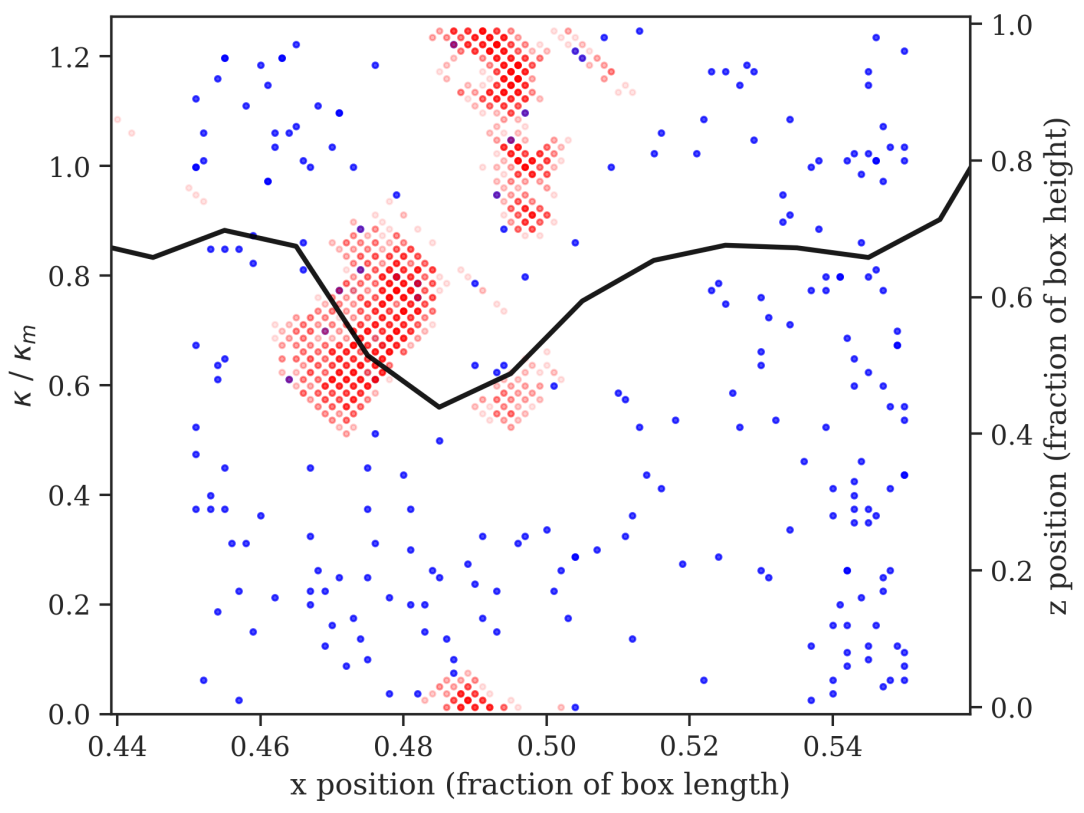


**FIG. S11**. Small system with *cdpa*=0.004

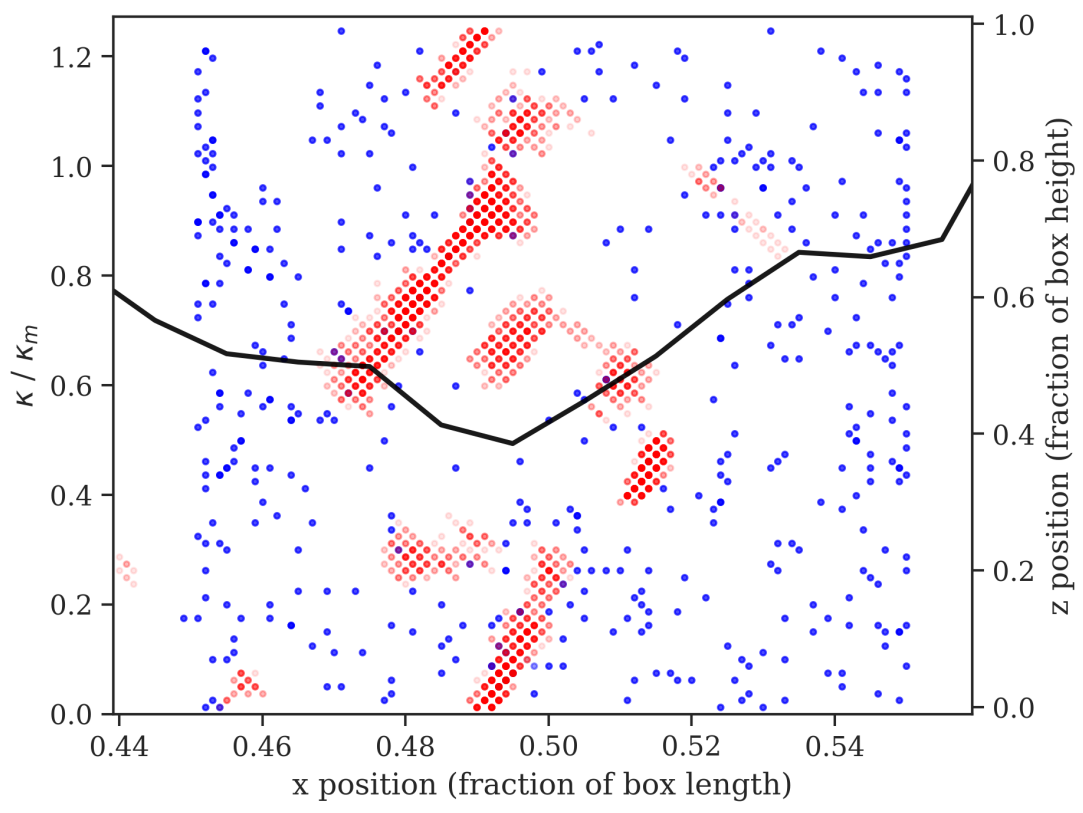


**FIG. S12**. Small system with *cdpa*=0.02

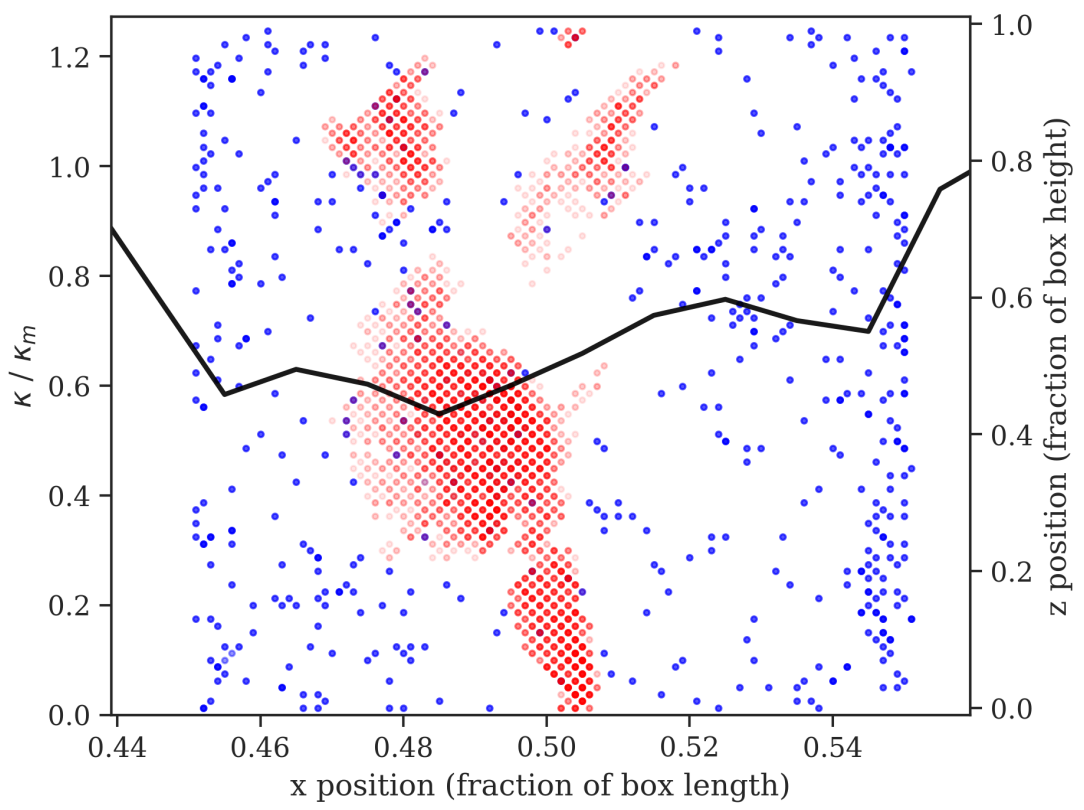


**FIG. S13**. Small system with *cdpa*=0.04

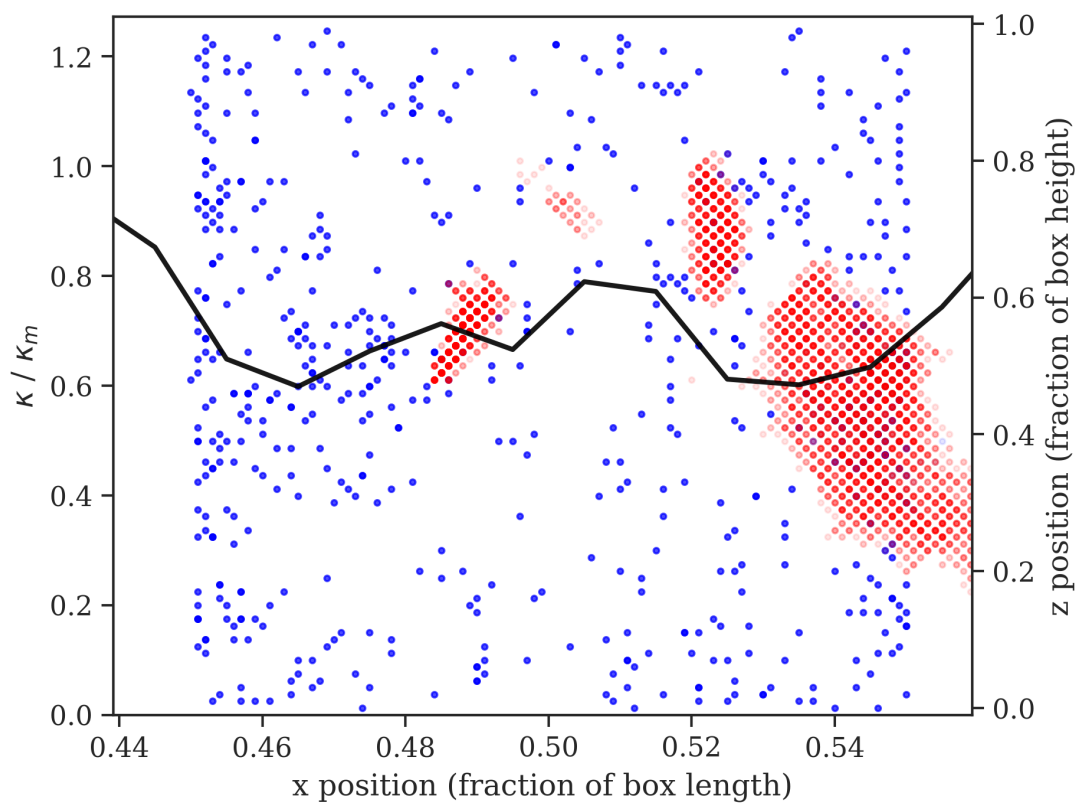


**FIG. S14**. Small system with *cdpa*=0.04

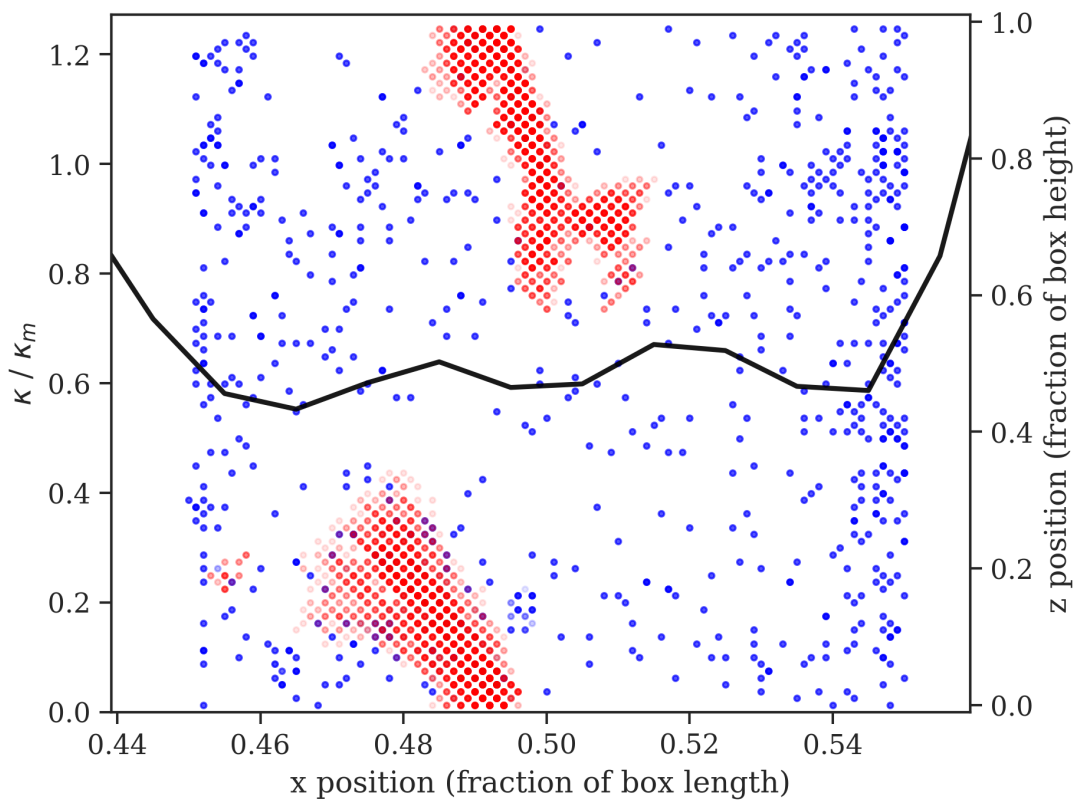


**FIG. S15**. Small system with *cdpa*=0.08

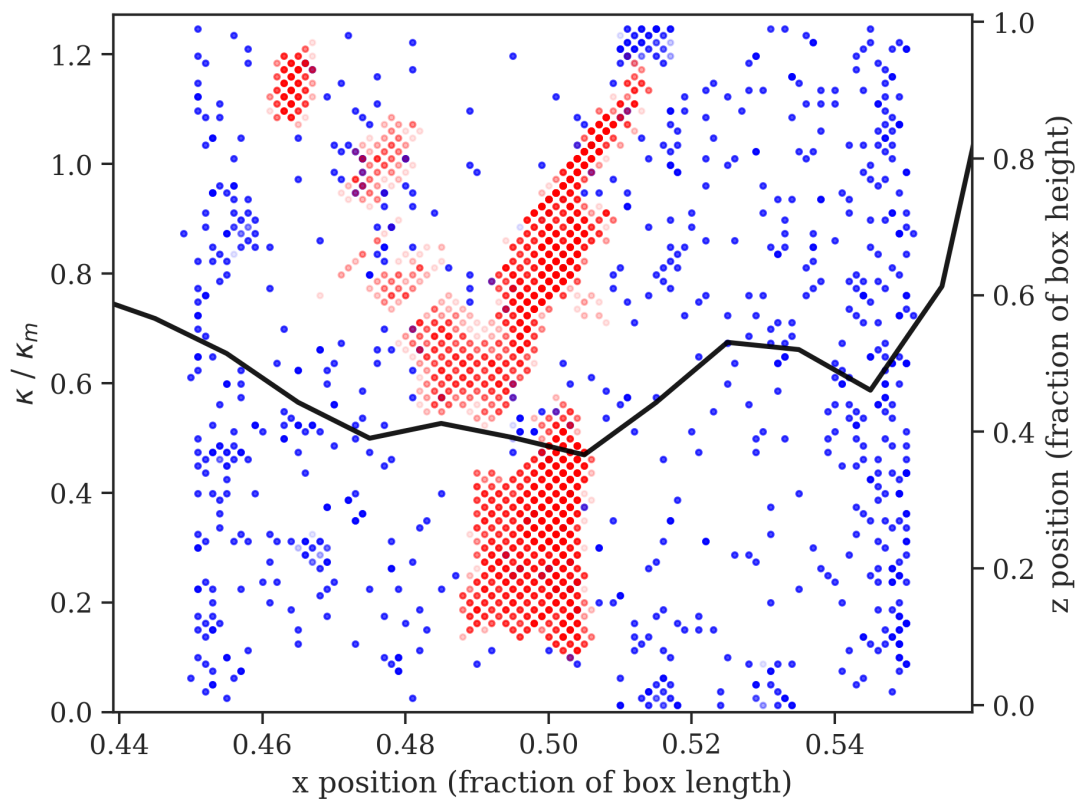


**FIG. S16**. Small system with *cdpa*=0.08

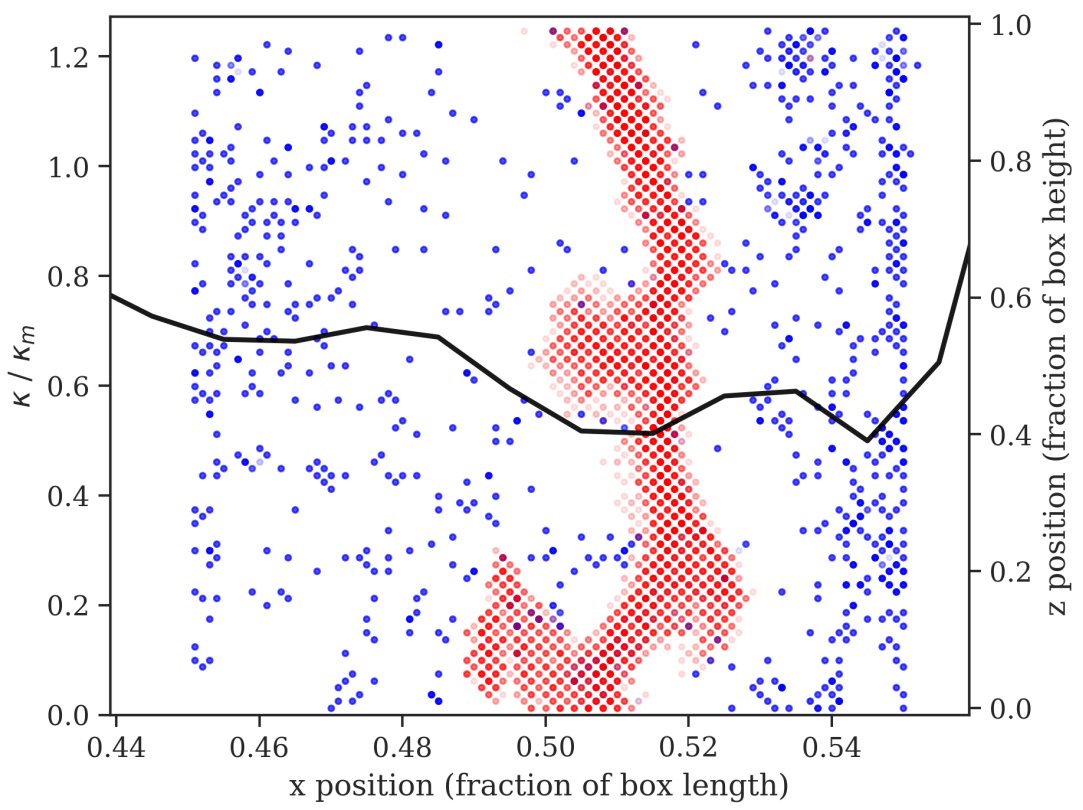


**FIG. S17**. Small system with *cdpa*=0.12

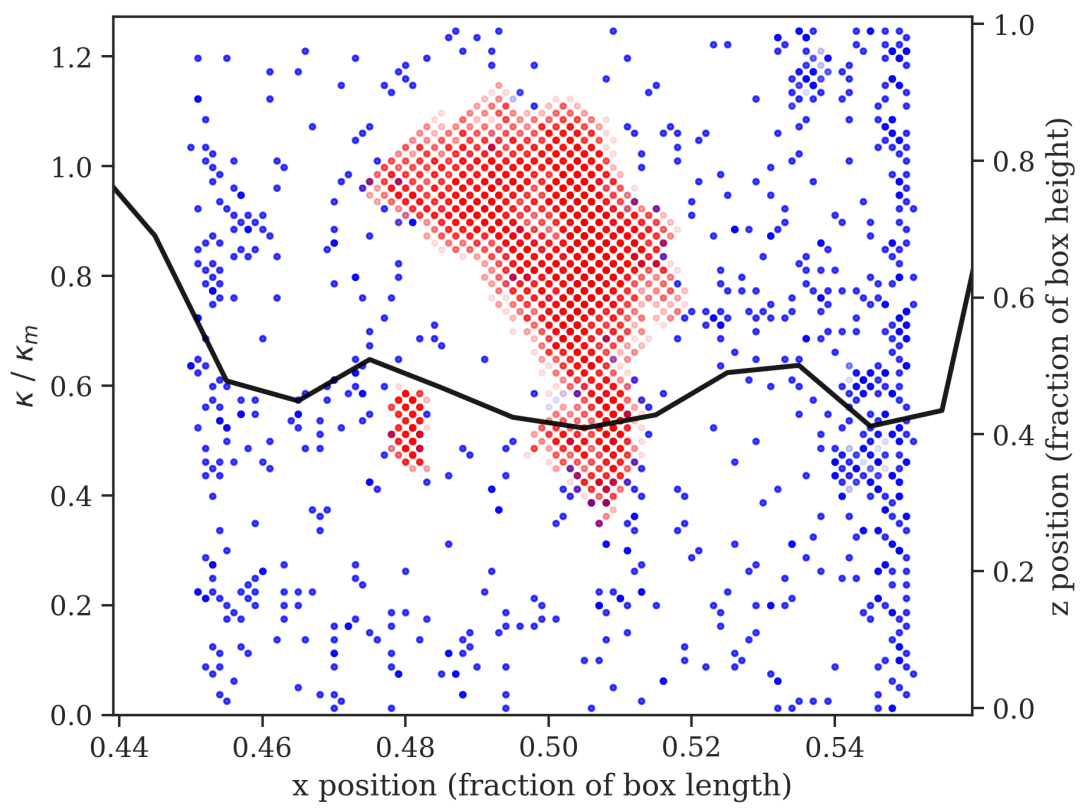


**FIG. S18**. Small system with *cdpa*=0.12

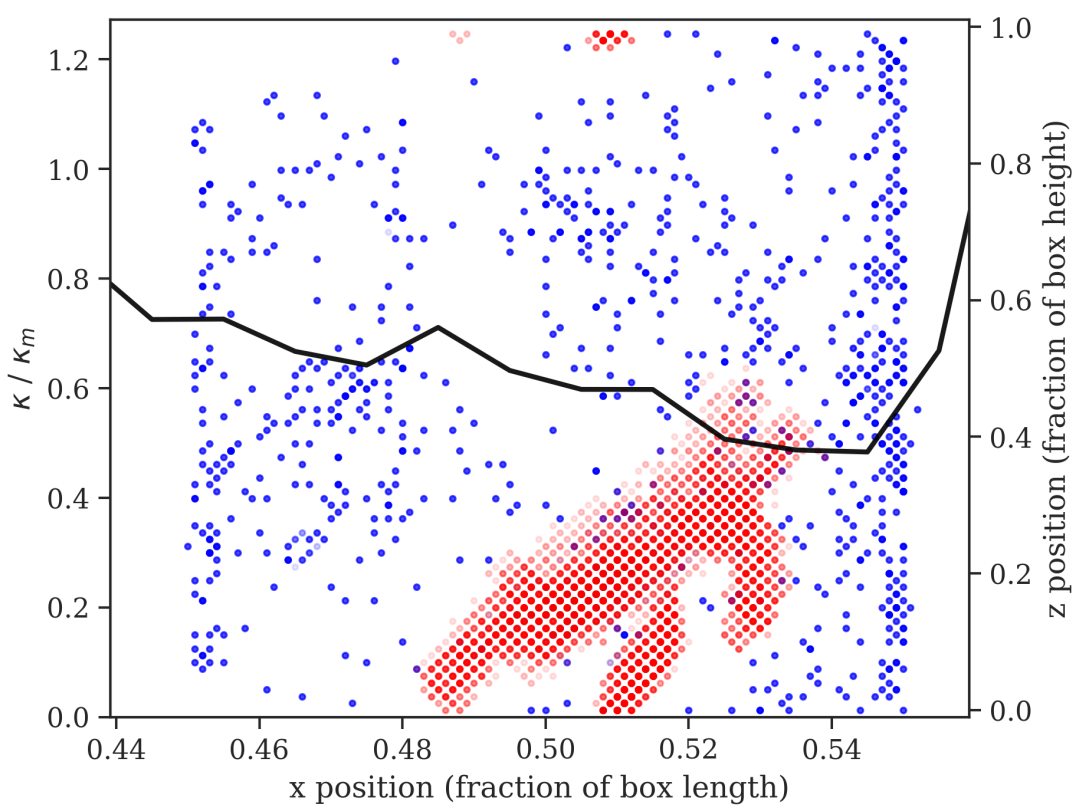


**FIG. S19**. Small system with *cdpa*=0.16

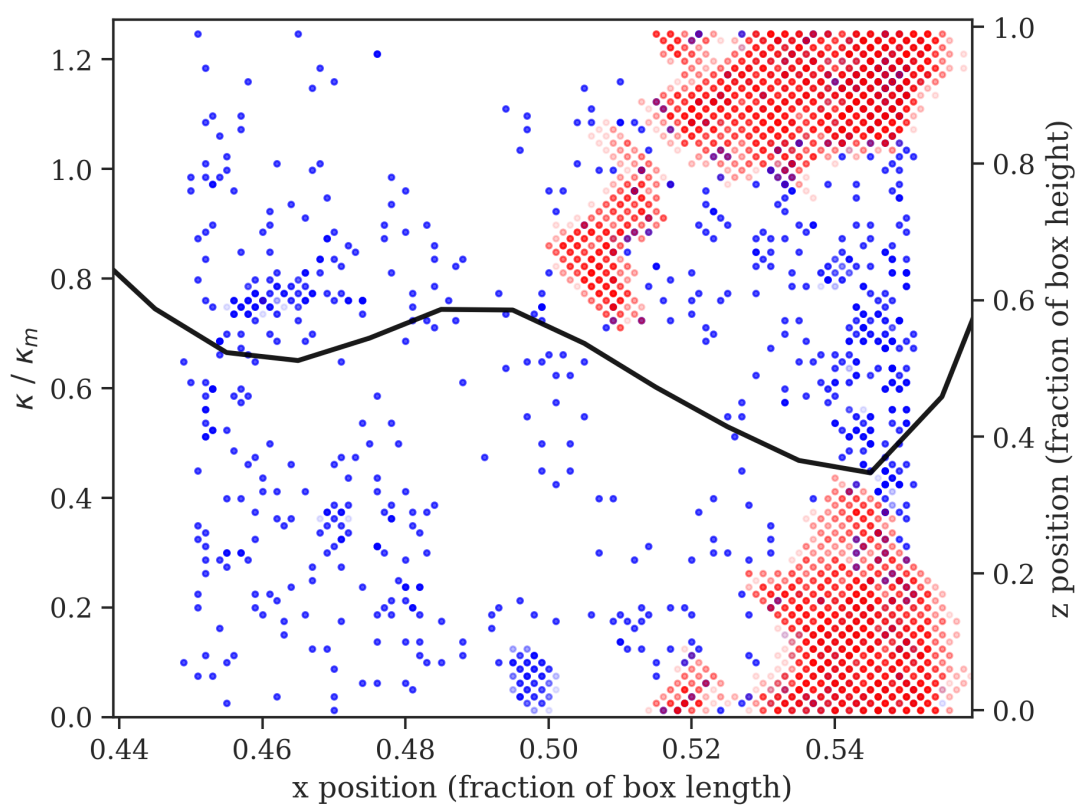


**FIG. S20**. Small system with *cdpa*=0.2

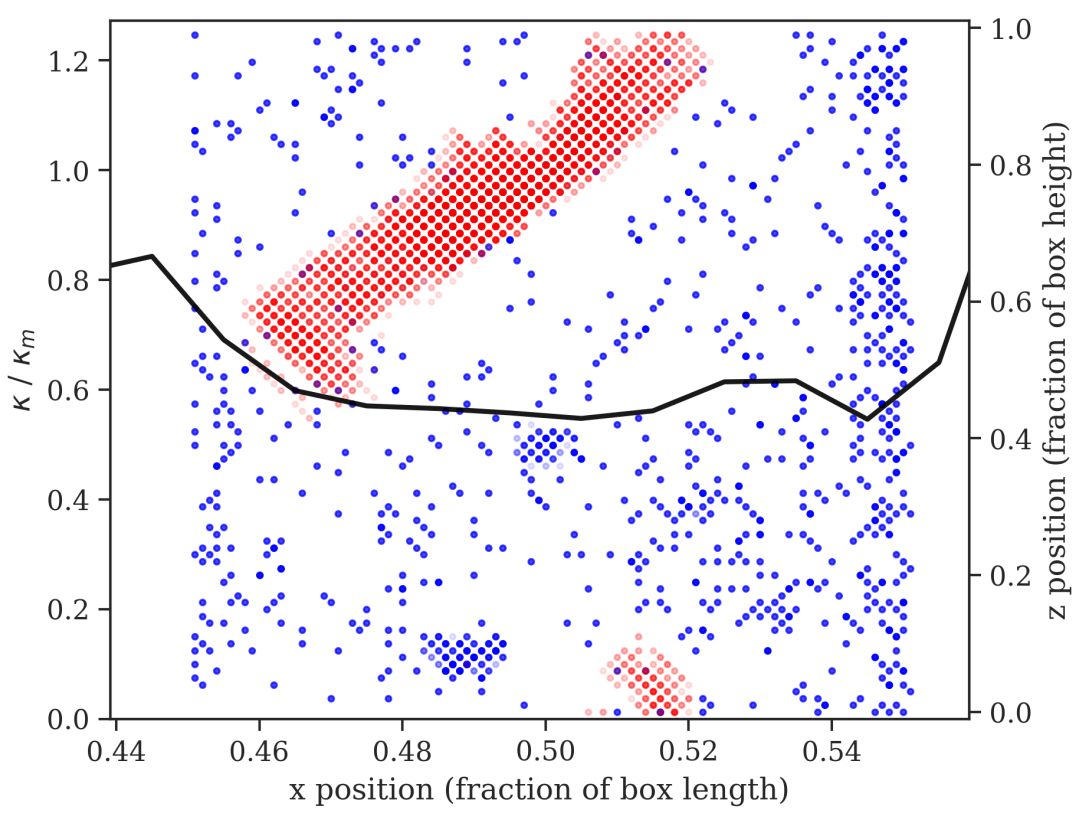


**FIG. S21**. Small system with *cdpa*=0.2

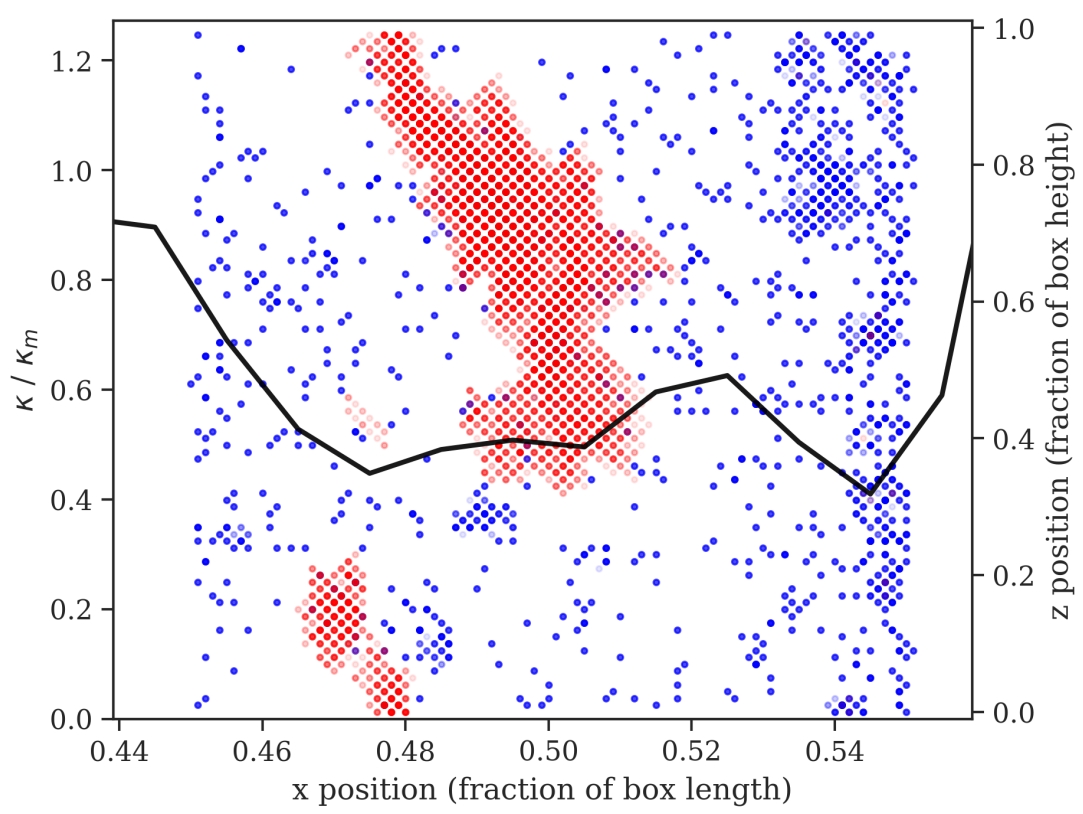


**FIG. S22**. Small system with *cdpa*=0.24

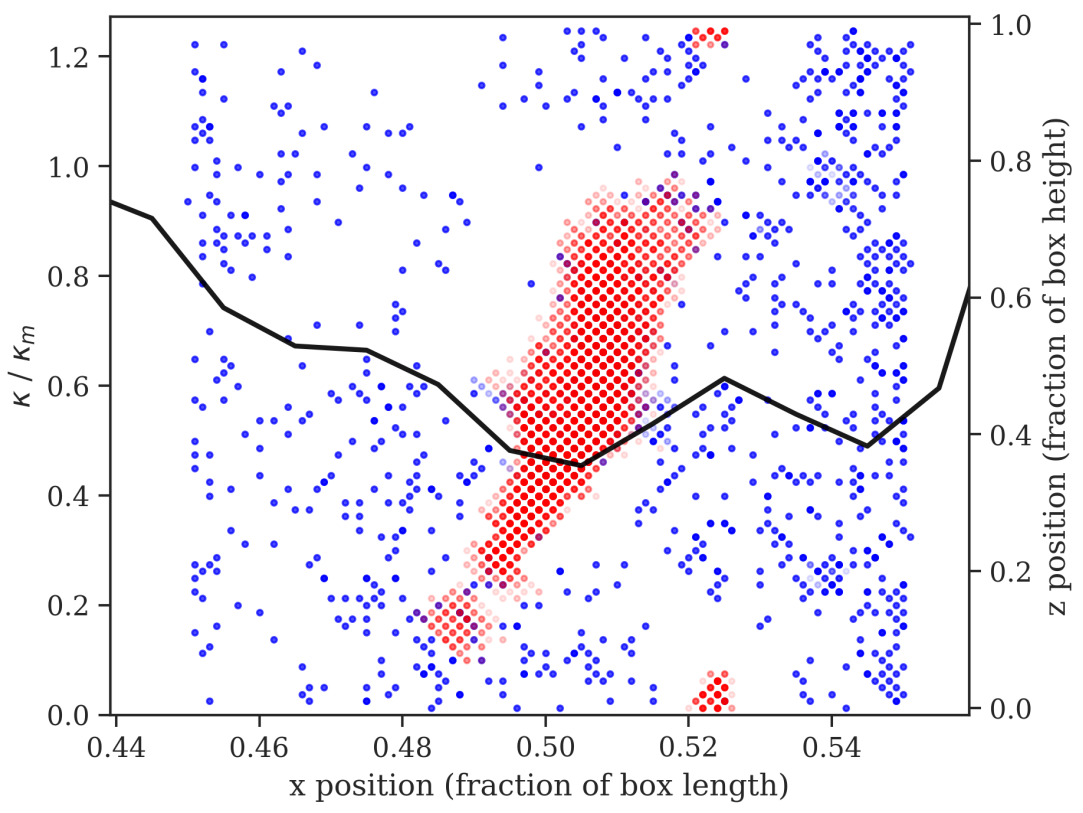


**FIG. S23**. Small system with *cdpa*=0.24

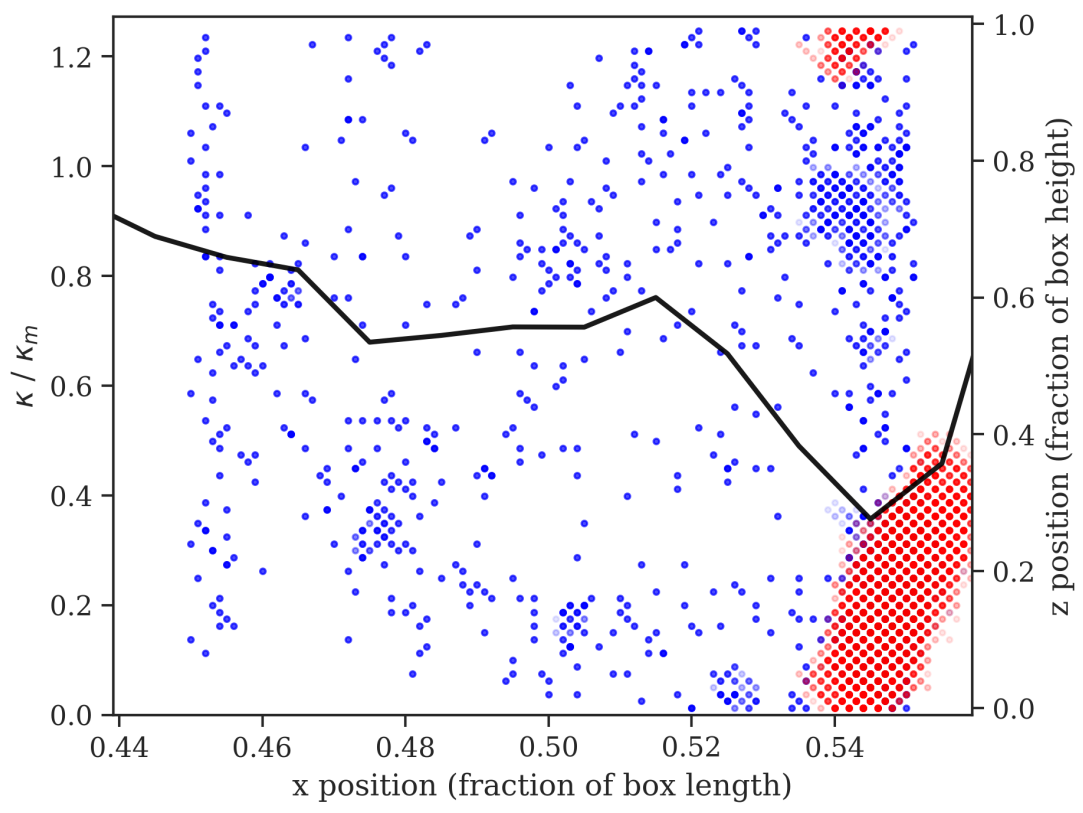


**FIG. S24**. Small system with *cdpa*=0.28

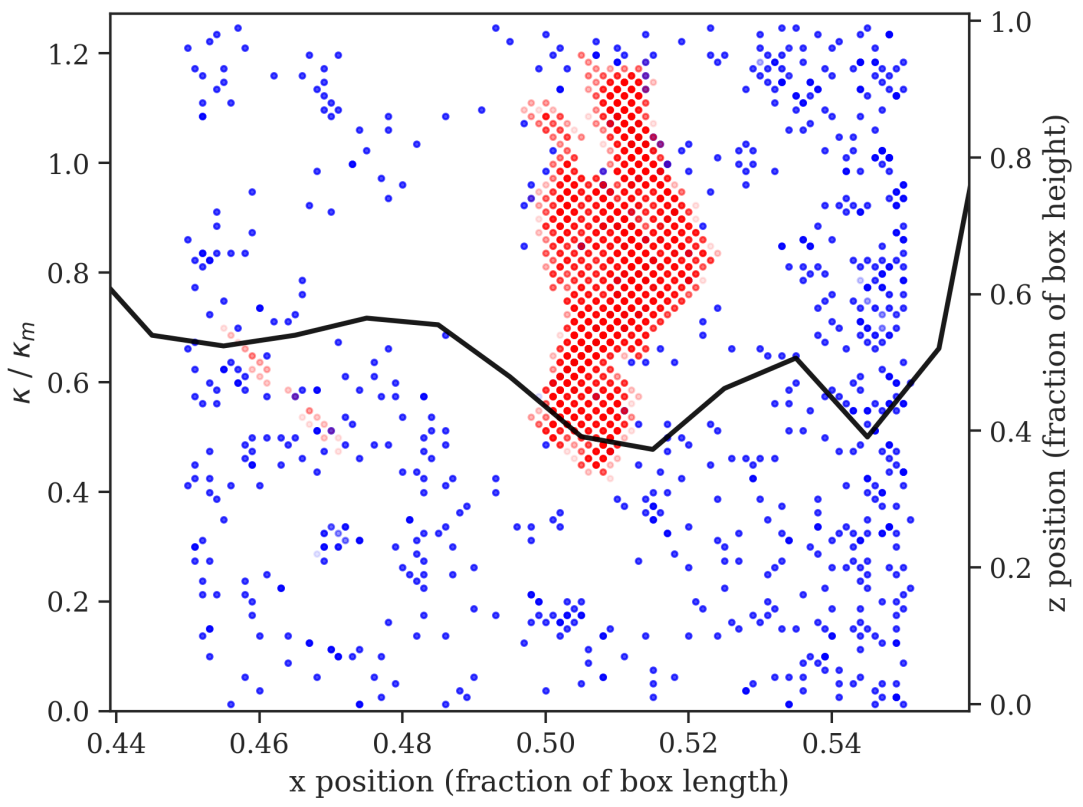


**FIG. S25**. Small system with *cdpa*=0.28

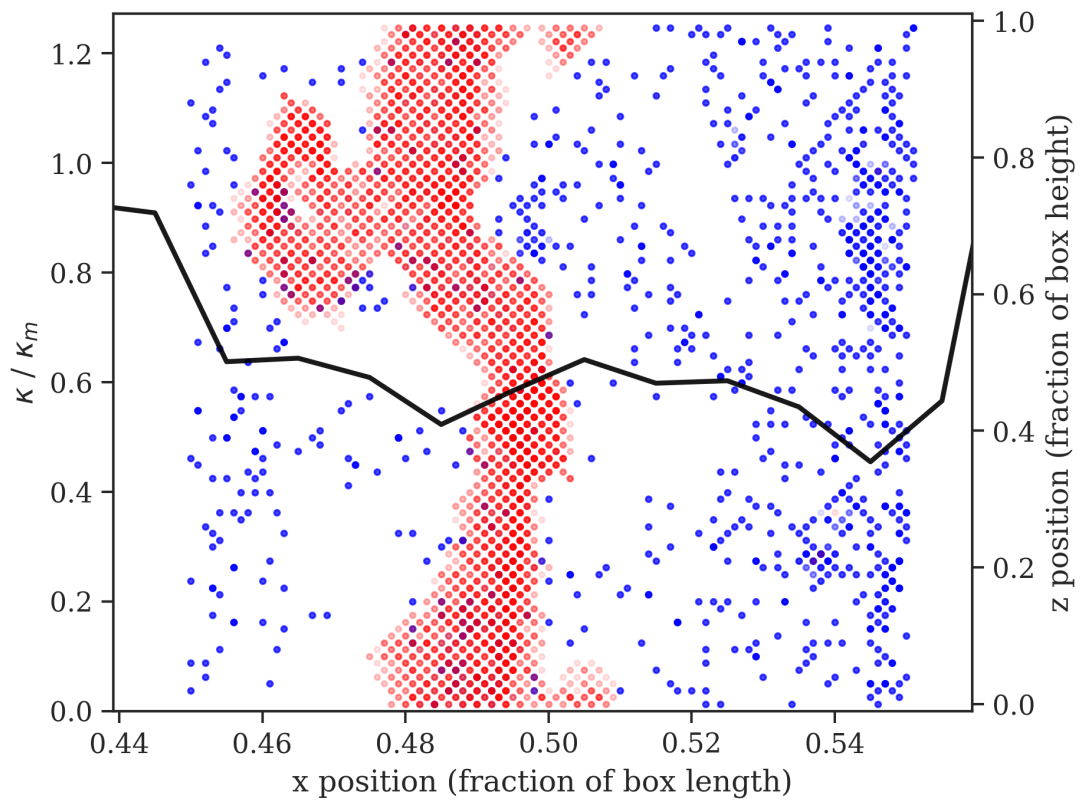


**FIG. S26**. Small system with *cdpa*=0.32

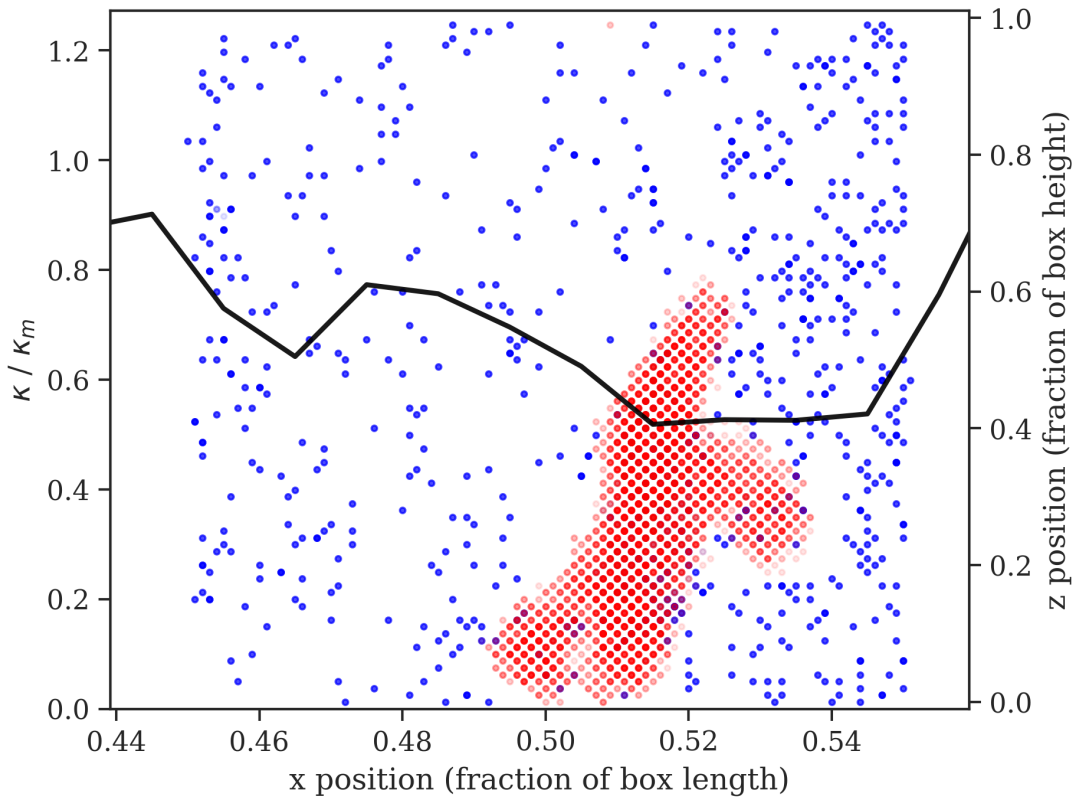


**FIG. S27**. Small system with *cdpa*=0.32

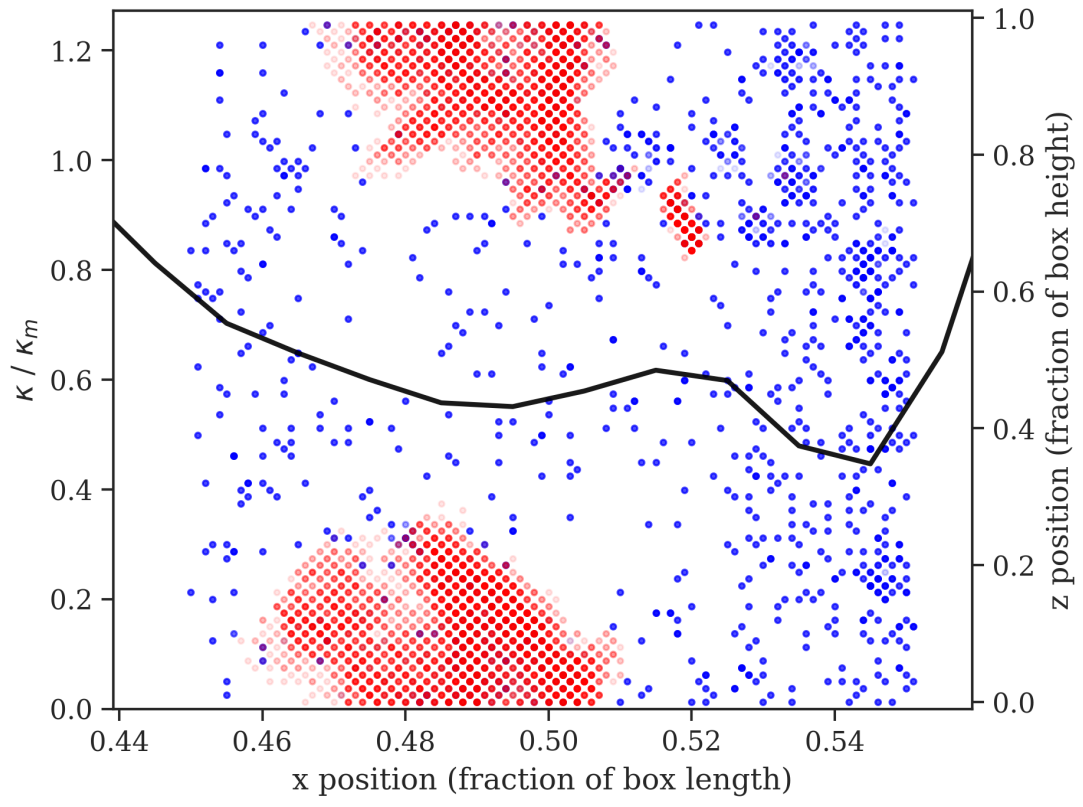


**FIG. S28**. Small system with *cdpa*=0.36

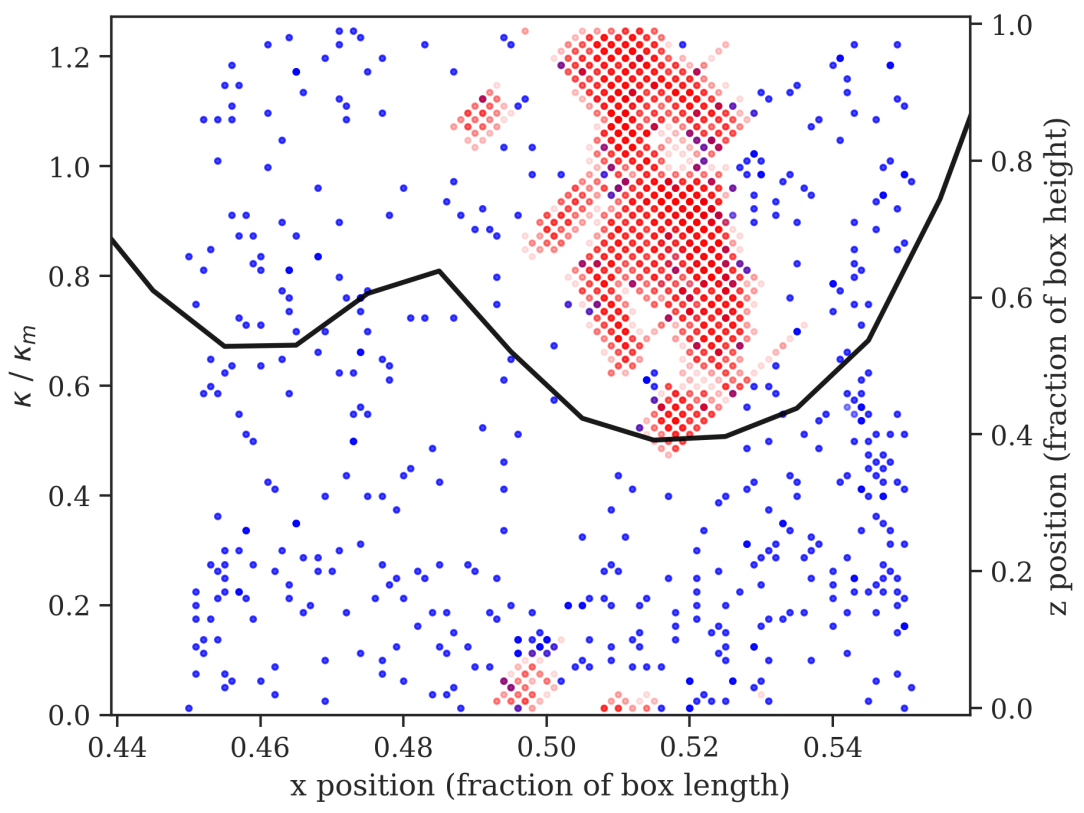


**FIG. S29**. Small system with *cdpa*=0.36

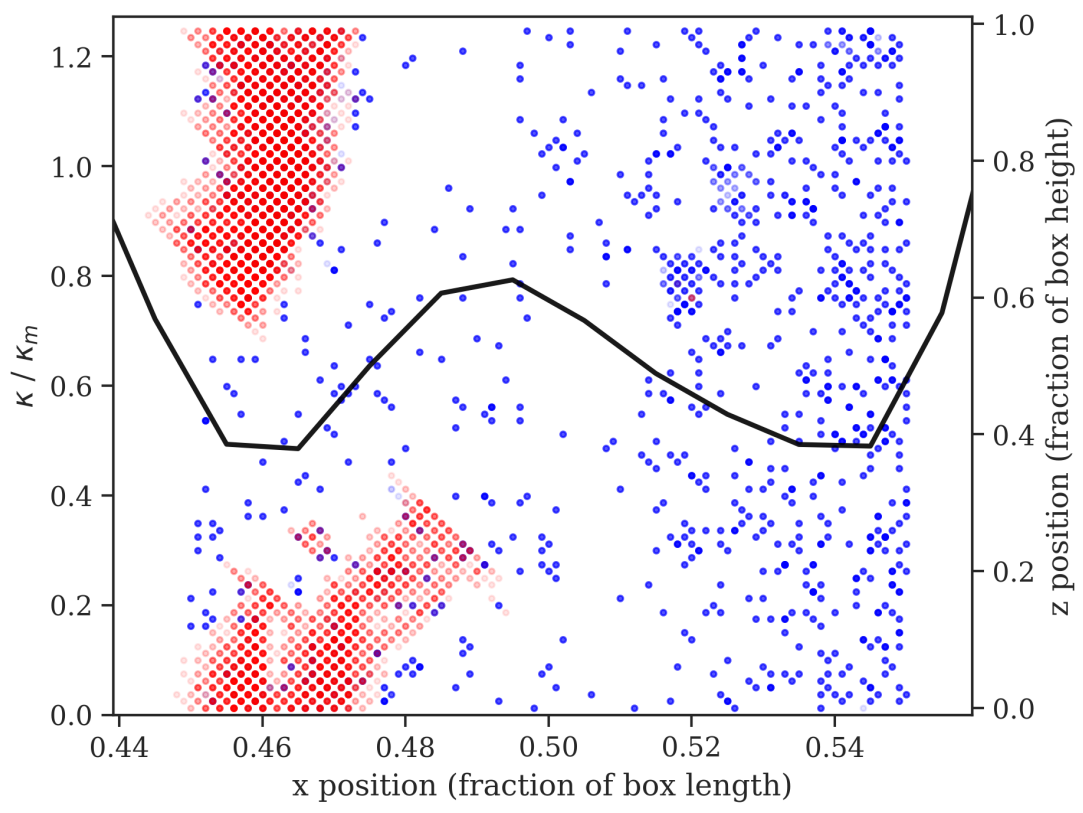


**FIG. S30**. Small system with *cdpa*=0.4

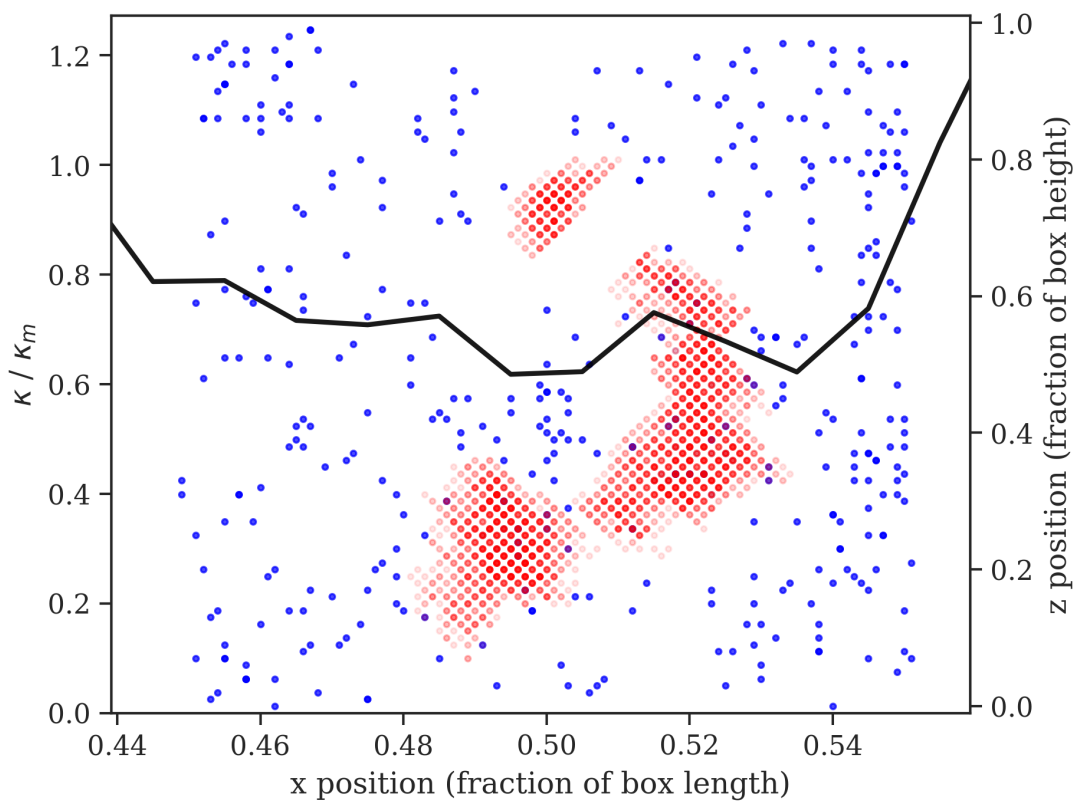


**FIG. S31**. Small system with *cdpa*=0.4

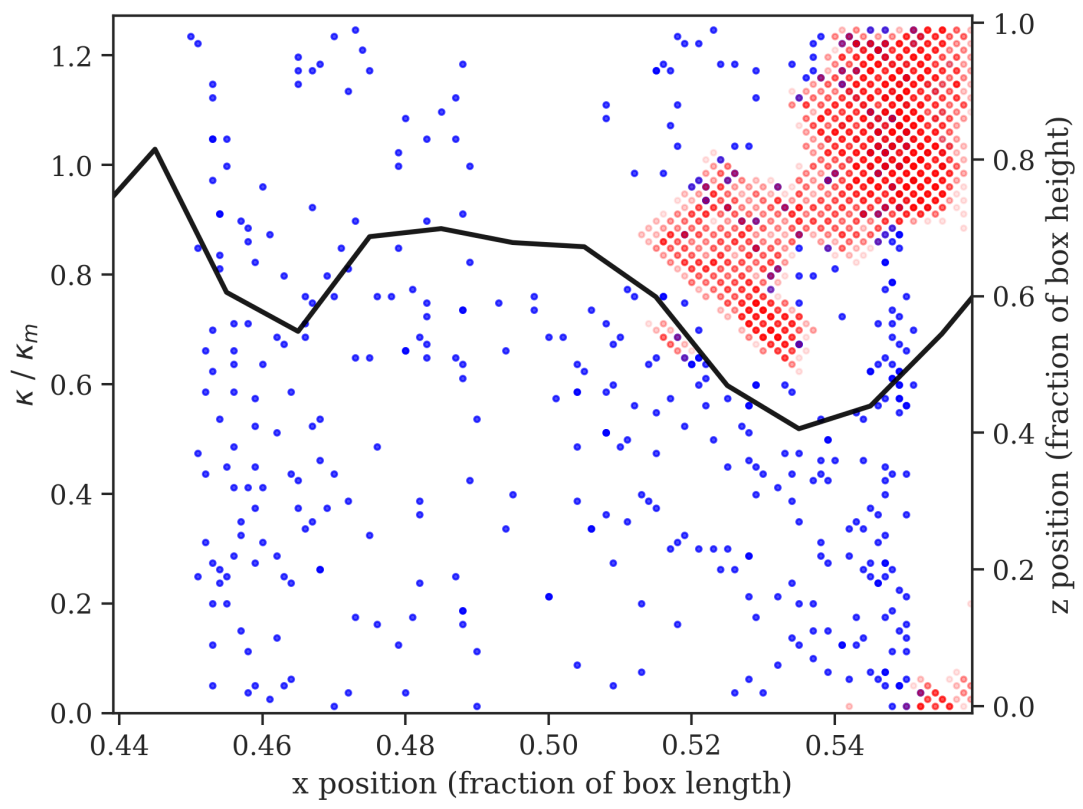


**FIG. S32**. Small system with *cdpa*=0.44

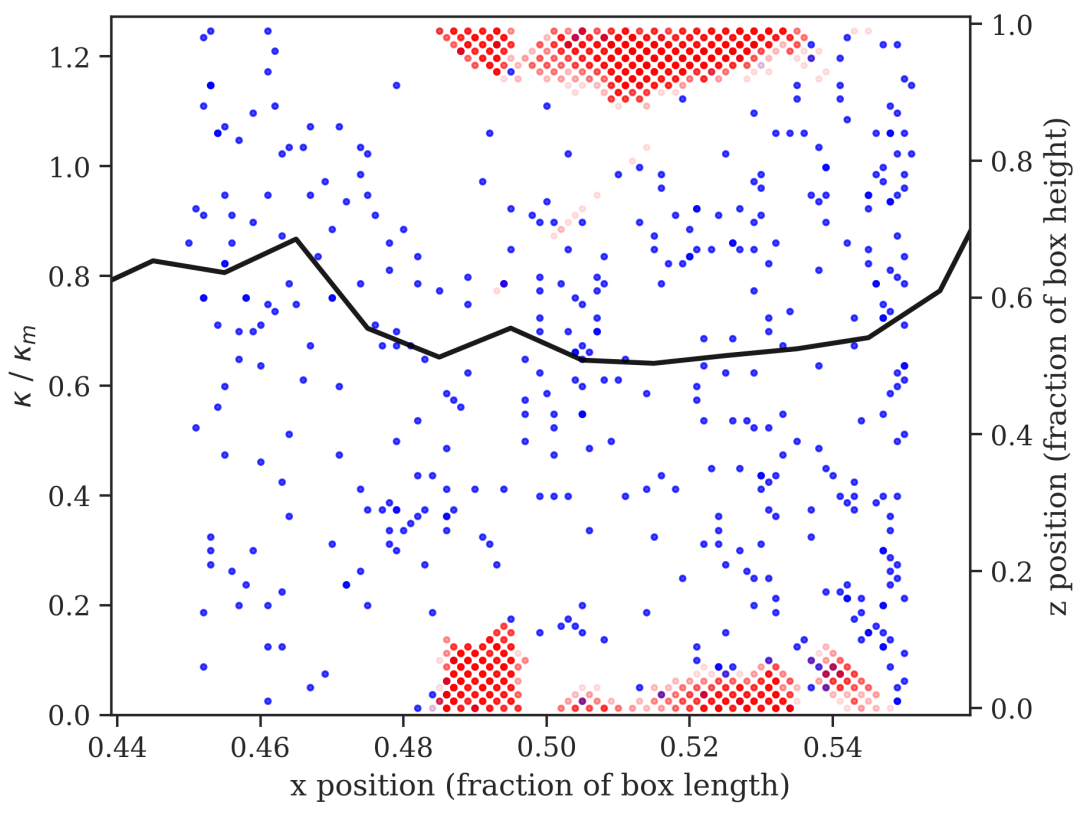


**FIG. S33**. Small system with *cdpa*=0.48

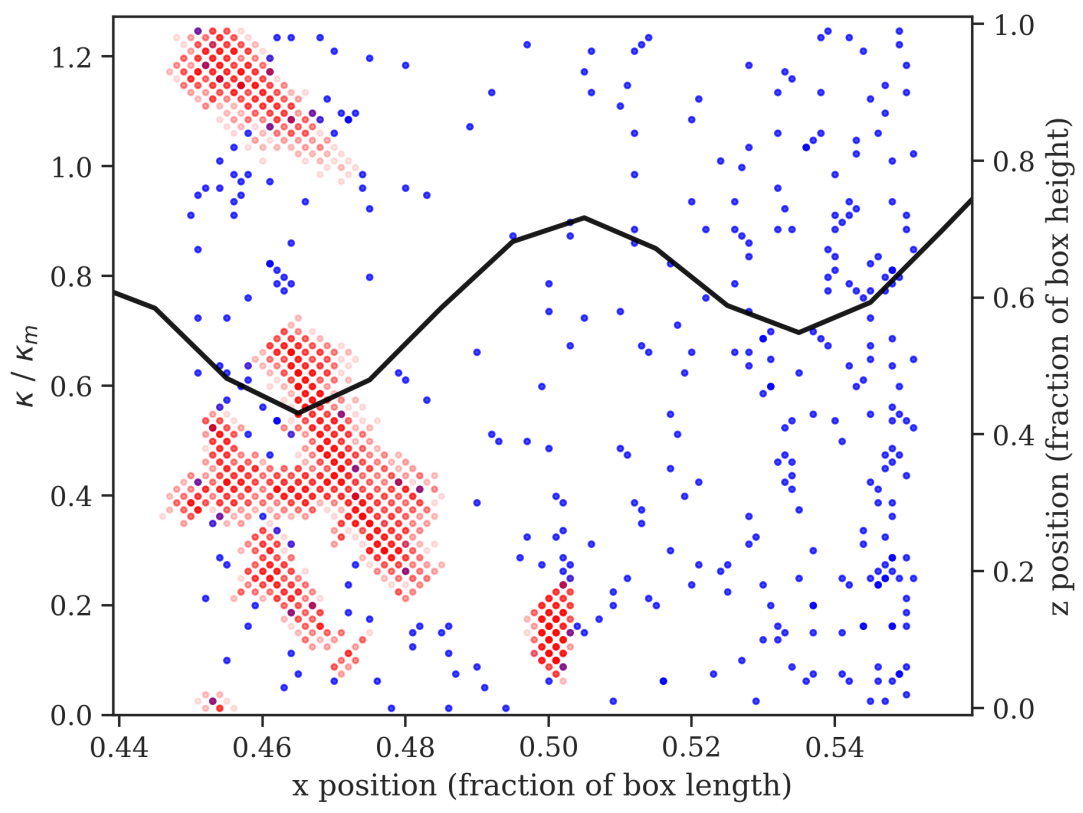


**FIG. S34**. Small system with *cdpa*=0.52

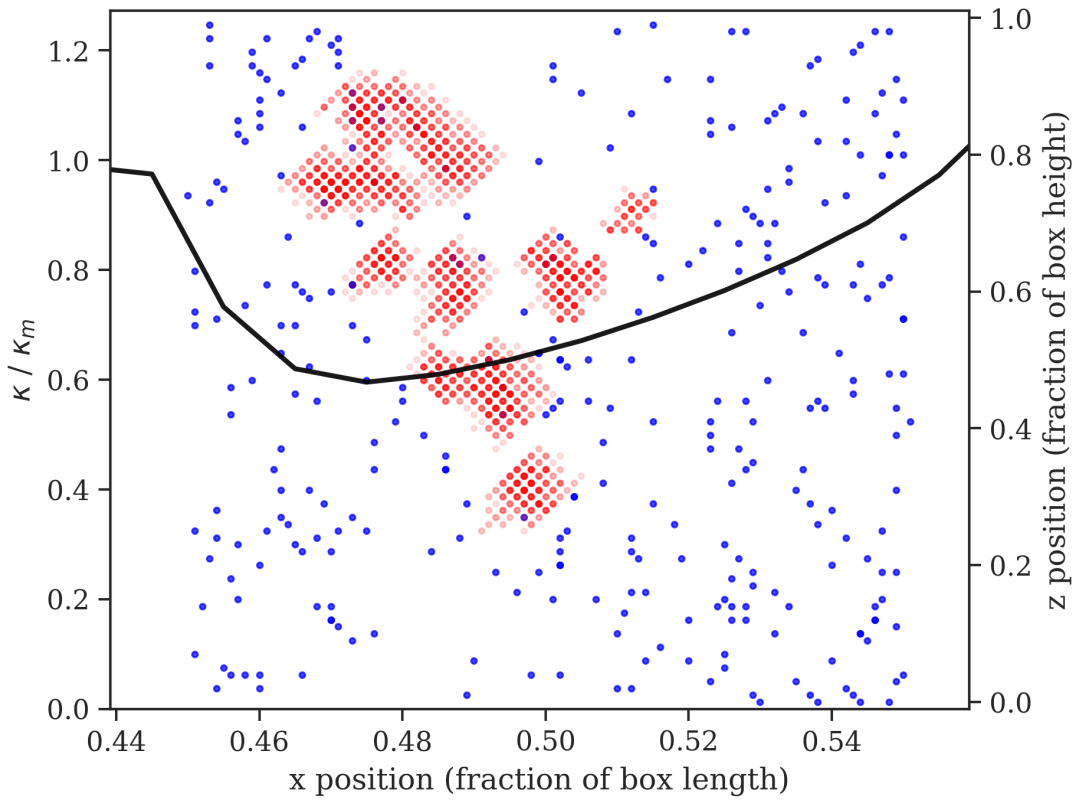


**FIG. S35**. Small system with *cdpa*=0.64

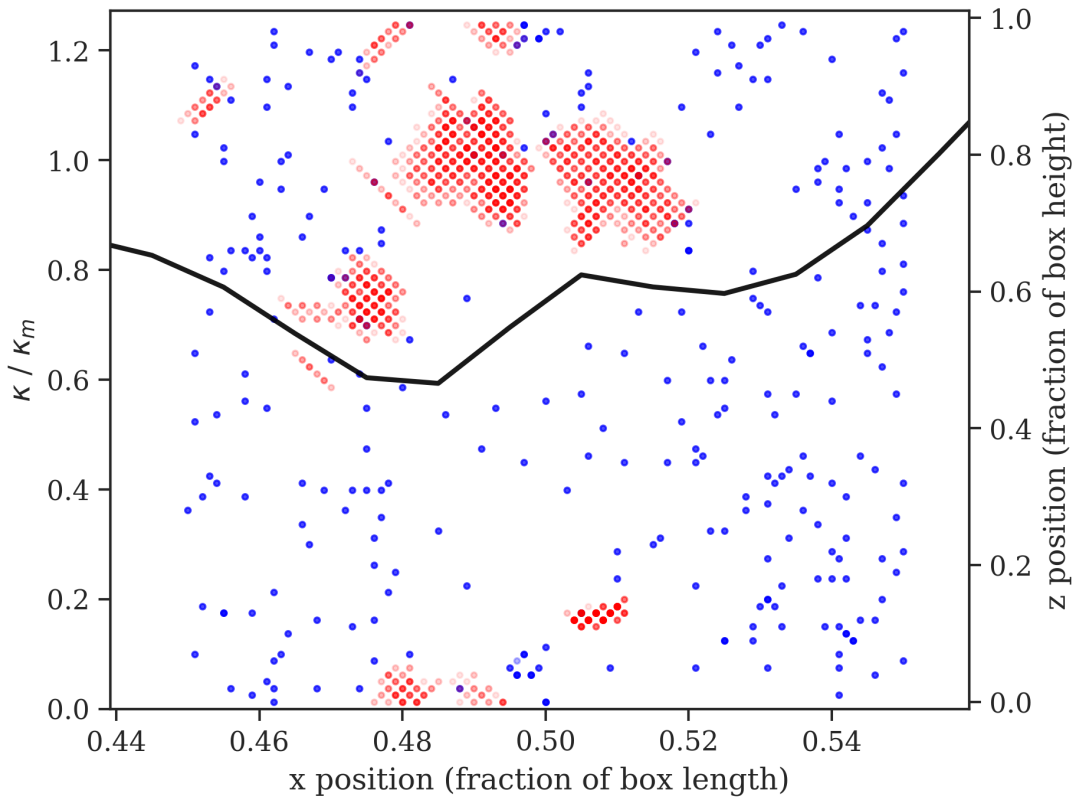


**FIG. S36**. Small system with *cdpa*=0.72

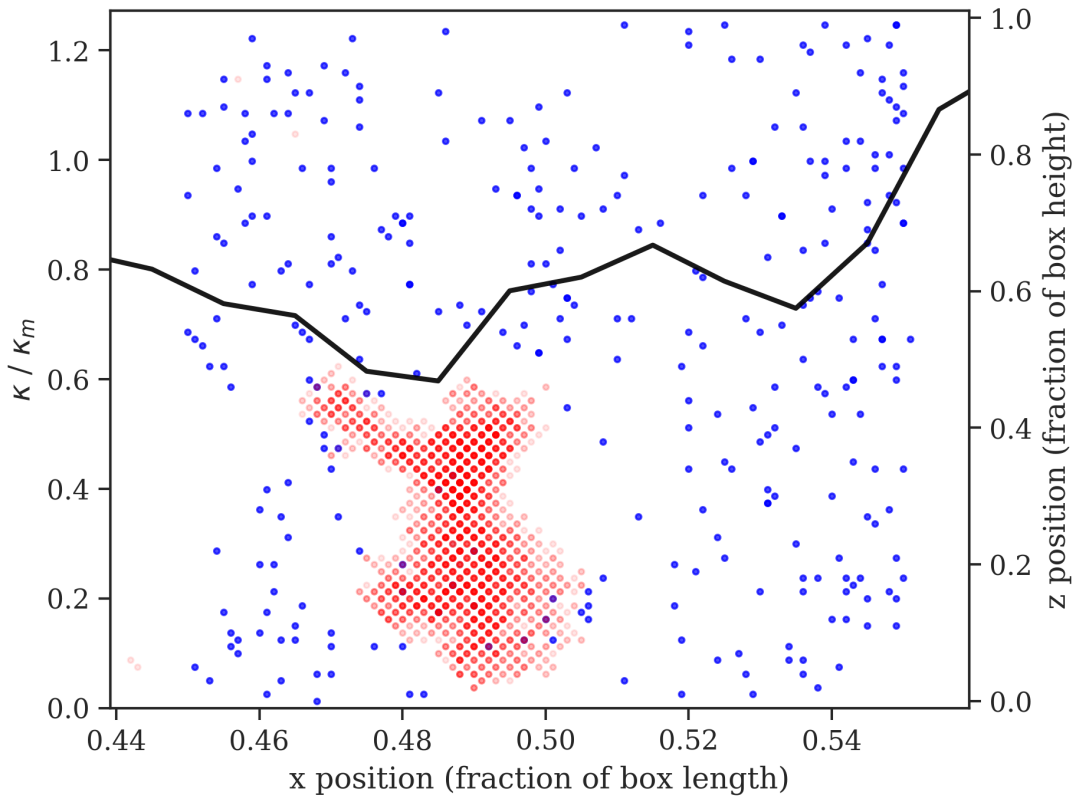


**FIG. S37**. Small system with *cdpa*=0.8

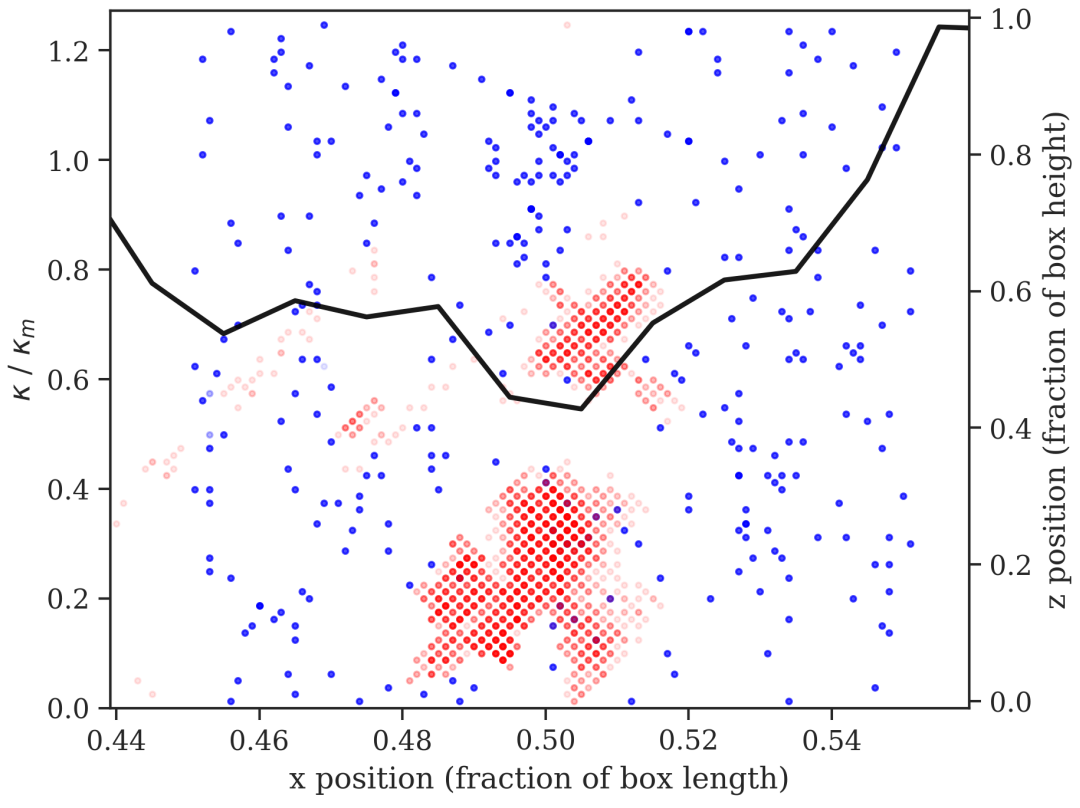


**FIG. S38**. Small system with *cdpa*=1.0

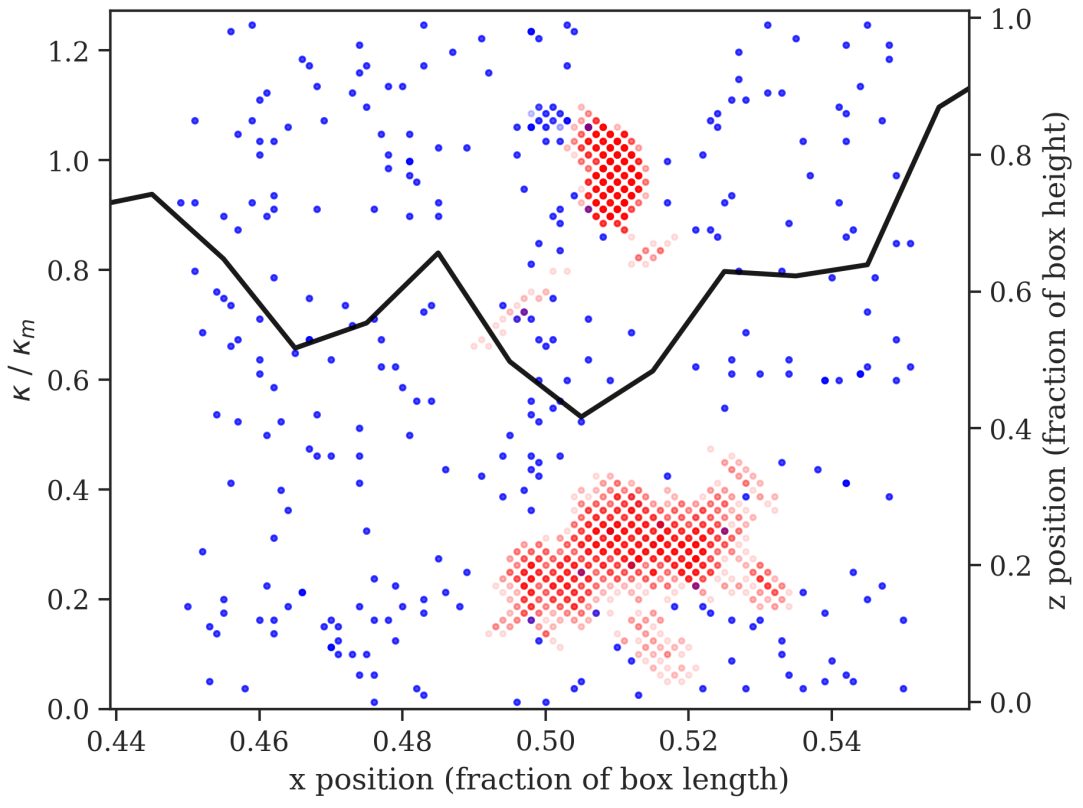


**FIG. S39**. Small system with *cdpa*=1.2

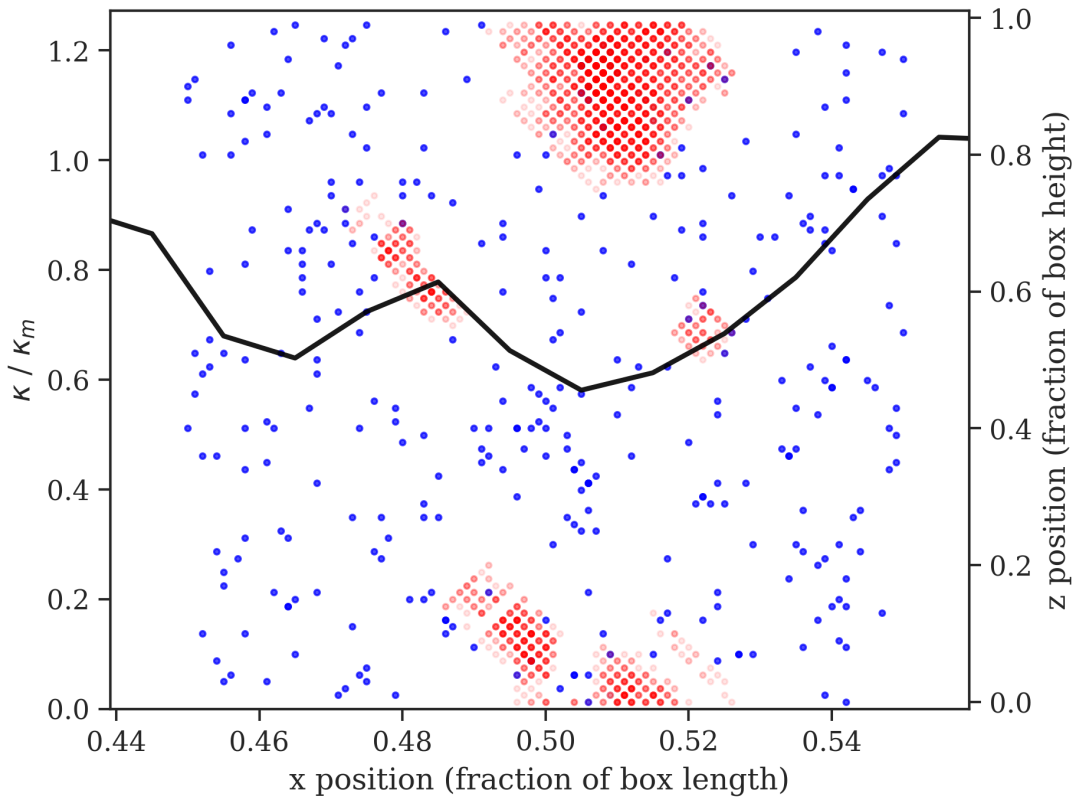


**FIG. S40**. Small system with *cdpa*=1.6

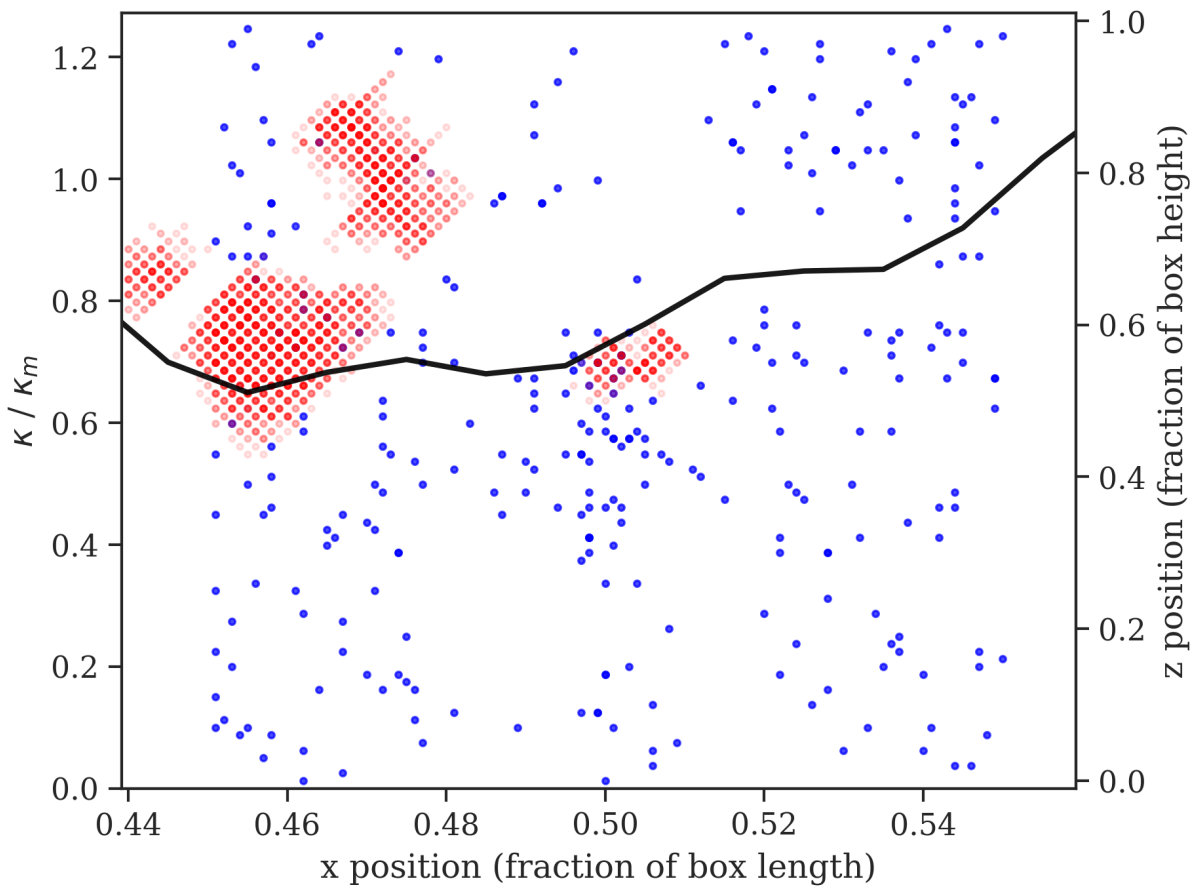


**FIG. S41**. Small system with *cdpa*=1.8

## 5. PARAMETER ESTIMATION VIA MULTIPLE LINEAR REGRESSION

Ordinary Least Squares (OLS) multiple linear regression was used to predict average $\kappa$ ($\bar{\kappa}$) for small and large systems separately, given independent variables $1/T$, $N_{\mathrm{Vac}}$, and $N_{\mathrm{SIA}}$ as

$$\bar{\kappa} = \hat{\beta}_0 + \hat{\beta}_1 \frac{1}{T} + \hat{\beta}_2 N_{\mathrm{Vac}} + \hat{\beta}_3 N_{\mathrm{SIA}} \ , \tag{1}$$

where estimates of coefficients $\hat{\beta}_{\mathrm{i}}$ minimized the sum of the squared errors between observations and predictions.

The OLS linear models are summarized in Table S1. The model for small systems ($R^2 = 0.528$) revealed that all three independent variables had statistically significant correlations with $\kappa$ ($p$-value $< 0.01$). The values of $R^2$ for the small and large models can be interpreted as the models explaining only 53% or 64% of the variance in the observations about their means, respectively, prompting more sophisticated approaches to data modeling.

**TABLE S1**. Summary of OLS multiple linear regression models of the target regions of the 32 small systems and 7 large systems. The 97.5% confidence interval based on the standard error for each independent variable is given in the final two columns. *Statistically significant with 99% confidence

| **Small Systems -** $R^2 = 0.528$ | | | | | | |
|---|---|---|---|---|---|---|
| | $\hat{\beta}$ | std err | $t$ | $p$ | [0.025 | 0.975] |
| Int | 8.2 | 1.214 | 6.76 | 0.00* | 5.8 | 10.6 |
| $1/T$ | 1183 | 327.3 | 3.61 | 0.00* | 539 | 1826 |
| $N_{\mathrm{Vac}}$ | -0.025 | 0.002 | -14.96 | 0.00* | -0.03 | -0.02 |
| $N_{\mathrm{SIA}}$ | -0.01 | 0.001 | -12.55 | 0.00* | -0.01 | -0.009 |

| **Large Systems -** $R^2 = 0.642$ | | | | | | |
|---|---|---|---|---|---|---|
| | $\hat{\beta}$ | std err | $t$ | $p$ | [0.025 | 0.975] |
| Int | 6.91 | 2.14 | 3.23 | 0.001 | 2.70 | 11.1 |
| $1/T$ | 3283 | 625 | 5.25 | 0.000* | 2053 | 4513 |
| $N_{\mathrm{Vac}}$ | -0.0622 | 0.003 | -20.5 | 0.000* | -0.068 | -0.056 |
| $N_{\mathrm{SIA}}$ | 0.002 | 0.001 | 1.35 | 0.178 | -0.001 | 0.004 |

## 6. ML MODEL COMPARISON

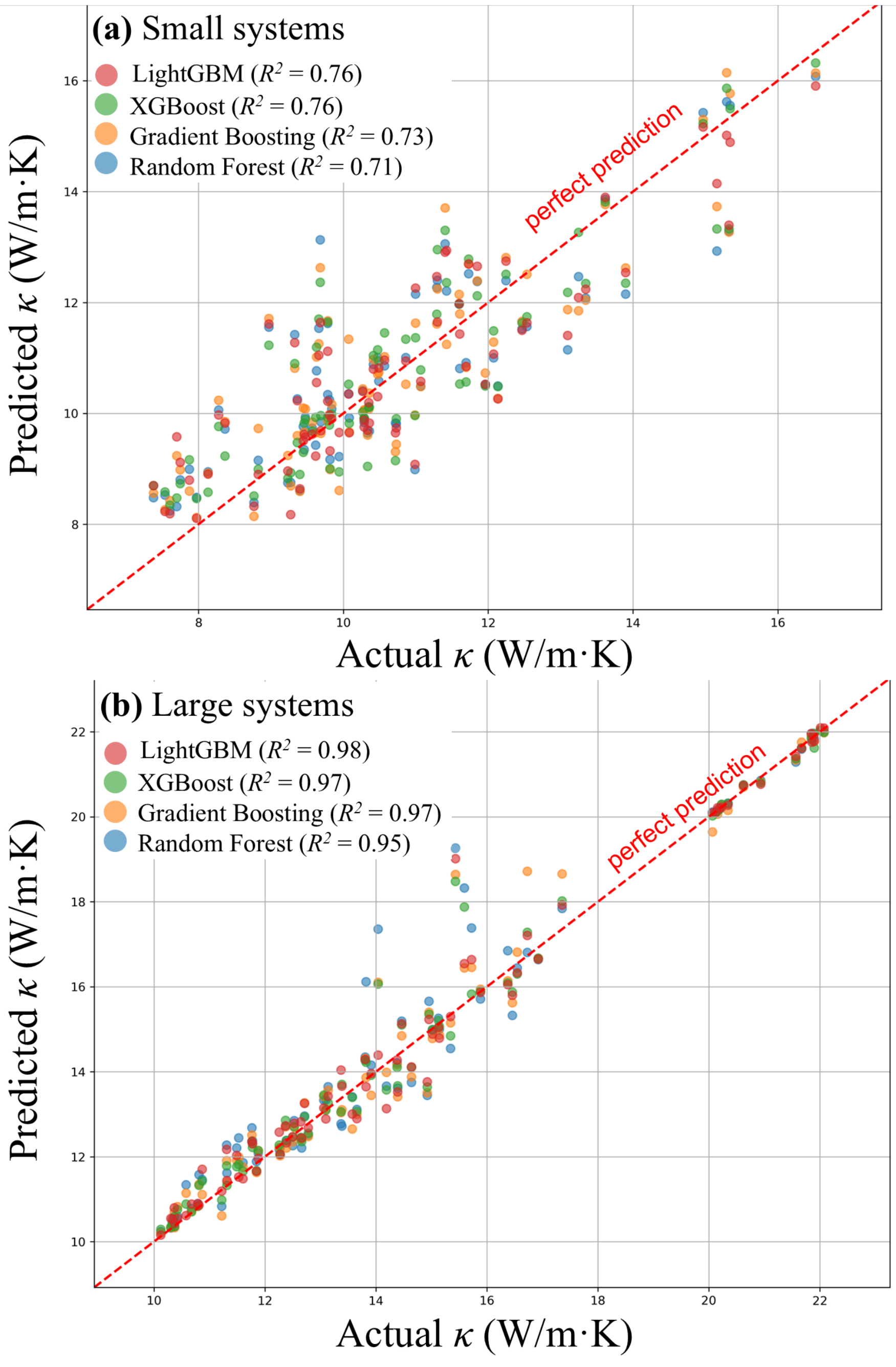


**FIG. S42**. Prediction of $\kappa$ for the different ML models in the (a) small and (b) large systems for 0.0 - 0.4 *cdpa*. With each re-running of the ML regression for testing, $R^2$ values varied slightly, with XGBoost or LightGBM always achieving the highest values, and the models performing better for the large systems than for the small systems.



\end{document}